\documentclass[prx,reprint,twocolumn,showkeys,superscriptaddress,noshowpacs,notitlepage,longbibliography,10pt]{revtex4-2}%
\usepackage{graphicx,bm,times}
\usepackage{amsmath}
\usepackage{mathtools}
\usepackage{amsfonts}
\usepackage{amssymb}
\usepackage{color}
\usepackage{hyperref}
\usepackage{braket}
\hypersetup{
	colorlinks = true,
	allcolors = {black}
}

\begin{document}

\title{{Exploring the energy spectrum of a four-terminal Josephson junction: Towards topological Andreev band structures}}

\author{Tommaso Antonelli}
\altaffiliation{These authors contributed equally}
\affiliation{IBM Research Europe—Zurich, 8803 R\"{u}schlikon, Switzerland}

\affiliation{Laboratory for Solid  State Physics, ETH Z\"{u}rich, 8093 Z\"{u}rich, Switzerland}

\author{Marco Coraiola}
\altaffiliation{These authors contributed equally}
\affiliation{IBM Research Europe—Zurich, 8803 R\"{u}schlikon, Switzerland}

\author{David Christian Ohnmacht}
\affiliation{Fachbereich Physik, Universität Konstanz, D-78457 Konstanz, Germany}

\author{Aleksandr E. Svetogorov}
\affiliation{Fachbereich Physik, Universität Konstanz, D-78457 Konstanz, Germany}

\author{Deividas Sabonis}
\affiliation{IBM Research Europe—Zurich, 8803 R\"{u}schlikon, Switzerland}

\author{Sofieke C. ten Kate}
\affiliation{IBM Research Europe—Zurich, 8803 R\"{u}schlikon, Switzerland}

\author{Erik Cheah}
\affiliation{Laboratory for Solid  State Physics, ETH Z\"{u}rich, 8093 Z\"{u}rich, Switzerland}

\author{Filip Krizek}
\affiliation{Laboratory for Solid  State Physics, ETH Z\"{u}rich, 8093 Z\"{u}rich, Switzerland}

\author{R\"{u}diger Schott}
\affiliation{Laboratory for Solid  State Physics, ETH Z\"{u}rich, 8093 Z\"{u}rich, Switzerland}

\author{Juan Carlos Cuevas}
\affiliation{Departamento de F\'{\i}sica Te\'orica de la Materia Condensada, Universidad Aut\'onoma de Madrid, 28049 Madrid, Spain}
\affiliation{Condensed Matter Physics Center (IFIMAC), Universidad Aut\'onoma de Madrid, 28049 Madrid, Spain}

\author{Wolfgang Belzig}
\affiliation{Fachbereich Physik, Universität Konstanz, D-78457 Konstanz, Germany}

\author{Werner Wegscheider}
\affiliation{Laboratory for Solid  State Physics, ETH Z\"{u}rich, 8093 Z\"{u}rich, Switzerland}

\author{Fabrizio Nichele}
\email{fni@zurich.ibm.com}
\affiliation{IBM Research Europe—Zurich, 8803 R\"{u}schlikon, Switzerland}

\begin{abstract}
	
Hybrid multiterminal Josephson junctions (JJs) are expected to harbor a novel class of Andreev bound states (ABSs), including topologically nontrivial states in four-terminal devices. In these systems, topological phases emerge when ABSs depend on at least three superconducting phase differences, resulting in a three-dimensional (3D) energy spectrum characterized by Weyl nodes at zero energy. Here, we realize a four-terminal JJ in a hybrid Al/InAs heterostructure, where ABSs form a synthetic 3D band structure. We probe the energy spectrum using tunneling spectroscopy and identify spectral features associated with the formation of a tri-Andreev molecule, a bound state whose energy depends on three superconducting phases and, therefore, is able to host topological ABSs. The experimental observations are well described by a numerical model. The calculations predict the appearance of four Weyl nodes at zero energy within a gap smaller than the experimental resolution. These topological states are theoretically predicted to remain stable within an extended region of the parameter space, well accessible by our device. These findings establish an experimental foundation to study high-dimensional synthetic band structures in multiterminal JJs, and to realize topological Andreev bands.

\end{abstract}

\keywords{Andreev bound state, multi-terminal Josephson junction, superconducting devices, Weyl nodes}

\date{\today}
\maketitle
\section{Introduction}

Josephson junctions (JJs) are key elements of superconducting circuits, used in quantum technology applications and fundamental research. In a superconducting-normal conductor-superconducting junction, supercurrent transport is mediated by Andreev bound states (ABSs), electron-hole superposition states confined within the normal area. The ABS energy depends on the phase difference between the superconducting wave functions of the two leads \cite{Andreev1964,Beenakker1991,Beenakker1991b,Furusaki1991}. Recently, these electronic modes have been the subject of extensive study \cite{Pillet2010,Bretheau2013a,Bretheau2013b,Lee2014,VanWoerkom2017,Hays2020,Tosi2019,Nichele2020}, including the coherent manipulation of ABSs \cite{Janvier2015,Hays2018,Hays2021,PitaVidal2023} and the exploration of topological superconductivity \cite{Mourik2012, Nichele2017, Fornieri2019, Ren2019}.

In multiterminal JJs (MTJJs), where three or more superconducting terminals are linked to a single normal scattering region, ABSs form synthetic band structures which are expected to host a wide range of properties not attainable in two-terminal devices. Among the most intriguing prospects is the potential to realize topologically nontrivial phases in the three-dimensional (3D) band structure of four-terminal JJs (4TJJs), with Weyl nodes arising in the energy spectrum \cite{Riwar2016,Eriksson2017,Xie2018,Klees2020,Xie2022,Repin2022,Teshler2023a}. Topological phases in these systems are inherently robust with respect to external perturbations \cite{Riwar2016}, making them particularly appealing for applications in quantum information processing \cite{Chen2021b,Boogers2022} and spintronics \cite{Chen2021}.

A first set of studies on MTJJs focused on their transport properties, including the signatures of Cooper pair quartets \cite{Freyn2011,Jonckheere2013,Pfeffer2014,Cohen2018,Huang2022} and the flow of supercurrents across multiple superconducting leads \cite{Draelos2019,Graziano2020,Pankratova2020,Arnault2021,Graziano2022,Gupta2023,Coraiola2024}.
Recently, MTJJs have been proposed as a platform to realize Andreev molecules---a system where ABSs hybridize due to the spatial overlap of their wave functions \cite{Pillet2019,Kornich2019,Keliri2023,Kocsis2023,Johannsen2024}, resulting in delocalized states that extend across all leads and exhibit nonlocal Josephson effect \cite{Pillet2019,Matsuo2022,Haxell2023,Matsuo2023b,Prosko2024}. In Andreev molecules, two primary coherent transport processes occur: double elastic cotunneling and double crossed Andreev reflection \cite{Deutscher2000,Freyn2011,Pillet2019}, both essential for the generation of Cooper pair multiplets \cite{Freyn2011,Jonckheere2013,Ohnmacht2024} and for engineering Kitaev chains \cite{Kitaev2001} in quantum dot arrays \cite{Sau2012,Leijnse2012,Fulga2013,Liu2022,Wang2022,Bordin2023,Bordin2024,vanDriel2024,Bordin2024b,Dvir2023,tenHaaf2024}.
Detailed understanding of Andreev band structures in multiterminal devices can be gained through local spectroscopy, which has been employed to probe hybridized ABSs \cite{Coraiola2023,Matsuo2023}, as well as spin-split energy levels and ground state parity transitions \cite{vanHeck2014,Coraiola2024b}. In these experiments, phase biasing allowed the exploration of ABS spectra as a function of up to two superconducting phase differences \cite{Lee2022,Coraiola2023,Coraiola2024b}. However, realizing topological phases with nontrivial Chern numbers strictly requires independent tuning of three phase degrees of freedom \cite{Riwar2016,Xie2017,Meyer2017,Klees2020}, a challenge that remains to be addressed.

\begin{figure*}[t!]
	\centering
	\includegraphics[width=\linewidth]{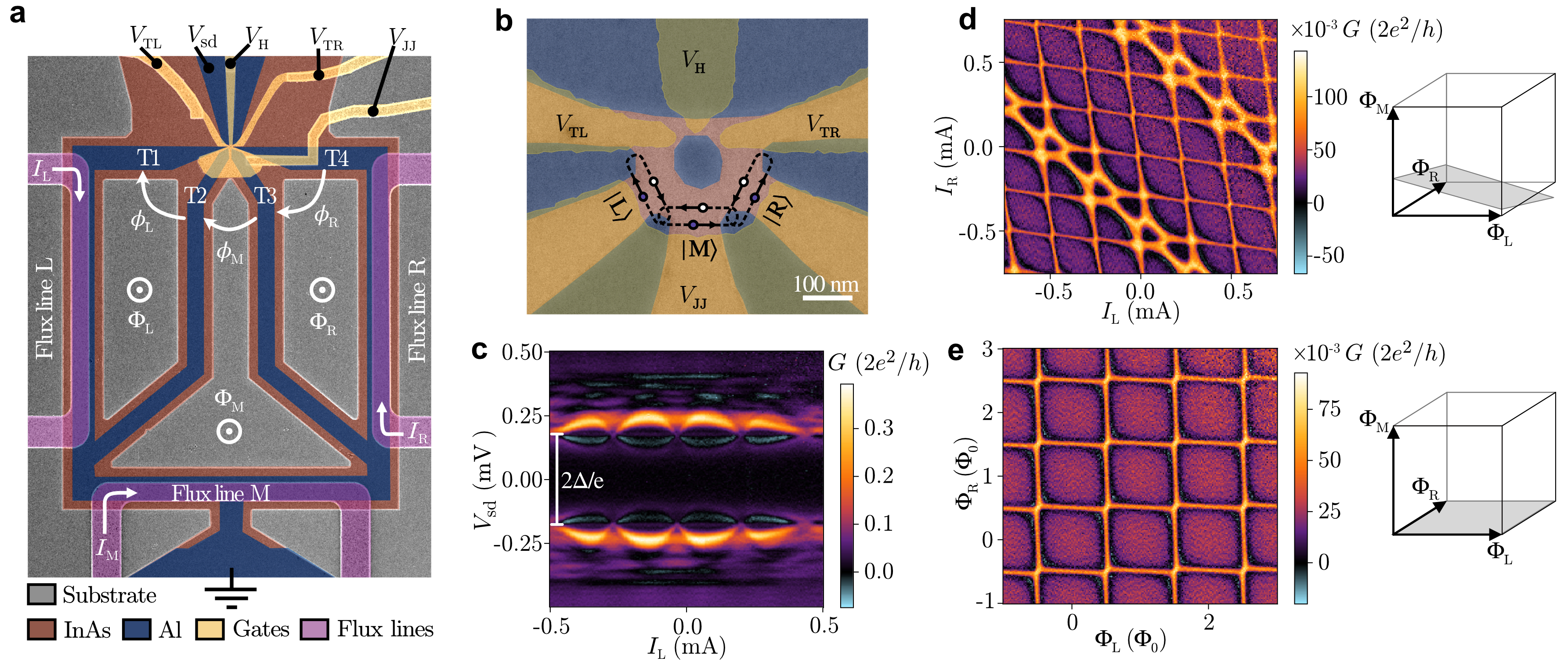}
	\caption{(a) False-colored scanning electron micrograph of the four-terminal device under study, showing the three-loop geometry. (b) Close-up view of the scattering area taken prior to the gate deposition. Gates structures are drawn on the image (yellow). (c) Tunneling spectroscopy measurement, showing the differential conductance $G$ as a function of bias $V_{\rm{sd}}$ and the current $I_{\rm{L}}$ flowing in flux line L. (d) Differential conductance map measured at $V_{\rm{sd}} = -175~\mu \rm{V}$ as a function of the flux-line currents $I_{\rm{L}}$ and $I_{\rm{R}}$. (e) As in (d), but measured as a function of the magnetic fluxes $\Phi_{\rm{L}}$ and $\Phi_{\rm{R}}$ after the current-to-flux remapping (see text). The schematics on the right of (d) and (e) represent the orientations of the corresponding maps with respect to the cubic unit cell in the 3D flux space.}
	\label{fig1}
\end{figure*}

In this work, we realize a 4TJJ where three phases are independently controlled through flux biasing. We probe the ABS energy spectrum of the system across the entire 3D phase space using tunneling spectroscopy. Moreover, we observe the simultaneous hybridization of three ABSs, i.e., the formation of a so called tri-Andreev molecule, when the three phases are tuned close to $\pi$.
Our findings are supported by a theoretical model, which qualitatively reproduces the main features of the measured Andreev spectra. Furthermore, our simulations indicate that the Andreev bands undergo a phase-controlled topological transition in which hybridization induces a band inversion accompanied by the appearance of Weyl nodes. Due to the finite resolution of the tunneling spectroscopy, with a linewidth of $\sim 15~\mu $eV, the gapless states (Weyl nodes) cannot be experimentally distinguished from the gapped states. Finally, we study the robustness of the topological phase under variations of experimentally addressable parameters, finding that the regime best describing our device is well within the topological region.
Overall, our work provides access to Andreev band structures in three synthetic dimensions, creating an experimental platform and practical guidelines for the realization of topological states in hybrid multiterminal devices.

\section{Experimental setup and 3D phase control}
The device under study, shown in Fig.~\ref{fig1}(a), consists of a 4TJJ embedded in a triple-loop geometry. It is realized in an InAs/Al heterostructure \cite{Shabani2016, Cheah2023} where the epitaxial Al layer is selectively removed to expose the III–V semiconductor below. Three flux-bias lines are patterned on top of a uniform dielectric layer to generate the external magnetic fluxes $\Phi_i$ $(i = \rm{L,M,R})$ threading the three interconnected superconducting loops (L, M, R). This enables us to control the phase differences $\phi_i$ between the four terminals (T1-T4). The latter couple with a common semiconducting region [see Fig.~\ref{fig1}(b)] where a superconducting island with diameter $\sim90$ nm is left at its center to partially screen the probe gate voltages. By design, the minimum distance between neighboring terminals is 50 nm, while the distance between T1 and T4 is 220 nm. All these lengths are small in comparison with the superconducting coherence length in the InAs 2DEG, estimated to be $600~\rm{nm}$ \cite{Haxell2023}. Four gate electrodes on the dielectric layer are energized by voltages $V_g$ ($g \in$  \{TL, TR, H, JJ\}), allowing for electrostatic tuning of the electron density in the InAs layer below. Tunneling spectroscopy of the scattering area is performed by measuring the differential conductance $G$ across a tunneling barrier, formed by depleting the InAs region below the gates $V_{\rm{TL}}$ and $V_{\rm{TR}}$ (see Supplemental Material \cite{SM}, Sec.~I for additional details). The device was measured in a dilution refrigerator with a base temperature of about $10~\mathrm{mK}$ using lock-in techniques, with a DC voltage bias $V_{\rm{sd}}$ and an AC voltage bias of amplitude $\delta V_{\rm{sd}} = 3~\mu \mathrm{V}$ applied between the probe and the four terminals. More information about materials, fabrication and measurement setup is available in Ref.~\cite{Coraiola2023}.

\begin{figure*}
	\centering
	\includegraphics[width=\linewidth]{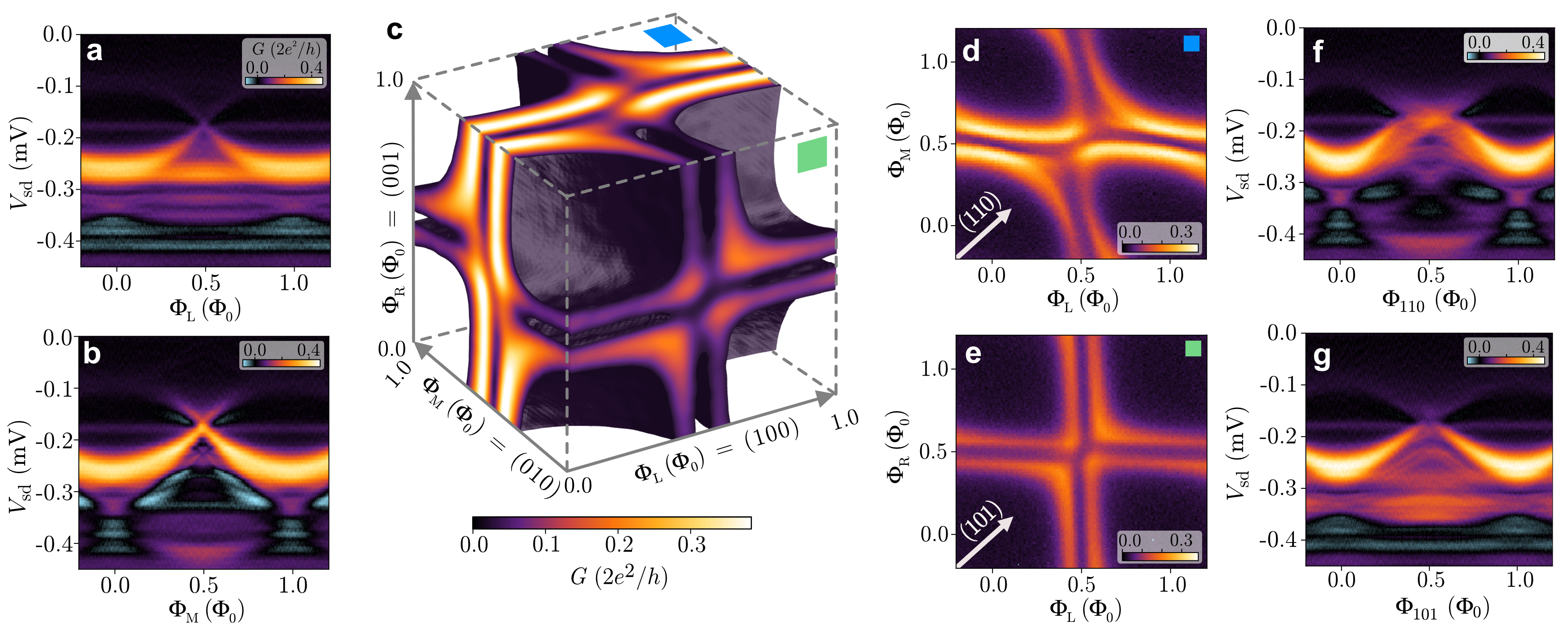}
	\caption{(a) Tunneling differential conductance $G$ measured as a function of DC voltage bias $V_{\rm{sd}}$ and magnetic flux $\Phi_{\rm{L}}$, with $\Phi_{\rm{M}} = \Phi_{\rm{R}} = 0$. (b) As in (a), but varying $\Phi_{\rm{M}}$ with $\Phi_{\rm{L}} = \Phi_{\rm{R}} = 0$. (c) Tunneling differential conductance measured at fixed $V_{\rm{sd}} = -200$ $\mu \rm{V}$ and plotted in the 3D flux unit cell for values $G \geq  0.095 \times 2e^2/h$. The unit cell axes are also labeled using the crystallographic-like notation used to define specific directions in the flux space. (d,e) Flux-flux maps of two cube faces [as indicated by the colored squares in (c)], measured at $V_{\rm{sd}} = -200$ $\mu \rm{V}$, showing avoided crossing induced by the formation of bi-Andreev molecules. (f,g)~$G$ measured as a function of $V_{\rm{sd}}$ along the directions (110) and (101), as indicated by the white arrows in (d) and (e), respectively.}
	\label{fig2}
\end{figure*}

Figure \ref{fig1}(c) shows a typical tunneling spectrum measured by varying the flux-line current $I_{\rm{L}}$ and the DC bias $V_{\rm{sd}}$. The flux-dependent ABS spectrum is visible outside a transport gap of $350~\mu \rm{V}  = 2\Delta/e$, that is due to the superconducting probe ($\Delta$ is the superconducting gap). Two flux-independent conductance peaks at $V_{\rm{sd}} = \pm \Delta /e$ highlight the probe gap edges and are attributed to multiple Andreev reflection processes. Assuming a BCS-like density of states (DOS) for the superconducting probe with peaks at energy $\pm \Delta$, $G$ is expected to have a resonance at a voltage $\pm (\Delta + E)/e$ when a peak is present in the DOS of the scattering area at energy $E$. Consequently, spectroscopic features observed at $eV_{\rm{sd}} = \pm \Delta/e = \pm 175~\mu \rm{eV}$ correspond to DOS peaks at zero energy ($E=0$) in the scattering area. When half of a superconducting flux quantum ($\Phi_0 = 2h / e$) induced by the left flux line penetrates loop L, the superconducting phase difference $\phi_{\rm{L}}$ is tuned to $\pi$ and the ABS energy approaches zero energy as for a highly transparent two-terminal JJ \cite{Beenakker1991}. Notably, the dispersion shows an additional modulation with a larger periodicity in $I_{\rm{L}}$, which is evident in the map shown in Fig.~\ref{fig1}(d) measured at constant $V_{\rm{sd}} =  - \Delta/e$ by varying $I_{\rm{L}}$ and $I_{\rm{R}}$. Here, we observe a slower modulation along the diagonal direction $I_{\rm{L}} = I_{\rm{R}}$, caused by the magnetic flux generated by lines L and R impinging through the middle loop.
A useful way to navigate within the 3D flux space is to consider the cubic unit cell defined by the three independent magnetic fluxes $\Phi_{\rm{L}}$, $\Phi_{\rm{M}}$, $\Phi_{\rm{R}}$, as schematically illustrated in the insets of Fig.~\ref{fig1}(d,e). Within this framework, the $I_{\rm{L}}$-$I_{\rm{R}}$ map follows a tilted plane sketched in gray, whose orientation is defined by the $3\times3$ mutual inductance matrix $M$:

\begin{equation}
\Phi = \begin{pmatrix}
\Phi_{\rm{L}} \\
\Phi_{\rm{M}} \\
\Phi_{\rm{R}}
\end{pmatrix}
=
M
\begin{pmatrix}
I_{\rm{L}} \\
I_{\rm{M}} \\
I_{\rm{R}}
\end{pmatrix}
+ \Phi_{(0,0,0)} ,
\label{EQ:phase virt}
\end{equation}
where $\Phi_{(0,0,0)}$ is an offset defining the corner of a unit cell.
In order to cut the unit cell in a controlled way, we compensate for the cross-coupling between loops and flux lines by simultaneously setting the three currents $I_{\rm{L}}$, $I_{\rm{M}}$ and $I_{\rm{R}}$ needed to reach a flux point $\Phi$, according to Eq.~(\ref{EQ:phase virt}). Figure \ref{fig1}(e) shows the differential conductance map measured in this way by sweeping the fluxes $\Phi_{\rm{L}}$ and $\Phi_{\rm{R}}$ and keeping $\Phi_{\rm{M}} = 0$. As a result, a periodic square net is obtained, demonstrating independent flux control over the three loops. More details about the current-to-flux remapping and how to extract the mutual inductance matrix elements are provided in Ref.~\cite{SM}, Sec.~II.

\begin{figure*}[t]
	\centering
	\includegraphics[width=\linewidth]{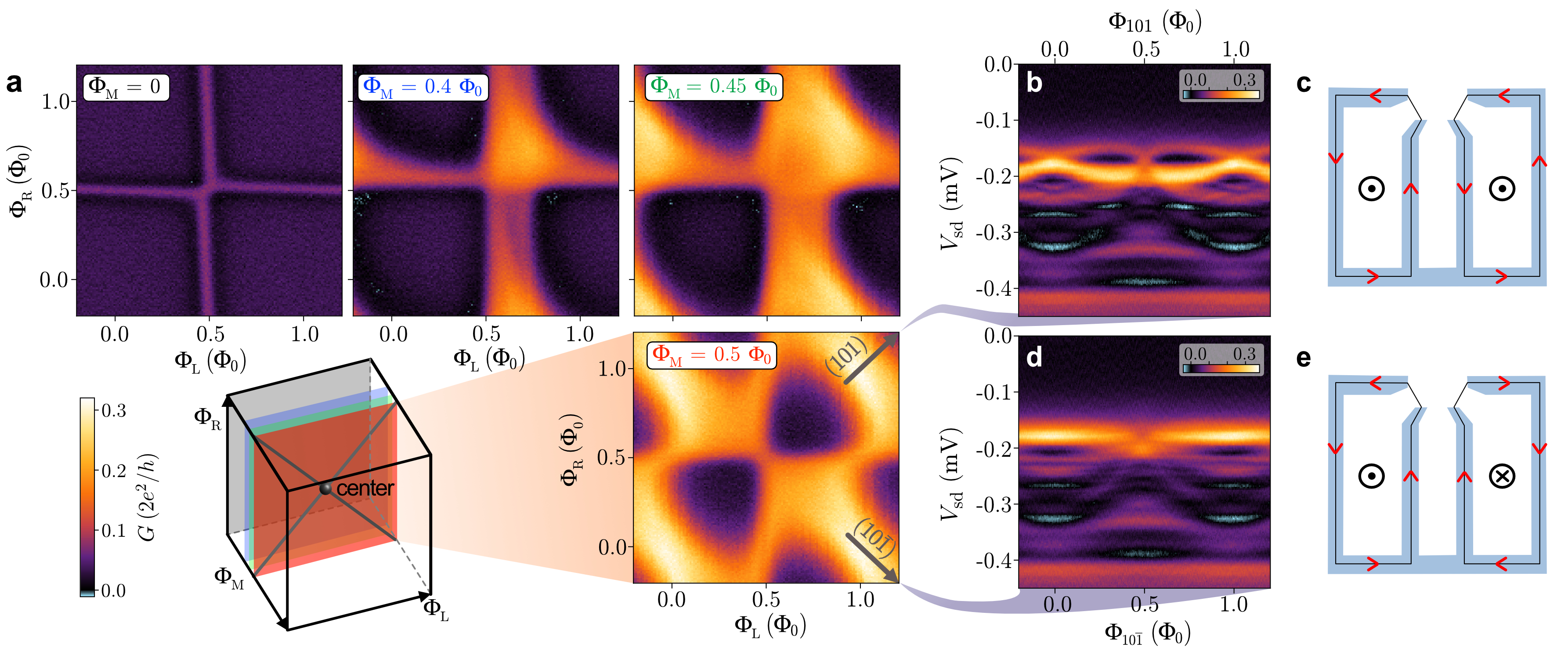}
	\caption{(a) Slicing the cubic unit cell (see schematic) by measuring $\Phi_{\rm{L}}$-$\Phi_{\rm{R}}$ maps at different values of $\Phi_{\rm{M}}$, with voltage bias $V_{\rm{sd}}$ kept fixed at $-175~\mu \rm{V}$. (b,c) Cuts measured as a function of $V_{\rm{sd}}$ along the $(101)$ and $(10\overline{1})$ direction (see gray arrows in (a) at $\Phi_M = 0.5~\Phi_0$). In the schematics on the right of the plots, the currents flowing in the two outer loops of device are sketched, revealing that mutual inductive effects on the center loop are expected along $\Phi_{101}$ and not along $\Phi_{10\overline{1}}$.}
	\label{fig3}
\end{figure*}

\section{ABS energy dispersion in the 3D flux space}
Having established a measurement protocol that allows for independent flux control, we systematically map the ABS energy spectrum in the 3D flux space. Owing to the periodicity of the spectrum, we can restrict our investigation to a single 3D flux unit cell. We start with a simple case where we sweep the flux threading a loop while keeping the other two fluxes at zero, as shown in Fig.~\ref{fig2}(a) for varying $\Phi_{\rm{L}}$. The energy spectrum reveals the dispersion of a highly transparent ABS, that we identify as the mode formed between the terminal pair T1-T2.
The large but finite transparency is expected to prevent the band from reaching zero energy, forming a minigap between the electron- and hole-like branches of the ABS spectrum~\cite{Beenakker1991b}. However, this small energy gap is not experimentally revolved due to the sizable spectral broadening, estimated to be $15~\mu$eV (see Ref.~\cite{SM}, Sec.~III).
Notably, the ABS dispersion does not reach the superconducting gap edge, but it is reduced to $\sim 0.4 \Delta/e$, potentially due to the repulsion with the lower-transmission states visible at $V_{\rm{sd}} < -0.3$ mV. Figure~\ref{fig2}(b) shows the dispersion of the ABSs formed between T2-T3, which are tuned by sweeping $\Phi_{\rm{M}}$. Similar to the previous case, the spectrum is dominated by a brighter highly transparent mode having the same reduced dispersion, and a low-transmission manifold of states at lower energies. The dispersion of the modes formed between T3 and T4 (shown in Ref.~\cite{SM}, Sec.~III) is qualitatively equivalent to the one along $\Phi_{\rm{L}}$. In the following, we focus on the three high-transmission ABSs that we label $|\rm{L}\rangle$, $|\rm{M}\rangle$ and $|\rm{R}\rangle$ as shown in Fig.~\ref{fig1}(b).

To have an overview on how such states disperse in the flux space, we map the differential conductance $G$ fixed at $V_{\rm{sd}} = -\Delta/e-25~\mu \rm{V}$ within the whole 3D unit cell. In Fig.~\ref{fig2}(c) we plot $G$ for values larger than $0.095 \times 2e^2/h$, i.e., where the DOS is nonzero. At this $V_{\rm{sd}}$ value, the Andreev dispersions are cut twice around $\Phi_i = 0.5~\Phi_0$, forming pairs of parallel conductance lines along the cube faces. At the center of each face, the ABSs do not cross each other, but they rather interact opening avoided crossings, as highlighted in the face maps of Fig.~\ref{fig2}(d,e). These avoided crossings are spectral signatures characteristic of bi-Andreev molecular states formed by the hybridization of two ABSs, which have recently been observed in three-terminal devices \cite{Coraiola2023}. Notably, the four-terminal device presented here acts as an effective three-terminal system on the cube faces, i.e., when one flux is kept fixed to zero. 

To better define directions in the 3D flux space, we introduce a crystallographic-like notation, where the three flux axes ($\Phi_{\rm{L}}$, $\Phi_{\rm{M}}$ and $\Phi_{\rm{R}}$) are denoted as $(100)$, $(010)$, and $(001)$, respectively. The hybridization lifts the degeneracy of the original two-terminal ABSs, splitting the dispersion in two bands as observed in the spectra in Figs.~\ref{fig2}(f) and \ref{fig2}(g), measured along the directions (110) and (101) (white arrows in Figs.~\ref{fig2}(d) and \ref{fig2}(e), respectively). A larger splitting is observed along the (110) direction (f) compared to (101) cut (g), indicating a stronger coupling between the nearest neighbor ABSs $\mid$L$\rangle$ and $\mid$M$\rangle$. A weaker hybridization is instead expected between $\mid$L$\rangle$ and $\mid$R$\rangle$, due to their smaller wavefunction overlap. In Fig.~S4 \cite{SM}, additional tunneling spectra show that the energy splitting is much smaller along the perpendicular directions $\Phi_{10\overline{1}}$ and $\Phi_{1\overline{1}0}$, as expected when the interacting ABSs have opposite phases ~\cite{Pillet2019, Johannsen2024,Coraiola2023}. Thus, our observations reveal a significant hybridization between all three ABSs, which couple in pairs to form bi-Andreev molecules on each face of the cubic unit cell.

\begin{figure*}[t]
	\centering
	\includegraphics[width=\linewidth]{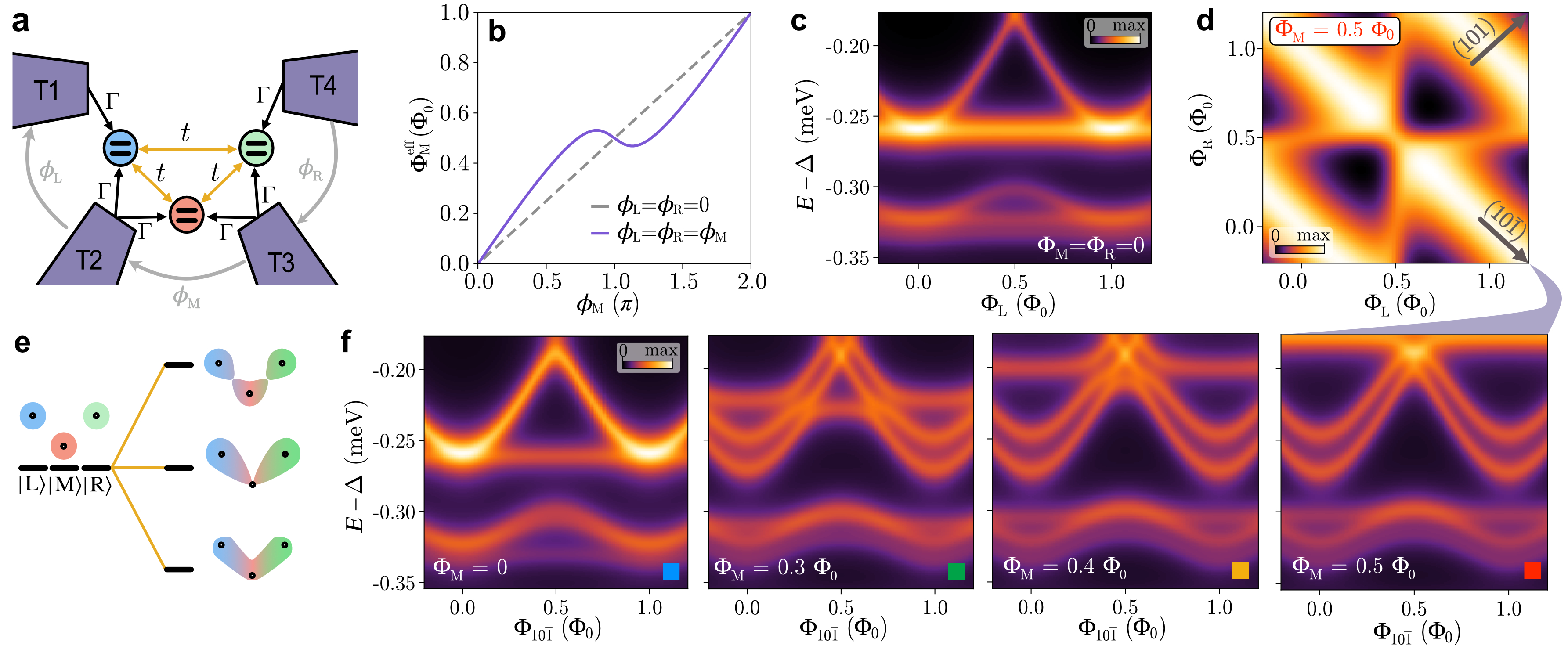}
	\caption{(a) Schematic of the model comprising three two-level quantum dots (blue, red, green) coupled to each other and to four superconducting terminals (T1-T4). Andreev bound states (ABSs) form in the quantum dots and have energies dispersing as a function of the superconducting phase differences $\phi_i$. The model parameters (see definitions in the text) are: $\epsilon = 0.005\Delta$, $\Gamma = 0.37\Delta$ and $t= 0.12\Delta$. (b) Phase-effective flux relation including mutual inductive coupling between the loops when the other phases $\phi_{\rm{L}}$ and $\phi_{\rm{R}}$ are kept constant to zero (dashed gray line) and when they are equal to $\phi_{\rm{M}}$ (solid purple line). (c) Simulated energy spectrum along $\Phi_{\rm{L}}$, with $\Phi_{\rm{M}} = \Phi_{\rm{R}} = 0$. The energy axis $E$ is shifted by $\Delta = 175~\mu \rm{eV}$ to align to the experimental data. (d) Simulated flux-flux map at $E = 0$ corresponding to the measurement shown in Fig.\ref{fig3}(a) for $\Phi_{\rm{M}} = 0.5~\Phi_0$. (e) Conceptual illustration of the hybridization among three degenerate energy levels resulting in bonding, nonbonding and antibonding states separated by energy gaps, as in a tri-atomic molecule. (f) Simulated energy spectra along $\Phi_{10\overline{1}}$ for different $\Phi_{\rm{M}}$ values, showing that hybridization between ABSs opens energy gaps at the crossing points between different levels. The colored squares refer to the arrows in Fig.~\ref{fig5}(f).}
	\label{fig4}
\end{figure*}

\section{Exploring the center of the unit cell}
The device configurations discussed so far reproduce the behavior of either two-terminal devices (when two phase differences are kept to zero, i.e., along the unit cell edges) or three-terminal ones (along the unit cell faces, where only one phase difference is kept to zero). Inside the unit cell, all the phase differences are nonzero, leading to a more complex ABS spectrum achievable only with four or more leads. To explore such configurations, we slice the cubic cell from the $\Phi_{\rm{L}}$-$\Phi_{\rm{R}}$ face to the center along parallel planes measured at different $\Phi_{\rm{M}}$ values and at $V_{\rm{sd}} = -\Delta/e$, as shown in Fig.~\ref{fig3}(a). At this $V_{\rm{sd}}$ value, we probe the energy spectrum near the maxima of the ABSs, resulting in one conductance line per state. By increasing $\Phi_{\rm{M}}$, the conductance becomes asymmetric around the center of the plot and develops a maximum at $\Phi_{\rm{L}} = \Phi_{\rm{R}} = 0.7~\Phi_0$. More detailed discussions of this figure are presented in Ref.~\cite{SM}, Sec.~V. Moving $\Phi_{\rm{M}}$ further to $0.5~\Phi_0$, the map recovers its inversion symmetry, featuring two lobes of low conductance around the center. In this configuration, one would expect to measure constant conductance across the entire plane, since $|\rm{M}\rangle$ is fixed at its energy maximum. Instead, the energy cut along $\Phi_{101}$ [Fig.~\ref{fig3}(b)] reveals that the state $|\rm{M}\rangle$ has a relatively weak dispersion just below $V_{\rm{sd}} = -0.175$ mV. 

We explain the weak influence of $\Phi_{\rm{L,R}}$ on this ABSs in terms of mutual inductive coupling between loops. When $\Phi_{\rm{L}}$ and $\Phi_{\rm{R}}$ are swept in-phase ($\Phi_{101}$), two opposite currents flow along the two long sides of loop M, as shown in Fig.~\ref{fig3}(c). These currents induce two parallel flux contributions to $\Phi_{\rm{M}}$, providing an additional phase difference between T2 and T3. When $\Phi_{\rm{L}}$ and $\Phi_{\rm{R}}$ are swept with opposite sign, i.e., along $\Phi_{10\overline{1}}$, the two currents flow in the same direction as sketched in Fig.~\ref{fig3}(e), and the two induced fluxes cancel each other out. Indeed, the spectrum measured along this direction [Fig.~\ref{fig3}(d)] shows that $|\rm{M}\rangle$ forms a flat band independent of the other two fluxes just below $V_{\rm{sd}} = -0.175$ mV.

In Fig.~\ref{fig3}(d), we also observe two dispersive bands representing the $|\rm{L}\rangle - |\rm{R}\rangle$ hybridized states and having their maxima at $\Phi_{10\overline{1}} \sim 0.5\Phi_0$. Here, they interact with the $|\rm{M}\rangle$-derived flat band which significantly decreases its energy to $V_{\rm{sd}} \sim -0.2$ mV. This indicates that an additional gap between the electron- and hole-like branches of the overall ABS spectrum is opened in addition to the minigap formed by the finite junction transparency. In the following, we show that this spectral feature marks the hybridization among three two-terminal ABSs occurring when they are tuned to the same energy, i.e., the formation of a tri-Andreev molecule.

\begin{figure*}
	\centering
	\includegraphics[width=\linewidth]{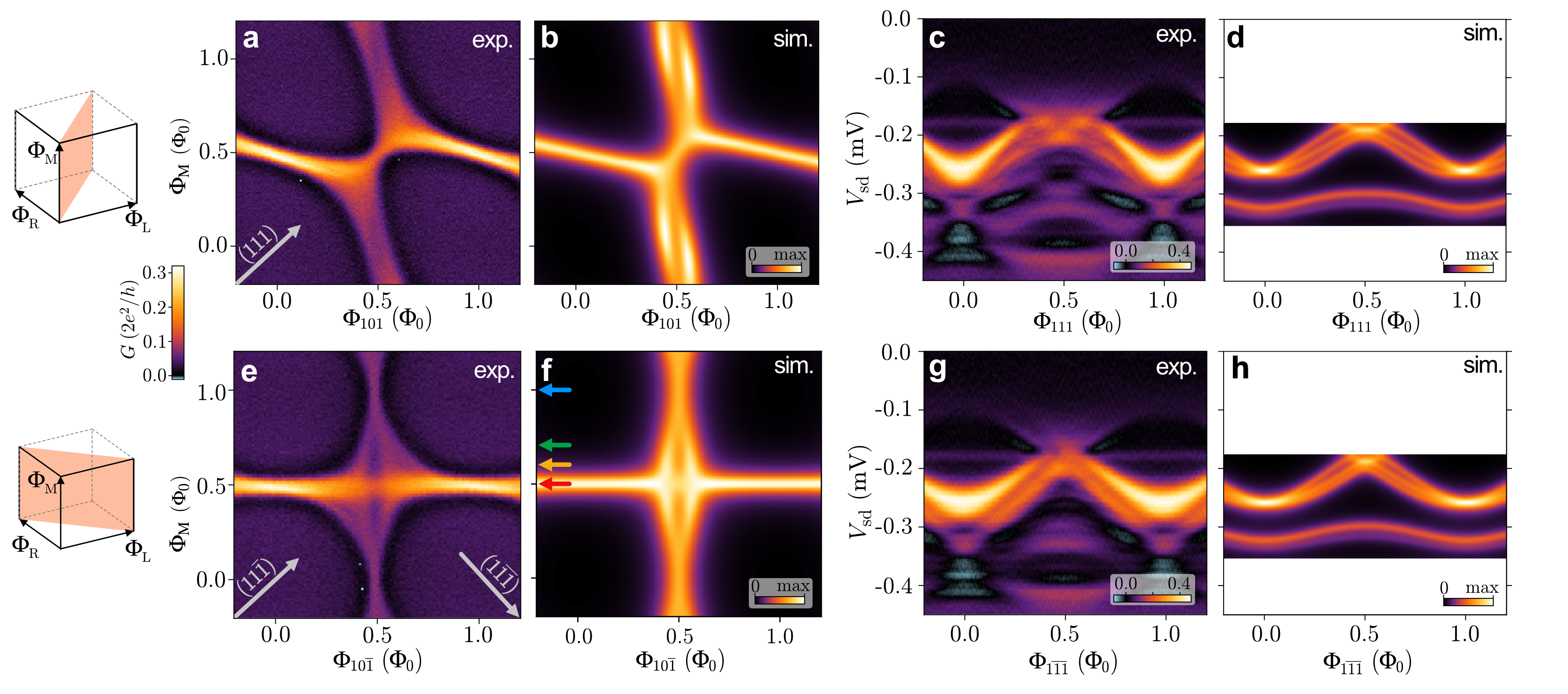}
	\caption{Experimental (a) and simulated (b) zero-energy plane (at fixed $V_{\rm{sd}}=-175~\mu \rm{V}$ or $E=0$, respectively) along a diagonal of the cubic unit cell (top-left schematic). Experimental (c) and simulated (d) energy spectrum along the diagonal direction of (a,b), i.e., along the flux direction $\Phi_{111}$ [gray arrow in (a)]. (e,f) As in (a,b), but along another diagonal plane of the cubic unit cell $\Phi_{10 \overline{1}}$-$\Phi_{\rm{M}}$ (bottom-left schematic). Experimental (g) and simulated (h) energy spectrum along the flux direction $\Phi_{1\overline{1}\overline{1}}$ [gray arrow in (e)]}
	\label{fig5}
\end{figure*}

\section{Theoretical model}
To better understand how the hybridization between ABSs reshapes the complex Andreev band structure observed in the previous paragraph, we develop a theoretical model schematically represented in Fig.~\ref{fig4}(a). The model features four superconducting leads coupled to a normal scattering region described by means of three coupled quantum dots. The superconducting phase differences $\phi_{\rm{L,M,R}}$ (or equivalently magnetic fluxes $\Phi_{\rm L,M,R} = \Phi_0\phi_{\rm L,M,R}/2\pi$) are defined between the leads using the same convention introduced in Fig.~\ref{fig1}(a). Each dot contains two noninteracting spin-degenerate levels of energy $\epsilon^{\pm}_{i}$ ($i=1,2,3$), and is coupled to the neighboring leads with a coupling strength characterized by the parameter $\Gamma$, as well as to the other dots, described by the parameter $t$. All dot-lead couplings and interdot couplings are assumed to be equal. We note that the choice of a quantum dot network provides a convenient description of a normal region hosting discrete ABSs, with the degree of hybridization among them determined by $t$. No Coulomb interaction or charging energy are introduced in the system, as the dots are strongly coupled to the leads. To compare with the experimental data, we compute the total DOS projected onto the three dots as a function of the energy $E$ and the three superconducting phase differences $\phi_{\rm{L,M,R}}$ (fluxes $\Phi_{\rm L,M,R}$), indicated in Fig.~\ref{fig4}(a). Furthermore, to include the effect of the mutual inductive coupling between the superconducting loops, which causes a nonlinear cross-dependence between the phases (magnetic fluxes), we remap them to effective phases $\phi_i^{\rm eff}$ (effective magnetic fluxes $\Phi_i^{\rm eff}$) using the relations:

\begin{gather*}
	\phi_{\rm{L,R}}^{\rm eff} =  \left[ \phi_{\rm{L,R}} + \alpha f\left(\phi_{\rm{M}}\right) \right] \\
	\phi_{\rm{M}}^{\rm eff} =  \left\{ \phi_{\rm{M}} + \alpha \left[ f(\phi_{\rm{L}}) + f(\phi_{\rm{R}})\right] \right\},
\end{gather*}
where $\phi_{i}^{\rm eff}/2\pi = \Phi_{i}^{\rm eff}/\Phi_0$ for $i = {\rm L,M,R}$. Here, $f(\phi) = -2\partial_{\phi}\left[1-\tau \sin^2(\phi/2)\right]^{-1/2} $, $\alpha = 0.2$ is the strength of the mutual coupling and $\tau = 0.9$ introduces nonsinusoidal character to $f(\phi)$. Additional details on the model are discussed in Ref.~\cite{SM}, Secs.~VI and VII. Figure \ref{fig4}(b) illustrates the influence of $\phi_{\rm{L,R}}$ on $\Phi_{\rm{M}}^{\rm eff}(\phi_{\rm{M}})$ as a consequence of the mutual coupling, highlighting a nonlinear behavior when $\phi_{\rm{L,R}}$ are varied together with $\phi_{\rm{M}}$.

The simulated DOS as a function of energy and of $\Phi_{\rm{L}}$ while $\Phi_{\rm{M}}=\Phi_{\rm{R}}=0$ is shown in Fig.~\ref{fig4}(c). The two energy levels in each of the three dots give rise to two distinct manifolds comprising three modes each. As expected, only one state per manifold has a significant energy dispersion, while the others remain mostly constant in energy. Similar to the tunneling spectra in Fig.~\ref{fig2}(a,b), the simulated band structure exhibits one resonance that approaches $E=0$ at $\Phi_{\rm{L}}=0.5\Phi_0$, forming a cusp consistent with an isolated high-transmission ABS.
Figure \ref{fig4}(d) displays the simulated $\Phi_{\rm{L}}$-$\Phi_{\rm{R}}$ plane for constant flux $\Phi_{\rm M}= 0.5\Phi_0$ and constant energy $E=0$, which corresponds to the measurement shown in Fig.~\ref{fig3}(a). The presence of two lobes of minimum DOS around $\Phi_{\rm{L}}=\Phi_{\rm{R}}=0.5\Phi_0$ is captured by the model as a result of the mutual coupling between the phases.

To focus on the effects of ABS hybridization, we consider the $\Phi_{10\overline1}$ direction (in which mutual inductance effects are negligible) and simulate the Andreev spectra as a function of $E$ for different values of $\Phi_{\rm{M}}$ [Fig.~\ref{fig4}(f)]. At $\Phi_{\rm{M}} = 0$, $\mid$M$\rangle$ remains flat at high energy, while both $\mid$R$\rangle$ and $\mid$L$\rangle$ form dispersing bands which are nearly degenerate in energy. By increasing $\Phi_{\rm{M}}$, the flat state approaches zero energy and forms avoided crossings with the dispersing states. These three nondegenerate ABSs resemble the energy levels of a tri-atomic molecule, where three molecular orbitals (bonding, nonbonding and antibonding) are split in energy, as conceptually depicted in Fig.~\ref{fig4}(e). At $\Phi_{\rm{M}}=0.5\Phi_0$, the state $\mid$M$\rangle$ reaches the energy closest to zero and remains constant around that energy. Here, its dispersion bends downwards from zero energy as observed in the corresponding measurement in Fig.~\ref{fig3}(c). These results support the interpretation that our device hosts three ABSs hybridizing among each other, forming a tri-Andreev molecule delocalized over the four superconducting terminals.

\begin{figure*}[t!]
	\centering
	\includegraphics[width=\linewidth]{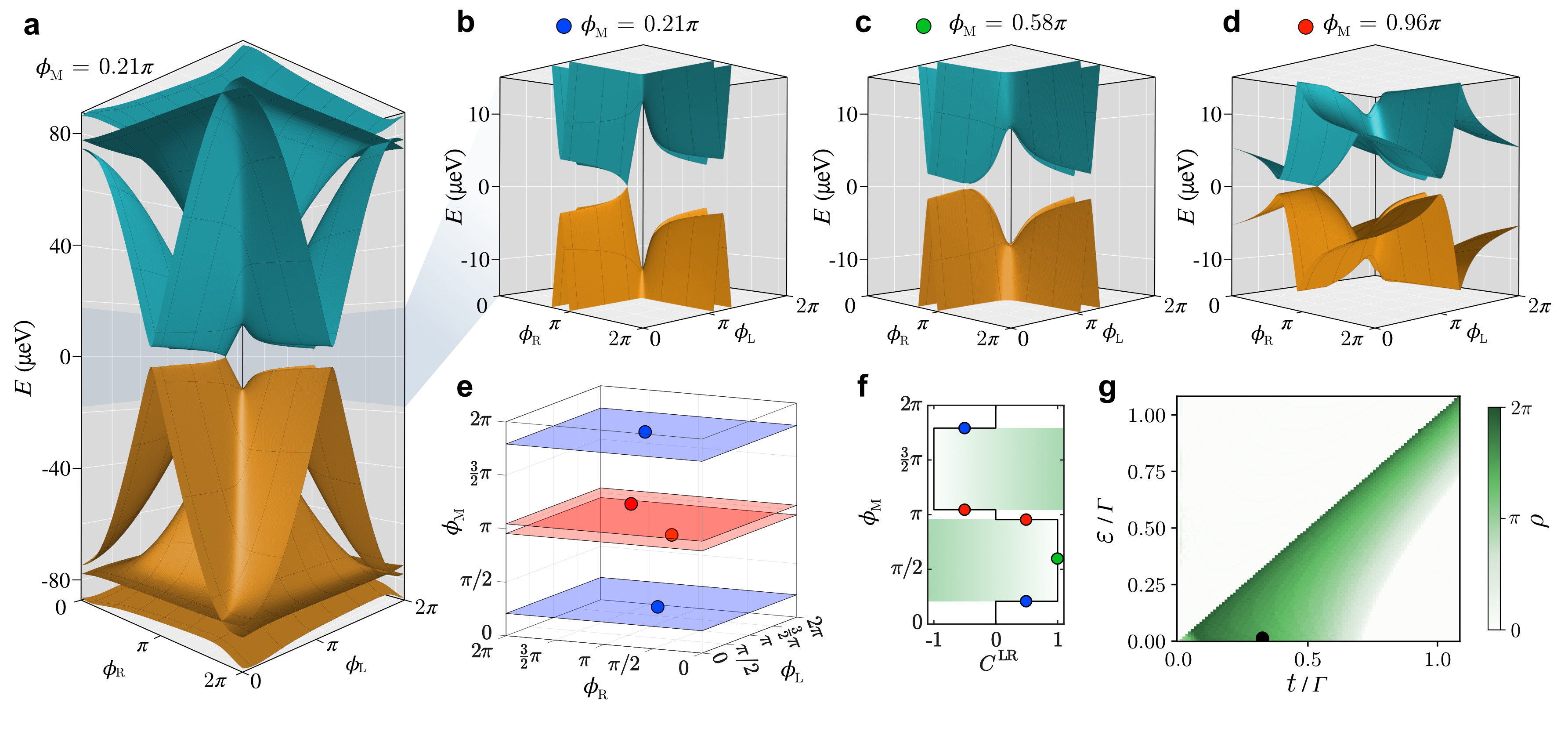}
	\caption{(a) ABS spectra extracted from the density of states as a function of the phase differences $\phi_{\rm L}$ and $\phi_{\rm R}$ for fixed $\phi_{\rm M}=0.21 \pi$. The energy gap between the highest band with $E < 0$ and the lowest band with $E>0$ vanishes in one point, where a Weyl node forms. (b) Enlargement of (a) around zero energy, highlighting the Weyl node. (c) As in (b), but for $\phi_{\rm M}=0.58 \pi$. A topological gap is present between the bands. (d) As in (b), but for $\phi_{\rm M}=0.96 \pi$, showing a second Weyl node with opposite topological charge where the gap closes. The parameters are the same as in Figs.~\ref{fig4} and \ref{fig5}. (e) Positions of the positively (blue) and negatively (red) charged Weyl nodes in the 3D phase space. (f) Chern number $C^{\rm{LR}}$ calculated as a function of $\phi_{\rm M}$ with $\tilde{\epsilon} \approx 0.01$ and $\tilde{t} \approx 0.32$, showing extensive topologically nontrivial regions where $|C^{\rm{LR}}|=1$ (green shading). (g) Topological phase diagram displaying the $\phi_{\rm M}$ range of the topological region, $\rho$, as a function of the model parameters $\tilde{\epsilon}$ and $\tilde{t}$. The diagram shows the robustness of the topological phase, that is present for $\rho > 0$. The black dot indicates the parametric point corresponding to the model used in Figs.~\ref{fig4} and \ref{fig5}.}
	\label{fig6}
\end{figure*}

\section{tri-Andreev molecule}

Supported by our theoretical model, we investigate more complex spectral features measured along the diagonals of the cubic unit cell, when all three fluxes are varied. Figure~\ref{fig5}(a) shows a map measured at $V_{\rm{sd}} = -\Delta/e$ along the $\Phi_{101}$-$\Phi_M$ plane sketched in the schematic on the left. As previously discussed in Sec.~III, $\mid$L$\rangle$ and $\mid$R$\rangle$ are already hybridized along the $\Phi_{101}$ direction, as also visible by the splitting of the conductance line at $\Phi_{\rm{M}} = 0 $ and $\Phi_{101} = 0.5\Phi_0$. When $\Phi_{\rm{M}}$ is tuned to $0.5\Phi_0$, these hybridized states mix with $\mid$M$\rangle$ forming an avoided crossing which is well reproduced by the simulation displayed in Fig.~\ref{fig5}(b). Taking an energy cut along the $(1,1,1)$ direction [gray arrow in (a)], we observe how the energy spectrum is affected by the hybridization, as shown in Fig. \ref{fig5}(c). Here, the bands split in energy forming an M-shape dispersion close to zero energy. Notably, such effect is much stronger compared to the one observed for the bi-Andreev molecules in Fig.~\ref{fig2}(f,g). By simulating the same energy spectrum [Fig.~\ref{fig5}(d)], we reproduce a similar M-shape of the topmost band and reveal the energy splitting between the three bands representing the bonding, nonbonding and antibonding molecular states illustrated in Fig.~\ref{fig4}(e).

As discussed in Sec.~III for the bi-Andreev molecules, the energy splitting induced by the hybridization is strongly anisotropic in the phase space. Figure~\ref{fig5}(e) shows the conductance map along the $\Phi_{10\overline{1}}$-$\Phi_M$ plane sketched in the schematic on the left. Here, the dispersions of $\mid$L$\rangle$ and $\mid$R$\rangle$ overlap with each other at $\Phi_{\rm{M}} = 0$, but split when $\Phi_{\rm{M}}$ is swept towards $0.5~\Phi_0$. Notably, the splitting appears larger along the $(11\bar1)$ direction compared to $(1\bar1\bar1)$, since $\mid$L$\rangle$ has a slightly larger transparency than $\mid$R$\rangle$ (see~\cite{SM}, Sec.~IV). Also, the horizontal conductance resonance representing the maximum of $\mid$M$\rangle$ significantly decreases in intensity approaching the center of the map. The simulation in Fig.~\ref{fig5}(f) reproduces the opening of a low conductance region at the crossing point, which ultimately derives from the splitting at $E=0$ observed in the simulated energy spectra in Fig.~\ref{fig4}(f). Indeed, those dispersions represent horizontal energy-dependent cuts of Fig.~\ref{fig5}(f), at $\Phi_{\rm{M}}$ values indicated by the colored arrows. Taking an energy-dependent cut along the direction $(1\bar1\bar1)$ [Fig.~\ref{fig5}(g)], we observe a sizable splitting into two bands, but smaller than the splitting observed along the $(111)$ direction. The overall hybridization strength is indeed reduced along the direction $(1\bar1\bar1)$ since two phases always have opposite values. Therefore, two bands are expected to remain nearly degenerate, as reproduced in the simulated spectrum in Fig.~\ref{fig5}(h). 

In summary, the dispersion of a tri-Andreev molecule is characterized by an anisotropic energy splitting, larger along the $(111)$ direction and weaker along its perpendicular direction $(1\bar1\bar1)$. This is directly reflected into the different shapes of the avoided crossings observed at the center of the cube diagonal conductance maps of Fig.~\ref{fig5}(a,e). The spectral signatures of a tri-Andreev molecule discussed here are also observed in a second device (see ~\cite{SM}, Sec.~VIII).

\section{Topological Andreev bands}
The hybridization between the three ABSs in our device leads to the formation of molecular-like states whose energy depends on all three superconducting phase differences. This property is one of the requirements for the formation of Weyl nodes, which would appear as zero-energy crossing points with linear dispersion as a function of all the three phases. In the following, we analyze the simulated Andreev band structure matching our experimental results. In Fig.~\ref{fig6}(a), we show the ABSs spectrum extracted from the maxima of the density of states for $\phi_{\rm M} = 0.21\pi$. Here, the two particle-hole symmetric bands closest to the Fermi level form a zero energy crossing at $\phi_{\rm L} = \phi_{\rm R} = 0.89\pi$ as highlighted in the zoomed-in plot in Fig.~\ref{fig6}(b). Increasing $\phi_{\rm M}$, we obtain gapped states [see Fig.~\ref{fig6}(c)] with a small energy gap. The gap is closed again at $\phi_{\rm M} = 0.96\pi$, where another zero-energy crossing appears at $\phi_{\rm L} \sim \phi_{\rm R} \sim 0.6\pi$ [Fig.~\ref{fig6}(d)]. Within the whole cubic unit cell we find four zero-energy crossings appearing in two pairs (red and blue points in Fig.~\ref{fig6}(e)), which are related by time-reversal symmetry. 

As shown in Ref.~\cite{Wang2012,Wang2013,Gavensky2023}, the full information about the topology of the ABS spectrum is encoded in the topological Hamiltonian, $H_{\rm top}$, given by the inverse of the central Green's functions $G_{\rm C}$ evaluated at zero energy, i.e., $H_{\rm top}= - G_{\rm C}^{-1}(E = 0)$. $H_{\rm top}$ depends on the dimensionless parameters $\tilde{\epsilon} = \epsilon/\Gamma$ and $\tilde{t} = t/\Gamma$ \cite{Klees2020,Teshler2023a} (see more details in Ref.~\cite{SM}, Sec.~VII). The values of $t$ and $\epsilon$ used for the DOS simulations matching our measurements correspond to $\tilde{\epsilon}\approx 0.01$ and $\tilde{t}\approx 0.32$. By diagonalizing $H_{\rm top}$, we obtain the eigenvectors $\ket{v_i}$ and the corresponding spin-degenerate eigenenergies. To establish the topological properties of the ABSs, we compute the topological invariant following the numerical method in Ref.~\cite{Fukui2005}. The Chern number is given by
\begin{equation}
    C_i^{\rm LR}(\phi_{\rm M}) = \frac{1}{2\pi} \int_{0}^{2\pi}\int_{0}^{2\pi} B_i^{\rm LR}(\phi_{\rm L}, \phi_{\rm R},\phi_{\rm M}) \, d\phi_{\rm L} d\phi_{\rm R},
\end{equation}
\noindent with the total Chern number $C^{\rm LR} = \sum_{i=1,2,3} C_i^{\rm LR}$ as the sum over all occupied bands and $B_i^{\rm LR} \equiv -2{\rm Im}\braket{\partial_{\phi_{\rm L}} v_i| \partial_{\phi_{\rm R}} v_i}$ represents the Berry curvature calculated at fixed $\phi_{\rm M}$. Figure \ref{fig6}(f) shows two distinct nontrivial topological phases having Chern number equal to $\pm 1$ within extensive $\phi_{\rm M}$ intervals indicated by the green areas. Therefore, the zero energy crossing points in (e) are positively (blue) and negatively (red) charged Weyl nodes appearing in the energy spectra (b,d) as the states cross the Fermi level. The eigenenergies of the topological Hamiltonian for the same parameters as in Fig.~\ref{fig6} are shown in Fig.~S7 \cite{SM}. The eigenenergies exhibit zero-energy states exactly when one local maximum of the DOS is at zero energy, as expected. Notably, the spectrum shown in Fig.~\ref{fig6}(c) represents a topological insulating phase with a small energy gap. Since such features are masked by the relatively large spectral broadening in our measurements, we cannot experimentally confirm the predicted topological phase transitions. 

To evaluate the robustness of the topological regime in our simulations, we calculate the extension $\rho$ of the region in $\phi_{\rm M}$ where the Chern number differs from zero [green areas in (f)] as a function of the two key parameters of our model, $\tilde{\epsilon}$ and $\tilde{t}$. The phase diagram in Fig.~\ref{fig6}(g) shows a first topological transition when the hybridization $\tilde{t}$ exceeds $\tilde{\epsilon}$. Intuitively, the hybridization has to be large enough to push one state through zero energy, a mechanism analogous to the band inversion driven by spin-orbit coupling occurring in topological insulators \cite{Hasan2010}. By further increasing $\tilde{t}$, the two opposite charged Weyl nodes get closer to each other gradually reducing $\rho$ to zero. At this point, the oppositely charged Weyl nodes annihilate with each other completely suppressing the topological phase. The parameters used for simulating our measurements corresponds to $\tilde{\epsilon} = 0.01$ and $\tilde{t} = 0.32$, which places our system well within the calculated topological region as indicated by the black dot in Fig.~\ref{fig6}(g).

\section{Discussion and conclusions}
In this work, we studied the hybridization between ABSs in a 4TJJ and demonstrated the formation of a tri-Andreev molecule. This state, whose energy is controlled by three superconducting phase differences, is expected to support topological Andreev bands. According to our model, Weyl nodes emerge when the hybridization shifts at least one ABS through zero energy, inducing an inversion between electron- and hole-like bands in certain regions of the phase space. By varying $\phi_{\rm{M}}$, the system undergoes four topological transitions marked by Weyl nodes as shown in Fig.\ref{fig6}(f). The Andreev bands as a function of $\phi_{\rm{L}}$-$\phi_{\rm{R}}$ have an energy gap ranging between 0 (at the Weyl nodes) and $\sim 7~\mu\rm{eV}$, depending on $\phi_{\rm{M}}$. The resolution of tunneling spectroscopy, approximately $15~\mu \rm{eV}$, prevents us from experimentally resolving any gapped states in the low-energy spectrum. Thus, the spectral detection of Weyl nodes would be facilitated by a larger minigap, which could be obtained through an increase of the superconducting gap $\Delta$ or by a enhanced device tunability. For example, a larger minigap would be obtained by making the coupling parameters [see $\Gamma$ in Fig.~\ref{fig4}(a)] asymmetric or by increasing $\epsilon$, while keeping a sufficiently large hybridization strength. The phase diagram shown in Fig.~\ref{fig6}(g) provides a useful guideline for engineering these parameters while preserving the topological properties of the device.

The large transparency and hybridization strength observed in our 4TJJ already fall well into the stability range of the topological state. Experimental techniques with higher energy resolution would be highly favorable for confirming the presence of Weyl nodes in the ABS spectrum, even in systems with such a large transparency. Microwave spectroscopy, in particular, offers sub-$\mu \rm{V}$ resolution, making it well suited for this purpose. Furthermore, hybrid InAs/Al heterostructures are readily integrated in circuit QED architectures~\cite{Larsen2015,deLange2015,Hays2018,Zellekens2022,Tosi2019,Hinderling2023}, offering a tangible prospect for studies of topological Andreev band structure. The employment of polarized microwave radiation can also make this technique sensitive to the Berry curvature \cite{Klees2020}.

In summary, we realized a phase-controlled 4TJJ and studied its energy spectrum using tunneling spectroscopy. The measurement protocol based on independent flux control developed here allows the systematic exploration of the Andreev spectrum in the 3D phase space. We identified the spectral signatures of a tri-Andreev molecule resulting from the simultaneous hybridization of three ABSs. A numerical model reproduced the key experimental observation, and suggested that the current generation of devices already hosts topological Andreev bands. In the light of our results, phase-tunable MTJJs offer new opportunities for studying topological phases in high-dimensional synthetic band structures and developing novel superconducting quantum circuits \cite{MatuteCanadas2024}.

\section{Data availability}
The data that support the findings of this article will be available on Zenodo.

\section{Acknowledgments}
We thank Manuel Hinderling for useful discussions.
We thank the Cleanroom Operations Team of the Binnig and Rohrer Nanotechnology Center (BRNC) for their help and support.
D.C.O., A.E.S.~and W.B.~acknowledge support by the Deutsche Forschungsgemeinschaft (DFG; German Research Foundation) via SFB 1432 (Project No. 425217212) and Project No. 467596333, and by the Excellence Strategy of the University of Konstanz via a Blue Sky project.
J.C.C.~thanks the  Spanish Ministry of Science and Innovation (Grant PID2020-114880GB-I00) for financial support and the DFG and SFB 1432 for sponsoring his stay at the University of Konstanz as a Mercator Fellow.
W.W.~acknowledges support from the Swiss National Science Foundation (grant number 200020\_207538).
F.N.~acknowledges support from the European Research Council (grant number 804273) and the Swiss National Science Foundation (grant number 200021\_201082).

\section{References}
\bibliography{bibliography.bib}

\begin{thebibliography}{86}%
\makeatletter
\providecommand \@ifxundefined [1]{%
 \@ifx{#1\undefined}
}%
\providecommand \@ifnum [1]{%
 \ifnum #1\expandafter \@firstoftwo
 \else \expandafter \@secondoftwo
 \fi
}%
\providecommand \@ifx [1]{%
 \ifx #1\expandafter \@firstoftwo
 \else \expandafter \@secondoftwo
 \fi
}%
\providecommand \natexlab [1]{#1}%
\providecommand \enquote  [1]{``#1''}%
\providecommand \bibnamefont  [1]{#1}%
\providecommand \bibfnamefont [1]{#1}%
\providecommand \citenamefont [1]{#1}%
\providecommand \href@noop [0]{\@secondoftwo}%
\providecommand \href [0]{\begingroup \@sanitize@url \@href}%
\providecommand \@href[1]{\@@startlink{#1}\@@href}%
\providecommand \@@href[1]{\endgroup#1\@@endlink}%
\providecommand \@sanitize@url [0]{\catcode `\\12\catcode `\$12\catcode
  `\&12\catcode `\#12\catcode `\^12\catcode `\_12\catcode `\%12\relax}%
\providecommand \@@startlink[1]{}%
\providecommand \@@endlink[0]{}%
\providecommand \url  [0]{\begingroup\@sanitize@url \@url }%
\providecommand \@url [1]{\endgroup\@href {#1}{\urlprefix }}%
\providecommand \urlprefix  [0]{URL }%
\providecommand \Eprint [0]{\href }%
\providecommand \doibase [0]{https://doi.org/}%
\providecommand \selectlanguage [0]{\@gobble}%
\providecommand \bibinfo  [0]{\@secondoftwo}%
\providecommand \bibfield  [0]{\@secondoftwo}%
\providecommand \translation [1]{[#1]}%
\providecommand \BibitemOpen [0]{}%
\providecommand \bibitemStop [0]{}%
\providecommand \bibitemNoStop [0]{.\EOS\space}%
\providecommand \EOS [0]{\spacefactor3000\relax}%
\providecommand \BibitemShut  [1]{\csname bibitem#1\endcsname}%
\let\auto@bib@innerbib\@empty
\bibitem [{\citenamefont {Andreev}(1964)}]{Andreev1964}%
  \BibitemOpen
  \bibfield  {author} {\bibinfo {author} {\bibfnamefont {A.~F.}\ \bibnamefont
  {Andreev}},\ }\bibfield  {title} {\bibinfo {title} {Thermal conductivity of
  the intermediate state of superconductors},\ }\href@noop {} {\bibfield
  {journal} {\bibinfo  {journal} {Sov. Phys. JETP}\ }\textbf {\bibinfo {volume}
  {19}},\ \bibinfo {pages} {1228} (\bibinfo {year} {1964})}\BibitemShut
  {NoStop}%
\bibitem [{\citenamefont {Beenakker}\ and\ \citenamefont {van
  Houten}(1991)}]{Beenakker1991}%
  \BibitemOpen
  \bibfield  {author} {\bibinfo {author} {\bibfnamefont {C.~W.~J.}\
  \bibnamefont {Beenakker}}\ and\ \bibinfo {author} {\bibfnamefont
  {H.}~\bibnamefont {van Houten}},\ }\bibfield  {title} {\bibinfo {title}
  {{J}osephson current through a superconducting quantum point contact shorter
  than the coherence length},\ }\href
  {https://doi.org/10.1103/PhysRevLett.66.3056} {\bibfield  {journal} {\bibinfo
   {journal} {Phys. Rev. Lett.}\ }\textbf {\bibinfo {volume} {66}},\ \bibinfo
  {pages} {3056} (\bibinfo {year} {1991})}\BibitemShut {NoStop}%
\bibitem [{\citenamefont {Beenakker}(1991)}]{Beenakker1991b}%
  \BibitemOpen
  \bibfield  {author} {\bibinfo {author} {\bibfnamefont {C.~W.~J.}\
  \bibnamefont {Beenakker}},\ }\bibfield  {title} {\bibinfo {title} {Universal
  limit of critical-current fluctuations in mesoscopic {J}osephson junctions},\
  }\href {https://doi.org/10.1103/PhysRevLett.67.3836} {\bibfield  {journal}
  {\bibinfo  {journal} {Phys. Rev. Lett.}\ }\textbf {\bibinfo {volume} {67}},\
  \bibinfo {pages} {3836} (\bibinfo {year} {1991})}\BibitemShut {NoStop}%
\bibitem [{\citenamefont {Furusaki}\ and\ \citenamefont
  {Tsukada}(1991)}]{Furusaki1991}%
  \BibitemOpen
  \bibfield  {author} {\bibinfo {author} {\bibfnamefont {A.}~\bibnamefont
  {Furusaki}}\ and\ \bibinfo {author} {\bibfnamefont {M.}~\bibnamefont
  {Tsukada}},\ }\bibfield  {title} {\bibinfo {title} {Current-carrying states
  in {J}osephson junctions},\ }\href
  {https://doi.org/10.1103/PhysRevB.43.10164} {\bibfield  {journal} {\bibinfo
  {journal} {Phys. Rev. B}\ }\textbf {\bibinfo {volume} {43}},\ \bibinfo
  {pages} {10164} (\bibinfo {year} {1991})}\BibitemShut {NoStop}%
\bibitem [{\citenamefont {Pillet}\ \emph {et~al.}(2010)\citenamefont {Pillet},
  \citenamefont {Quay}, \citenamefont {Morfin}, \citenamefont {Bena},
  \citenamefont {Yeyati},\ and\ \citenamefont {Joyez}}]{Pillet2010}%
  \BibitemOpen
  \bibfield  {author} {\bibinfo {author} {\bibfnamefont {J.-D.}\ \bibnamefont
  {Pillet}}, \bibinfo {author} {\bibfnamefont {C.~H.~L.}\ \bibnamefont {Quay}},
  \bibinfo {author} {\bibfnamefont {P.}~\bibnamefont {Morfin}}, \bibinfo
  {author} {\bibfnamefont {C.}~\bibnamefont {Bena}}, \bibinfo {author}
  {\bibfnamefont {A.~L.}\ \bibnamefont {Yeyati}},\ and\ \bibinfo {author}
  {\bibfnamefont {P.}~\bibnamefont {Joyez}},\ }\bibfield  {title} {\bibinfo
  {title} {Andreev bound states in supercurrent-carrying carbon nanotubes
  revealed},\ }\href {https://doi.org/10.1038/nphys1811} {\bibfield  {journal}
  {\bibinfo  {journal} {Nat. Phys.}\ }\textbf {\bibinfo {volume} {6}},\
  \bibinfo {pages} {965} (\bibinfo {year} {2010})}\BibitemShut {NoStop}%
\bibitem [{\citenamefont {Bretheau}\ \emph
  {et~al.}(2013{\natexlab{a}})\citenamefont {Bretheau}, \citenamefont {Girit},
  \citenamefont {Pothier}, \citenamefont {Esteve},\ and\ \citenamefont
  {Urbina}}]{Bretheau2013a}%
  \BibitemOpen
  \bibfield  {author} {\bibinfo {author} {\bibfnamefont {L.}~\bibnamefont
  {Bretheau}}, \bibinfo {author} {\bibfnamefont {{\c{C}}.~{\"O}.}\ \bibnamefont
  {Girit}}, \bibinfo {author} {\bibfnamefont {H.}~\bibnamefont {Pothier}},
  \bibinfo {author} {\bibfnamefont {D.}~\bibnamefont {Esteve}},\ and\ \bibinfo
  {author} {\bibfnamefont {C.}~\bibnamefont {Urbina}},\ }\bibfield  {title}
  {\bibinfo {title} {Exciting {A}ndreev pairs in a superconducting atomic
  contact},\ }\href {https://doi.org/10.1038/nature12315} {\bibfield  {journal}
  {\bibinfo  {journal} {Nature}\ }\textbf {\bibinfo {volume} {499}},\ \bibinfo
  {pages} {312} (\bibinfo {year} {2013}{\natexlab{a}})}\BibitemShut {NoStop}%
\bibitem [{\citenamefont {Bretheau}\ \emph
  {et~al.}(2013{\natexlab{b}})\citenamefont {Bretheau}, \citenamefont {Girit},
  \citenamefont {Urbina}, \citenamefont {Esteve},\ and\ \citenamefont
  {Pothier}}]{Bretheau2013b}%
  \BibitemOpen
  \bibfield  {author} {\bibinfo {author} {\bibfnamefont {L.}~\bibnamefont
  {Bretheau}}, \bibinfo {author} {\bibfnamefont {{\c{C}}.~{\"O}.}\ \bibnamefont
  {Girit}}, \bibinfo {author} {\bibfnamefont {C.}~\bibnamefont {Urbina}},
  \bibinfo {author} {\bibfnamefont {D.}~\bibnamefont {Esteve}},\ and\ \bibinfo
  {author} {\bibfnamefont {H.}~\bibnamefont {Pothier}},\ }\bibfield  {title}
  {\bibinfo {title} {Supercurrent spectroscopy of {A}ndreev states},\ }\href
  {https://doi.org/10.1103/PhysRevX.3.041034} {\bibfield  {journal} {\bibinfo
  {journal} {Phys. Rev. X}\ }\textbf {\bibinfo {volume} {3}},\ \bibinfo {pages}
  {041034} (\bibinfo {year} {2013}{\natexlab{b}})}\BibitemShut {NoStop}%
\bibitem [{\citenamefont {Lee}\ \emph {et~al.}(2014)\citenamefont {Lee},
  \citenamefont {Jiang}, \citenamefont {Houzet}, \citenamefont {Aguado},
  \citenamefont {Lieber},\ and\ \citenamefont {Franceschi}}]{Lee2014}%
  \BibitemOpen
  \bibfield  {author} {\bibinfo {author} {\bibfnamefont {E.~J.~H.}\
  \bibnamefont {Lee}}, \bibinfo {author} {\bibfnamefont {X.}~\bibnamefont
  {Jiang}}, \bibinfo {author} {\bibfnamefont {M.}~\bibnamefont {Houzet}},
  \bibinfo {author} {\bibfnamefont {R.}~\bibnamefont {Aguado}}, \bibinfo
  {author} {\bibfnamefont {C.~M.}\ \bibnamefont {Lieber}},\ and\ \bibinfo
  {author} {\bibfnamefont {S.~D.}\ \bibnamefont {Franceschi}},\ }\bibfield
  {title} {\bibinfo {title} {Spin-resolved {A}ndreev levels and parity
  crossings in hybrid superconductor{\textendash}semiconductor
  nanostructures},\ }\href {https://doi.org/10.1038/nnano.2013.267} {\bibfield
  {journal} {\bibinfo  {journal} {Nat. Nanotechnol.}\ }\textbf {\bibinfo
  {volume} {9}},\ \bibinfo {pages} {79} (\bibinfo {year} {2014})}\BibitemShut
  {NoStop}%
\bibitem [{\citenamefont {van Woerkom}\ \emph {et~al.}(2017)\citenamefont {van
  Woerkom}, \citenamefont {Proutski}, \citenamefont {van Heck}, \citenamefont
  {Bouman}, \citenamefont {V{\"{a}}yrynen}, \citenamefont {Glazman},
  \citenamefont {Krogstrup}, \citenamefont {Nyg{\aa}rd}, \citenamefont
  {Kouwenhoven},\ and\ \citenamefont {Geresdi}}]{VanWoerkom2017}%
  \BibitemOpen
  \bibfield  {author} {\bibinfo {author} {\bibfnamefont {D.~J.}\ \bibnamefont
  {van Woerkom}}, \bibinfo {author} {\bibfnamefont {A.}~\bibnamefont
  {Proutski}}, \bibinfo {author} {\bibfnamefont {B.}~\bibnamefont {van Heck}},
  \bibinfo {author} {\bibfnamefont {D.}~\bibnamefont {Bouman}}, \bibinfo
  {author} {\bibfnamefont {J.~I.}\ \bibnamefont {V{\"{a}}yrynen}}, \bibinfo
  {author} {\bibfnamefont {L.~I.}\ \bibnamefont {Glazman}}, \bibinfo {author}
  {\bibfnamefont {P.}~\bibnamefont {Krogstrup}}, \bibinfo {author}
  {\bibfnamefont {J.}~\bibnamefont {Nyg{\aa}rd}}, \bibinfo {author}
  {\bibfnamefont {L.~P.}\ \bibnamefont {Kouwenhoven}},\ and\ \bibinfo {author}
  {\bibfnamefont {A.}~\bibnamefont {Geresdi}},\ }\bibfield  {title} {\bibinfo
  {title} {{Microwave spectroscopy of spinful {A}ndreev bound states in
  ballistic semiconductor {J}osephson junctions}},\ }\href
  {https://doi.org/10.1038/nphys4150} {\bibfield  {journal} {\bibinfo
  {journal} {Nat. Phys.}\ }\textbf {\bibinfo {volume} {13}},\ \bibinfo {pages}
  {876} (\bibinfo {year} {2017})}\BibitemShut {NoStop}%
\bibitem [{\citenamefont {Hays}\ \emph {et~al.}(2020)\citenamefont {Hays},
  \citenamefont {Fatemi}, \citenamefont {Serniak}, \citenamefont {Bouman},
  \citenamefont {Diamond}, \citenamefont {de~Lange}, \citenamefont {Krogstrup},
  \citenamefont {Nyg{\aa}rd}, \citenamefont {Geresdi},\ and\ \citenamefont
  {Devoret}}]{Hays2020}%
  \BibitemOpen
  \bibfield  {author} {\bibinfo {author} {\bibfnamefont {M.}~\bibnamefont
  {Hays}}, \bibinfo {author} {\bibfnamefont {V.}~\bibnamefont {Fatemi}},
  \bibinfo {author} {\bibfnamefont {K.}~\bibnamefont {Serniak}}, \bibinfo
  {author} {\bibfnamefont {D.}~\bibnamefont {Bouman}}, \bibinfo {author}
  {\bibfnamefont {S.}~\bibnamefont {Diamond}}, \bibinfo {author} {\bibfnamefont
  {G.}~\bibnamefont {de~Lange}}, \bibinfo {author} {\bibfnamefont
  {P.}~\bibnamefont {Krogstrup}}, \bibinfo {author} {\bibfnamefont
  {J.}~\bibnamefont {Nyg{\aa}rd}}, \bibinfo {author} {\bibfnamefont
  {A.}~\bibnamefont {Geresdi}},\ and\ \bibinfo {author} {\bibfnamefont {M.~H.}\
  \bibnamefont {Devoret}},\ }\bibfield  {title} {\bibinfo {title} {Continuous
  monitoring of a trapped superconducting spin},\ }\href
  {https://doi.org/10.1038/s41567-020-0952-3} {\bibfield  {journal} {\bibinfo
  {journal} {Nat. Phys.}\ }\textbf {\bibinfo {volume} {16}},\ \bibinfo {pages}
  {1103} (\bibinfo {year} {2020})}\BibitemShut {NoStop}%
\bibitem [{\citenamefont {Tosi}\ \emph {et~al.}(2019)\citenamefont {Tosi},
  \citenamefont {Metzger}, \citenamefont {Goffman}, \citenamefont {Urbina},
  \citenamefont {Pothier}, \citenamefont {Park}, \citenamefont {Yeyati},
  \citenamefont {Nyg\aa{}rd},\ and\ \citenamefont {Krogstrup}}]{Tosi2019}%
  \BibitemOpen
  \bibfield  {author} {\bibinfo {author} {\bibfnamefont {L.}~\bibnamefont
  {Tosi}}, \bibinfo {author} {\bibfnamefont {C.}~\bibnamefont {Metzger}},
  \bibinfo {author} {\bibfnamefont {M.~F.}\ \bibnamefont {Goffman}}, \bibinfo
  {author} {\bibfnamefont {C.}~\bibnamefont {Urbina}}, \bibinfo {author}
  {\bibfnamefont {H.}~\bibnamefont {Pothier}}, \bibinfo {author} {\bibfnamefont
  {S.}~\bibnamefont {Park}}, \bibinfo {author} {\bibfnamefont {A.~L.}\
  \bibnamefont {Yeyati}}, \bibinfo {author} {\bibfnamefont {J.}~\bibnamefont
  {Nyg\aa{}rd}},\ and\ \bibinfo {author} {\bibfnamefont {P.}~\bibnamefont
  {Krogstrup}},\ }\bibfield  {title} {\bibinfo {title} {Spin-orbit splitting of
  {A}ndreev states revealed by microwave spectroscopy},\ }\href
  {https://doi.org/10.1103/PhysRevX.9.011010} {\bibfield  {journal} {\bibinfo
  {journal} {Phys. Rev. X}\ }\textbf {\bibinfo {volume} {9}},\ \bibinfo {pages}
  {011010} (\bibinfo {year} {2019})}\BibitemShut {NoStop}%
\bibitem [{\citenamefont {Nichele}\ \emph {et~al.}(2020)\citenamefont
  {Nichele}, \citenamefont {Portol\'es}, \citenamefont {Fornieri},
  \citenamefont {Whiticar}, \citenamefont {Drachmann}, \citenamefont {Gronin},
  \citenamefont {Wang}, \citenamefont {Gardner}, \citenamefont {Thomas},
  \citenamefont {Hatke}, \citenamefont {Manfra},\ and\ \citenamefont
  {Marcus}}]{Nichele2020}%
  \BibitemOpen
  \bibfield  {author} {\bibinfo {author} {\bibfnamefont {F.}~\bibnamefont
  {Nichele}}, \bibinfo {author} {\bibfnamefont {E.}~\bibnamefont {Portol\'es}},
  \bibinfo {author} {\bibfnamefont {A.}~\bibnamefont {Fornieri}}, \bibinfo
  {author} {\bibfnamefont {A.~M.}\ \bibnamefont {Whiticar}}, \bibinfo {author}
  {\bibfnamefont {A.~C.~C.}\ \bibnamefont {Drachmann}}, \bibinfo {author}
  {\bibfnamefont {S.}~\bibnamefont {Gronin}}, \bibinfo {author} {\bibfnamefont
  {T.}~\bibnamefont {Wang}}, \bibinfo {author} {\bibfnamefont {G.~C.}\
  \bibnamefont {Gardner}}, \bibinfo {author} {\bibfnamefont {C.}~\bibnamefont
  {Thomas}}, \bibinfo {author} {\bibfnamefont {A.~T.}\ \bibnamefont {Hatke}},
  \bibinfo {author} {\bibfnamefont {M.~J.}\ \bibnamefont {Manfra}},\ and\
  \bibinfo {author} {\bibfnamefont {C.~M.}\ \bibnamefont {Marcus}},\ }\bibfield
   {title} {\bibinfo {title} {Relating {A}ndreev bound states and supercurrents
  in hybrid {J}osephson junctions},\ }\href
  {https://doi.org/10.1103/PhysRevLett.124.226801} {\bibfield  {journal}
  {\bibinfo  {journal} {Phys. Rev. Lett.}\ }\textbf {\bibinfo {volume} {124}},\
  \bibinfo {pages} {226801} (\bibinfo {year} {2020})}\BibitemShut {NoStop}%
\bibitem [{\citenamefont {Janvier}\ \emph {et~al.}(2015)\citenamefont
  {Janvier}, \citenamefont {Tosi}, \citenamefont {Bretheau}, \citenamefont
  {Girit}, \citenamefont {Stern}, \citenamefont {Bertet}, \citenamefont
  {Joyez}, \citenamefont {Vion}, \citenamefont {Esteve}, \citenamefont
  {Goffman}, \citenamefont {Pothier},\ and\ \citenamefont
  {Urbina}}]{Janvier2015}%
  \BibitemOpen
  \bibfield  {author} {\bibinfo {author} {\bibfnamefont {C.}~\bibnamefont
  {Janvier}}, \bibinfo {author} {\bibfnamefont {L.}~\bibnamefont {Tosi}},
  \bibinfo {author} {\bibfnamefont {L.}~\bibnamefont {Bretheau}}, \bibinfo
  {author} {\bibfnamefont {C.~O.}\ \bibnamefont {Girit}}, \bibinfo {author}
  {\bibfnamefont {M.}~\bibnamefont {Stern}}, \bibinfo {author} {\bibfnamefont
  {P.}~\bibnamefont {Bertet}}, \bibinfo {author} {\bibfnamefont
  {P.}~\bibnamefont {Joyez}}, \bibinfo {author} {\bibfnamefont
  {D.}~\bibnamefont {Vion}}, \bibinfo {author} {\bibfnamefont {D.}~\bibnamefont
  {Esteve}}, \bibinfo {author} {\bibfnamefont {M.~F.}\ \bibnamefont {Goffman}},
  \bibinfo {author} {\bibfnamefont {H.}~\bibnamefont {Pothier}},\ and\ \bibinfo
  {author} {\bibfnamefont {C.}~\bibnamefont {Urbina}},\ }\bibfield  {title}
  {\bibinfo {title} {Coherent manipulation of {A}ndreev states in
  superconducting atomic contacts},\ }\href
  {https://doi.org/10.1126/science.aab2179} {\bibfield  {journal} {\bibinfo
  {journal} {Science}\ }\textbf {\bibinfo {volume} {349}},\ \bibinfo {pages}
  {1199} (\bibinfo {year} {2015})}\BibitemShut {NoStop}%
\bibitem [{\citenamefont {Hays}\ \emph {et~al.}(2018)\citenamefont {Hays},
  \citenamefont {de~Lange}, \citenamefont {Serniak}, \citenamefont {van
  Woerkom}, \citenamefont {Bouman}, \citenamefont {Krogstrup}, \citenamefont
  {Nyg\aa{}rd}, \citenamefont {Geresdi},\ and\ \citenamefont
  {Devoret}}]{Hays2018}%
  \BibitemOpen
  \bibfield  {author} {\bibinfo {author} {\bibfnamefont {M.}~\bibnamefont
  {Hays}}, \bibinfo {author} {\bibfnamefont {G.}~\bibnamefont {de~Lange}},
  \bibinfo {author} {\bibfnamefont {K.}~\bibnamefont {Serniak}}, \bibinfo
  {author} {\bibfnamefont {D.~J.}\ \bibnamefont {van Woerkom}}, \bibinfo
  {author} {\bibfnamefont {D.}~\bibnamefont {Bouman}}, \bibinfo {author}
  {\bibfnamefont {P.}~\bibnamefont {Krogstrup}}, \bibinfo {author}
  {\bibfnamefont {J.}~\bibnamefont {Nyg\aa{}rd}}, \bibinfo {author}
  {\bibfnamefont {A.}~\bibnamefont {Geresdi}},\ and\ \bibinfo {author}
  {\bibfnamefont {M.~H.}\ \bibnamefont {Devoret}},\ }\bibfield  {title}
  {\bibinfo {title} {Direct microwave measurement of {A}ndreev-bound-state
  dynamics in a semiconductor-nanowire {J}osephson junction},\ }\href
  {https://doi.org/10.1103/PhysRevLett.121.047001} {\bibfield  {journal}
  {\bibinfo  {journal} {Phys. Rev. Lett.}\ }\textbf {\bibinfo {volume} {121}},\
  \bibinfo {pages} {047001} (\bibinfo {year} {2018})}\BibitemShut {NoStop}%
\bibitem [{\citenamefont {Hays}\ \emph {et~al.}(2021)\citenamefont {Hays},
  \citenamefont {Fatemi}, \citenamefont {Bouman}, \citenamefont {Cerrillo},
  \citenamefont {Diamond}, \citenamefont {Serniak}, \citenamefont {Connolly},
  \citenamefont {Krogstrup}, \citenamefont {Nygård}, \citenamefont {Yeyati},
  \citenamefont {Geresdi},\ and\ \citenamefont {Devoret}}]{Hays2021}%
  \BibitemOpen
  \bibfield  {author} {\bibinfo {author} {\bibfnamefont {M.}~\bibnamefont
  {Hays}}, \bibinfo {author} {\bibfnamefont {V.}~\bibnamefont {Fatemi}},
  \bibinfo {author} {\bibfnamefont {D.}~\bibnamefont {Bouman}}, \bibinfo
  {author} {\bibfnamefont {J.}~\bibnamefont {Cerrillo}}, \bibinfo {author}
  {\bibfnamefont {S.}~\bibnamefont {Diamond}}, \bibinfo {author} {\bibfnamefont
  {K.}~\bibnamefont {Serniak}}, \bibinfo {author} {\bibfnamefont
  {T.}~\bibnamefont {Connolly}}, \bibinfo {author} {\bibfnamefont
  {P.}~\bibnamefont {Krogstrup}}, \bibinfo {author} {\bibfnamefont
  {J.}~\bibnamefont {Nygård}}, \bibinfo {author} {\bibfnamefont {A.~L.}\
  \bibnamefont {Yeyati}}, \bibinfo {author} {\bibfnamefont {A.}~\bibnamefont
  {Geresdi}},\ and\ \bibinfo {author} {\bibfnamefont {M.~H.}\ \bibnamefont
  {Devoret}},\ }\bibfield  {title} {\bibinfo {title} {Coherent manipulation of
  an {A}ndreev spin qubit},\ }\href {https://doi.org/10.1126/science.abf0345}
  {\bibfield  {journal} {\bibinfo  {journal} {Science}\ }\textbf {\bibinfo
  {volume} {373}},\ \bibinfo {pages} {430} (\bibinfo {year}
  {2021})}\BibitemShut {NoStop}%
\bibitem [{\citenamefont {Pita-Vidal}\ \emph {et~al.}(2023)\citenamefont
  {Pita-Vidal}, \citenamefont {Bargerbos}, \citenamefont {{\v{Z}}itko},
  \citenamefont {Splitthoff}, \citenamefont {Gr\"{u}nhaupt}, \citenamefont
  {Wesdorp}, \citenamefont {Liu}, \citenamefont {Kouwenhoven}, \citenamefont
  {Aguado}, \citenamefont {van Heck}, \citenamefont {Kou},\ and\ \citenamefont
  {Andersen}}]{PitaVidal2023}%
  \BibitemOpen
  \bibfield  {author} {\bibinfo {author} {\bibfnamefont {M.}~\bibnamefont
  {Pita-Vidal}}, \bibinfo {author} {\bibfnamefont {A.}~\bibnamefont
  {Bargerbos}}, \bibinfo {author} {\bibfnamefont {R.}~\bibnamefont
  {{\v{Z}}itko}}, \bibinfo {author} {\bibfnamefont {L.~J.}\ \bibnamefont
  {Splitthoff}}, \bibinfo {author} {\bibfnamefont {L.}~\bibnamefont
  {Gr\"{u}nhaupt}}, \bibinfo {author} {\bibfnamefont {J.~J.}\ \bibnamefont
  {Wesdorp}}, \bibinfo {author} {\bibfnamefont {Y.}~\bibnamefont {Liu}},
  \bibinfo {author} {\bibfnamefont {L.~P.}\ \bibnamefont {Kouwenhoven}},
  \bibinfo {author} {\bibfnamefont {R.}~\bibnamefont {Aguado}}, \bibinfo
  {author} {\bibfnamefont {B.}~\bibnamefont {van Heck}}, \bibinfo {author}
  {\bibfnamefont {A.}~\bibnamefont {Kou}},\ and\ \bibinfo {author}
  {\bibfnamefont {C.~K.}\ \bibnamefont {Andersen}},\ }\bibfield  {title}
  {\bibinfo {title} {Direct manipulation of a superconducting spin qubit
  strongly coupled to a transmon qubit},\ }\href
  {https://doi.org/10.1038/s41567-023-02071-x} {\bibfield  {journal} {\bibinfo
  {journal} {Nat. Phys.}\ }\textbf {\bibinfo {volume} {19}},\ \bibinfo {pages}
  {1110} (\bibinfo {year} {2023})}\BibitemShut {NoStop}%
\bibitem [{\citenamefont {Mourik}\ \emph {et~al.}(2012)\citenamefont {Mourik},
  \citenamefont {Zuo}, \citenamefont {Frolov}, \citenamefont {Plissard},
  \citenamefont {Bakkers},\ and\ \citenamefont {Kouwenhoven}}]{Mourik2012}%
  \BibitemOpen
  \bibfield  {author} {\bibinfo {author} {\bibfnamefont {V.}~\bibnamefont
  {Mourik}}, \bibinfo {author} {\bibfnamefont {K.}~\bibnamefont {Zuo}},
  \bibinfo {author} {\bibfnamefont {S.~M.}\ \bibnamefont {Frolov}}, \bibinfo
  {author} {\bibfnamefont {S.~R.}\ \bibnamefont {Plissard}}, \bibinfo {author}
  {\bibfnamefont {E.~P. A.~M.}\ \bibnamefont {Bakkers}},\ and\ \bibinfo
  {author} {\bibfnamefont {L.~P.}\ \bibnamefont {Kouwenhoven}},\ }\bibfield
  {title} {\bibinfo {title} {Signatures of {M}ajorana fermions in hybrid
  superconductor-semiconductor nanowire devices},\ }\href
  {https://doi.org/10.1126/science.1222360} {\bibfield  {journal} {\bibinfo
  {journal} {Science}\ }\textbf {\bibinfo {volume} {336}},\ \bibinfo {pages}
  {1003–1007} (\bibinfo {year} {2012})}\BibitemShut {NoStop}%
\bibitem [{\citenamefont {Nichele}\ \emph {et~al.}(2017)\citenamefont
  {Nichele}, \citenamefont {Drachmann}, \citenamefont {Whiticar}, \citenamefont
  {O'Farrell}, \citenamefont {Suominen}, \citenamefont {Fornieri},
  \citenamefont {Wang}, \citenamefont {Gardner}, \citenamefont {Thomas},
  \citenamefont {Hatke}, \citenamefont {Krogstrup}, \citenamefont {Manfra},
  \citenamefont {Flensberg},\ and\ \citenamefont {Marcus}}]{Nichele2017}%
  \BibitemOpen
  \bibfield  {author} {\bibinfo {author} {\bibfnamefont {F.}~\bibnamefont
  {Nichele}}, \bibinfo {author} {\bibfnamefont {A.~C.~C.}\ \bibnamefont
  {Drachmann}}, \bibinfo {author} {\bibfnamefont {A.~M.}\ \bibnamefont
  {Whiticar}}, \bibinfo {author} {\bibfnamefont {E.~C.~T.}\ \bibnamefont
  {O'Farrell}}, \bibinfo {author} {\bibfnamefont {H.~J.}\ \bibnamefont
  {Suominen}}, \bibinfo {author} {\bibfnamefont {A.}~\bibnamefont {Fornieri}},
  \bibinfo {author} {\bibfnamefont {T.}~\bibnamefont {Wang}}, \bibinfo {author}
  {\bibfnamefont {G.~C.}\ \bibnamefont {Gardner}}, \bibinfo {author}
  {\bibfnamefont {C.}~\bibnamefont {Thomas}}, \bibinfo {author} {\bibfnamefont
  {A.~T.}\ \bibnamefont {Hatke}}, \bibinfo {author} {\bibfnamefont
  {P.}~\bibnamefont {Krogstrup}}, \bibinfo {author} {\bibfnamefont {M.~J.}\
  \bibnamefont {Manfra}}, \bibinfo {author} {\bibfnamefont {K.}~\bibnamefont
  {Flensberg}},\ and\ \bibinfo {author} {\bibfnamefont {C.~M.}\ \bibnamefont
  {Marcus}},\ }\bibfield  {title} {\bibinfo {title} {Scaling of {M}ajorana
  zero-bias conductance peaks},\ }\href
  {https://doi.org/10.1103/PhysRevLett.119.136803} {\bibfield  {journal}
  {\bibinfo  {journal} {Phys. Rev. Lett.}\ }\textbf {\bibinfo {volume} {119}},\
  \bibinfo {pages} {136803} (\bibinfo {year} {2017})}\BibitemShut {NoStop}%
\bibitem [{\citenamefont {Fornieri}\ \emph {et~al.}(2019)\citenamefont
  {Fornieri}, \citenamefont {Whiticar}, \citenamefont {Setiawan}, \citenamefont
  {Portol{\'{e}}s}, \citenamefont {Drachmann}, \citenamefont {Keselman},
  \citenamefont {Gronin}, \citenamefont {Thomas}, \citenamefont {Wang},
  \citenamefont {Kallaher}, \citenamefont {Gardner}, \citenamefont {Berg},
  \citenamefont {Manfra}, \citenamefont {Stern}, \citenamefont {Marcus},\ and\
  \citenamefont {Nichele}}]{Fornieri2019}%
  \BibitemOpen
  \bibfield  {author} {\bibinfo {author} {\bibfnamefont {A.}~\bibnamefont
  {Fornieri}}, \bibinfo {author} {\bibfnamefont {A.~M.}\ \bibnamefont
  {Whiticar}}, \bibinfo {author} {\bibfnamefont {F.}~\bibnamefont {Setiawan}},
  \bibinfo {author} {\bibfnamefont {E.}~\bibnamefont {Portol{\'{e}}s}},
  \bibinfo {author} {\bibfnamefont {A.~C.~C.}\ \bibnamefont {Drachmann}},
  \bibinfo {author} {\bibfnamefont {A.}~\bibnamefont {Keselman}}, \bibinfo
  {author} {\bibfnamefont {S.}~\bibnamefont {Gronin}}, \bibinfo {author}
  {\bibfnamefont {C.}~\bibnamefont {Thomas}}, \bibinfo {author} {\bibfnamefont
  {T.}~\bibnamefont {Wang}}, \bibinfo {author} {\bibfnamefont {R.}~\bibnamefont
  {Kallaher}}, \bibinfo {author} {\bibfnamefont {G.~C.}\ \bibnamefont
  {Gardner}}, \bibinfo {author} {\bibfnamefont {E.}~\bibnamefont {Berg}},
  \bibinfo {author} {\bibfnamefont {M.~J.}\ \bibnamefont {Manfra}}, \bibinfo
  {author} {\bibfnamefont {A.}~\bibnamefont {Stern}}, \bibinfo {author}
  {\bibfnamefont {C.~M.}\ \bibnamefont {Marcus}},\ and\ \bibinfo {author}
  {\bibfnamefont {F.}~\bibnamefont {Nichele}},\ }\bibfield  {title} {\bibinfo
  {title} {{Evidence of topological superconductivity in planar {J}osephson
  junctions}},\ }\href {https://doi.org/10.1038/s41586-019-1068-8} {\bibfield
  {journal} {\bibinfo  {journal} {Nature}\ }\textbf {\bibinfo {volume} {569}},\
  \bibinfo {pages} {89} (\bibinfo {year} {2019})}\BibitemShut {NoStop}%
\bibitem [{\citenamefont {Ren}\ \emph {et~al.}(2019)\citenamefont {Ren},
  \citenamefont {Pientka}, \citenamefont {Hart}, \citenamefont {Pierce},
  \citenamefont {Kosowsky}, \citenamefont {Lunczer}, \citenamefont {Schlereth},
  \citenamefont {Scharf}, \citenamefont {Hankiewicz}, \citenamefont
  {Molenkamp}, \citenamefont {Halperin},\ and\ \citenamefont
  {Yacoby}}]{Ren2019}%
  \BibitemOpen
  \bibfield  {author} {\bibinfo {author} {\bibfnamefont {H.}~\bibnamefont
  {Ren}}, \bibinfo {author} {\bibfnamefont {F.}~\bibnamefont {Pientka}},
  \bibinfo {author} {\bibfnamefont {S.}~\bibnamefont {Hart}}, \bibinfo {author}
  {\bibfnamefont {A.~T.}\ \bibnamefont {Pierce}}, \bibinfo {author}
  {\bibfnamefont {M.}~\bibnamefont {Kosowsky}}, \bibinfo {author}
  {\bibfnamefont {L.}~\bibnamefont {Lunczer}}, \bibinfo {author} {\bibfnamefont
  {R.}~\bibnamefont {Schlereth}}, \bibinfo {author} {\bibfnamefont
  {B.}~\bibnamefont {Scharf}}, \bibinfo {author} {\bibfnamefont {E.~M.}\
  \bibnamefont {Hankiewicz}}, \bibinfo {author} {\bibfnamefont {L.~W.}\
  \bibnamefont {Molenkamp}}, \bibinfo {author} {\bibfnamefont {B.~I.}\
  \bibnamefont {Halperin}},\ and\ \bibinfo {author} {\bibfnamefont
  {A.}~\bibnamefont {Yacoby}},\ }\bibfield  {title} {\bibinfo {title}
  {{Topological superconductivity in a phase-controlled {J}osephson
  junction}},\ }\href {https://doi.org/10.1038/s41586-019-1148-9} {\bibfield
  {journal} {\bibinfo  {journal} {Nature}\ ,\ \bibinfo {pages} {3}} (\bibinfo
  {year} {2019})}\BibitemShut {NoStop}%
\bibitem [{\citenamefont {Riwar}\ \emph {et~al.}(2016)\citenamefont {Riwar},
  \citenamefont {Houzet}, \citenamefont {Meyer},\ and\ \citenamefont
  {Nazarov}}]{Riwar2016}%
  \BibitemOpen
  \bibfield  {author} {\bibinfo {author} {\bibfnamefont {R.-P.}\ \bibnamefont
  {Riwar}}, \bibinfo {author} {\bibfnamefont {M.}~\bibnamefont {Houzet}},
  \bibinfo {author} {\bibfnamefont {J.~S.}\ \bibnamefont {Meyer}},\ and\
  \bibinfo {author} {\bibfnamefont {Y.~V.}\ \bibnamefont {Nazarov}},\
  }\bibfield  {title} {\bibinfo {title} {Multi-terminal {J}osephson junctions
  as topological matter},\ }\href {https://doi.org/10.1038/ncomms11167}
  {\bibfield  {journal} {\bibinfo  {journal} {Nat. Commun.}\ }\textbf {\bibinfo
  {volume} {7}},\ \bibinfo {pages} {11167} (\bibinfo {year}
  {2016})}\BibitemShut {NoStop}%
\bibitem [{\citenamefont {Eriksson}\ \emph {et~al.}(2017)\citenamefont
  {Eriksson}, \citenamefont {Riwar}, \citenamefont {Houzet}, \citenamefont
  {Meyer},\ and\ \citenamefont {Nazarov}}]{Eriksson2017}%
  \BibitemOpen
  \bibfield  {author} {\bibinfo {author} {\bibfnamefont {E.}~\bibnamefont
  {Eriksson}}, \bibinfo {author} {\bibfnamefont {R.-P.}\ \bibnamefont {Riwar}},
  \bibinfo {author} {\bibfnamefont {M.}~\bibnamefont {Houzet}}, \bibinfo
  {author} {\bibfnamefont {J.~S.}\ \bibnamefont {Meyer}},\ and\ \bibinfo
  {author} {\bibfnamefont {Y.~V.}\ \bibnamefont {Nazarov}},\ }\bibfield
  {title} {\bibinfo {title} {Topological transconductance quantization in a
  four-terminal {J}osephson junction},\ }\href
  {https://doi.org/10.1103/PhysRevB.95.075417} {\bibfield  {journal} {\bibinfo
  {journal} {Phys. Rev. B}\ }\textbf {\bibinfo {volume} {95}},\ \bibinfo
  {pages} {075417} (\bibinfo {year} {2017})}\BibitemShut {NoStop}%
\bibitem [{\citenamefont {Xie}\ \emph {et~al.}(2018)\citenamefont {Xie},
  \citenamefont {Vavilov},\ and\ \citenamefont {Levchenko}}]{Xie2018}%
  \BibitemOpen
  \bibfield  {author} {\bibinfo {author} {\bibfnamefont {H.-Y.}\ \bibnamefont
  {Xie}}, \bibinfo {author} {\bibfnamefont {M.~G.}\ \bibnamefont {Vavilov}},\
  and\ \bibinfo {author} {\bibfnamefont {A.}~\bibnamefont {Levchenko}},\
  }\bibfield  {title} {\bibinfo {title} {Weyl nodes in {A}ndreev spectra of
  multiterminal {J}osephson junctions: {C}hern numbers, conductances, and
  supercurrents},\ }\href {https://doi.org/10.1103/PhysRevB.97.035443}
  {\bibfield  {journal} {\bibinfo  {journal} {Phys. Rev. B}\ }\textbf {\bibinfo
  {volume} {97}},\ \bibinfo {pages} {035443} (\bibinfo {year}
  {2018})}\BibitemShut {NoStop}%
\bibitem [{\citenamefont {Klees}\ \emph {et~al.}(2020)\citenamefont {Klees},
  \citenamefont {Rastelli}, \citenamefont {Cuevas},\ and\ \citenamefont
  {Belzig}}]{Klees2020}%
  \BibitemOpen
  \bibfield  {author} {\bibinfo {author} {\bibfnamefont {R.~L.}\ \bibnamefont
  {Klees}}, \bibinfo {author} {\bibfnamefont {G.}~\bibnamefont {Rastelli}},
  \bibinfo {author} {\bibfnamefont {J.~C.}\ \bibnamefont {Cuevas}},\ and\
  \bibinfo {author} {\bibfnamefont {W.}~\bibnamefont {Belzig}},\ }\bibfield
  {title} {\bibinfo {title} {Microwave spectroscopy reveals the quantum
  geometric tensor of topological {J}osephson matter},\ }\href
  {https://doi.org/10.1103/PhysRevLett.124.197002} {\bibfield  {journal}
  {\bibinfo  {journal} {Phys. Rev. Lett.}\ }\textbf {\bibinfo {volume} {124}},\
  \bibinfo {pages} {197002} (\bibinfo {year} {2020})}\BibitemShut {NoStop}%
\bibitem [{\citenamefont {Xie}\ \emph {et~al.}(2022)\citenamefont {Xie},
  \citenamefont {Hasan},\ and\ \citenamefont {Levchenko}}]{Xie2022}%
  \BibitemOpen
  \bibfield  {author} {\bibinfo {author} {\bibfnamefont {H.-Y.}\ \bibnamefont
  {Xie}}, \bibinfo {author} {\bibfnamefont {J.}~\bibnamefont {Hasan}},\ and\
  \bibinfo {author} {\bibfnamefont {A.}~\bibnamefont {Levchenko}},\ }\bibfield
  {title} {\bibinfo {title} {Non-{A}belian monopoles in the multiterminal
  {J}osephson effect},\ }\href {https://doi.org/10.1103/PhysRevB.105.L241404}
  {\bibfield  {journal} {\bibinfo  {journal} {Phys. Rev. B}\ }\textbf {\bibinfo
  {volume} {105}},\ \bibinfo {pages} {L241404} (\bibinfo {year}
  {2022})}\BibitemShut {NoStop}%
\bibitem [{\citenamefont {Repin}\ and\ \citenamefont
  {Nazarov}(2022)}]{Repin2022}%
  \BibitemOpen
  \bibfield  {author} {\bibinfo {author} {\bibfnamefont {E.~V.}\ \bibnamefont
  {Repin}}\ and\ \bibinfo {author} {\bibfnamefont {Y.~V.}\ \bibnamefont
  {Nazarov}},\ }\bibfield  {title} {\bibinfo {title} {Weyl points in
  multiterminal hybrid superconductor-semiconductor nanowire devices},\ }\href
  {https://doi.org/10.1103/PhysRevB.105.L041405} {\bibfield  {journal}
  {\bibinfo  {journal} {Phys. Rev. B}\ }\textbf {\bibinfo {volume} {105}},\
  \bibinfo {pages} {L041405} (\bibinfo {year} {2022})}\BibitemShut {NoStop}%
\bibitem [{\citenamefont {Teshler}\ \emph {et~al.}(2023)\citenamefont
  {Teshler}, \citenamefont {Weisbrich}, \citenamefont {Sturm}, \citenamefont
  {Klees}, \citenamefont {Rastelli},\ and\ \citenamefont
  {Belzig}}]{Teshler2023a}%
  \BibitemOpen
  \bibfield  {author} {\bibinfo {author} {\bibfnamefont {L.}~\bibnamefont
  {Teshler}}, \bibinfo {author} {\bibfnamefont {H.}~\bibnamefont {Weisbrich}},
  \bibinfo {author} {\bibfnamefont {J.}~\bibnamefont {Sturm}}, \bibinfo
  {author} {\bibfnamefont {R.~L.}\ \bibnamefont {Klees}}, \bibinfo {author}
  {\bibfnamefont {G.}~\bibnamefont {Rastelli}},\ and\ \bibinfo {author}
  {\bibfnamefont {W.}~\bibnamefont {Belzig}},\ }\bibfield  {title} {\bibinfo
  {title} {Ground state topology of a four-terminal superconducting double
  quantum dot},\ }\href {https://doi.org/10.21468/SciPostPhys.15.5.214}
  {\bibfield  {journal} {\bibinfo  {journal} {SciPost Phys.}\ }\textbf
  {\bibinfo {volume} {15}},\ \bibinfo {pages} {214} (\bibinfo {year} {2023})},\
  \Eprint {https://arxiv.org/abs/2304.11982} {arXiv:2304.11982 [cond-mat]}
  \BibitemShut {NoStop}%
\bibitem [{\citenamefont {Chen}\ and\ \citenamefont
  {Nazarov}(2021{\natexlab{a}})}]{Chen2021b}%
  \BibitemOpen
  \bibfield  {author} {\bibinfo {author} {\bibfnamefont {Y.}~\bibnamefont
  {Chen}}\ and\ \bibinfo {author} {\bibfnamefont {Y.~V.}\ \bibnamefont
  {Nazarov}},\ }\bibfield  {title} {\bibinfo {title} {Spin {W}eyl quantum unit:
  {A} theoretical proposal},\ }\href
  {https://doi.org/10.1103/PhysRevB.103.045410} {\bibfield  {journal} {\bibinfo
   {journal} {Phys. Rev. B}\ }\textbf {\bibinfo {volume} {103}},\ \bibinfo
  {pages} {045410} (\bibinfo {year} {2021}{\natexlab{a}})}\BibitemShut
  {NoStop}%
\bibitem [{\citenamefont {Boogers}\ \emph {et~al.}(2022)\citenamefont
  {Boogers}, \citenamefont {Erdmanis},\ and\ \citenamefont
  {Nazarov}}]{Boogers2022}%
  \BibitemOpen
  \bibfield  {author} {\bibinfo {author} {\bibfnamefont {V.}~\bibnamefont
  {Boogers}}, \bibinfo {author} {\bibfnamefont {J.}~\bibnamefont {Erdmanis}},\
  and\ \bibinfo {author} {\bibfnamefont {Y.}~\bibnamefont {Nazarov}},\
  }\bibfield  {title} {\bibinfo {title} {Holonomic quantum manipulation in the
  {W}eyl disk},\ }\href {https://doi.org/10.1103/PhysRevB.105.235437}
  {\bibfield  {journal} {\bibinfo  {journal} {Phys. Rev. B}\ }\textbf {\bibinfo
  {volume} {105}},\ \bibinfo {pages} {235437} (\bibinfo {year}
  {2022})}\BibitemShut {NoStop}%
\bibitem [{\citenamefont {Chen}\ and\ \citenamefont
  {Nazarov}(2021{\natexlab{b}})}]{Chen2021}%
  \BibitemOpen
  \bibfield  {author} {\bibinfo {author} {\bibfnamefont {Y.}~\bibnamefont
  {Chen}}\ and\ \bibinfo {author} {\bibfnamefont {Y.~V.}\ \bibnamefont
  {Nazarov}},\ }\bibfield  {title} {\bibinfo {title} {Spintronics with a {W}eyl
  point in superconducting nanostructures},\ }\href
  {https://doi.org/10.1103/PhysRevB.103.165424} {\bibfield  {journal} {\bibinfo
   {journal} {Phys. Rev. B}\ }\textbf {\bibinfo {volume} {103}},\ \bibinfo
  {pages} {165424} (\bibinfo {year} {2021}{\natexlab{b}})}\BibitemShut
  {NoStop}%
\bibitem [{\citenamefont {Freyn}\ \emph {et~al.}(2011)\citenamefont {Freyn},
  \citenamefont {Dou\ifmmode~\mbox{\c{c}}\else \c{c}\fi{}ot}, \citenamefont
  {Feinberg},\ and\ \citenamefont {M\'elin}}]{Freyn2011}%
  \BibitemOpen
  \bibfield  {author} {\bibinfo {author} {\bibfnamefont {A.}~\bibnamefont
  {Freyn}}, \bibinfo {author} {\bibfnamefont {B.}~\bibnamefont
  {Dou\ifmmode~\mbox{\c{c}}\else \c{c}\fi{}ot}}, \bibinfo {author}
  {\bibfnamefont {D.}~\bibnamefont {Feinberg}},\ and\ \bibinfo {author}
  {\bibfnamefont {R.}~\bibnamefont {M\'elin}},\ }\bibfield  {title} {\bibinfo
  {title} {Production of nonlocal quartets and phase-sensitive entanglement in
  a superconducting beam splitter},\ }\href
  {https://doi.org/10.1103/PhysRevLett.106.257005} {\bibfield  {journal}
  {\bibinfo  {journal} {Phys. Rev. Lett.}\ }\textbf {\bibinfo {volume} {106}},\
  \bibinfo {pages} {257005} (\bibinfo {year} {2011})}\BibitemShut {NoStop}%
\bibitem [{\citenamefont {Jonckheere}\ \emph {et~al.}(2013)\citenamefont
  {Jonckheere}, \citenamefont {Rech}, \citenamefont {Martin}, \citenamefont
  {Dou\ifmmode~\mbox{\c{c}}\else \c{c}\fi{}ot}, \citenamefont {Feinberg},\ and\
  \citenamefont {M\'elin}}]{Jonckheere2013}%
  \BibitemOpen
  \bibfield  {author} {\bibinfo {author} {\bibfnamefont {T.}~\bibnamefont
  {Jonckheere}}, \bibinfo {author} {\bibfnamefont {J.}~\bibnamefont {Rech}},
  \bibinfo {author} {\bibfnamefont {T.}~\bibnamefont {Martin}}, \bibinfo
  {author} {\bibfnamefont {B.}~\bibnamefont {Dou\ifmmode~\mbox{\c{c}}\else
  \c{c}\fi{}ot}}, \bibinfo {author} {\bibfnamefont {D.}~\bibnamefont
  {Feinberg}},\ and\ \bibinfo {author} {\bibfnamefont {R.}~\bibnamefont
  {M\'elin}},\ }\bibfield  {title} {\bibinfo {title} {Multipair dc {J}osephson
  resonances in a biased all-superconducting bijunction},\ }\href
  {https://doi.org/10.1103/PhysRevB.87.214501} {\bibfield  {journal} {\bibinfo
  {journal} {Phys. Rev. B}\ }\textbf {\bibinfo {volume} {87}},\ \bibinfo
  {pages} {214501} (\bibinfo {year} {2013})}\BibitemShut {NoStop}%
\bibitem [{\citenamefont {Pfeffer}\ \emph {et~al.}(2014)\citenamefont
  {Pfeffer}, \citenamefont {Duvauchelle}, \citenamefont {Courtois},
  \citenamefont {M\'elin}, \citenamefont {Feinberg},\ and\ \citenamefont
  {Lefloch}}]{Pfeffer2014}%
  \BibitemOpen
  \bibfield  {author} {\bibinfo {author} {\bibfnamefont {A.~H.}\ \bibnamefont
  {Pfeffer}}, \bibinfo {author} {\bibfnamefont {J.~E.}\ \bibnamefont
  {Duvauchelle}}, \bibinfo {author} {\bibfnamefont {H.}~\bibnamefont
  {Courtois}}, \bibinfo {author} {\bibfnamefont {R.}~\bibnamefont {M\'elin}},
  \bibinfo {author} {\bibfnamefont {D.}~\bibnamefont {Feinberg}},\ and\
  \bibinfo {author} {\bibfnamefont {F.}~\bibnamefont {Lefloch}},\ }\bibfield
  {title} {\bibinfo {title} {Subgap structure in the conductance of a
  three-terminal {J}osephson junction},\ }\href
  {https://doi.org/10.1103/PhysRevB.90.075401} {\bibfield  {journal} {\bibinfo
  {journal} {Phys. Rev. B}\ }\textbf {\bibinfo {volume} {90}},\ \bibinfo
  {pages} {075401} (\bibinfo {year} {2014})}\BibitemShut {NoStop}%
\bibitem [{\citenamefont {Cohen}\ \emph {et~al.}(2018)\citenamefont {Cohen},
  \citenamefont {Ronen}, \citenamefont {Kang}, \citenamefont {Heiblum},
  \citenamefont {Feinberg}, \citenamefont {M{\'{e}}lin},\ and\ \citenamefont
  {Shtrikman}}]{Cohen2018}%
  \BibitemOpen
  \bibfield  {author} {\bibinfo {author} {\bibfnamefont {Y.}~\bibnamefont
  {Cohen}}, \bibinfo {author} {\bibfnamefont {Y.}~\bibnamefont {Ronen}},
  \bibinfo {author} {\bibfnamefont {J.-H.}\ \bibnamefont {Kang}}, \bibinfo
  {author} {\bibfnamefont {M.}~\bibnamefont {Heiblum}}, \bibinfo {author}
  {\bibfnamefont {D.}~\bibnamefont {Feinberg}}, \bibinfo {author}
  {\bibfnamefont {R.}~\bibnamefont {M{\'{e}}lin}},\ and\ \bibinfo {author}
  {\bibfnamefont {H.}~\bibnamefont {Shtrikman}},\ }\bibfield  {title} {\bibinfo
  {title} {Nonlocal supercurrent of quartets in a three-terminal {J}osephson
  junction},\ }\href {https://doi.org/10.1073/pnas.1800044115} {\bibfield
  {journal} {\bibinfo  {journal} {Proc. Natl. Acad. Sci. U.S.A.}\ }\textbf
  {\bibinfo {volume} {115}},\ \bibinfo {pages} {6991} (\bibinfo {year}
  {2018})}\BibitemShut {NoStop}%
\bibitem [{\citenamefont {Huang}\ \emph {et~al.}(2022)\citenamefont {Huang},
  \citenamefont {Ronen}, \citenamefont {M{\'e}lin}, \citenamefont {Feinberg},
  \citenamefont {Watanabe}, \citenamefont {Taniguchi},\ and\ \citenamefont
  {Kim}}]{Huang2022}%
  \BibitemOpen
  \bibfield  {author} {\bibinfo {author} {\bibfnamefont {K.-F.}\ \bibnamefont
  {Huang}}, \bibinfo {author} {\bibfnamefont {Y.}~\bibnamefont {Ronen}},
  \bibinfo {author} {\bibfnamefont {R.}~\bibnamefont {M{\'e}lin}}, \bibinfo
  {author} {\bibfnamefont {D.}~\bibnamefont {Feinberg}}, \bibinfo {author}
  {\bibfnamefont {K.}~\bibnamefont {Watanabe}}, \bibinfo {author}
  {\bibfnamefont {T.}~\bibnamefont {Taniguchi}},\ and\ \bibinfo {author}
  {\bibfnamefont {P.}~\bibnamefont {Kim}},\ }\bibfield  {title} {\bibinfo
  {title} {Evidence for 4e charge of {C}ooper quartets in a biased
  multi-terminal graphene-based {J}osephson junction},\ }\href
  {https://doi.org/10.1038/s41467-022-30732-7} {\bibfield  {journal} {\bibinfo
  {journal} {Nat. Commun.}\ }\textbf {\bibinfo {volume} {13}},\ \bibinfo
  {pages} {3032} (\bibinfo {year} {2022})}\BibitemShut {NoStop}%
\bibitem [{\citenamefont {Draelos}\ \emph {et~al.}(2019)\citenamefont
  {Draelos}, \citenamefont {Wei}, \citenamefont {Seredinski}, \citenamefont
  {Li}, \citenamefont {Mehta}, \citenamefont {Watanabe}, \citenamefont
  {Taniguchi}, \citenamefont {Borzenets}, \citenamefont {Amet},\ and\
  \citenamefont {Finkelstein}}]{Draelos2019}%
  \BibitemOpen
  \bibfield  {author} {\bibinfo {author} {\bibfnamefont {A.~W.}\ \bibnamefont
  {Draelos}}, \bibinfo {author} {\bibfnamefont {M.-T.}\ \bibnamefont {Wei}},
  \bibinfo {author} {\bibfnamefont {A.}~\bibnamefont {Seredinski}}, \bibinfo
  {author} {\bibfnamefont {H.}~\bibnamefont {Li}}, \bibinfo {author}
  {\bibfnamefont {Y.}~\bibnamefont {Mehta}}, \bibinfo {author} {\bibfnamefont
  {K.}~\bibnamefont {Watanabe}}, \bibinfo {author} {\bibfnamefont
  {T.}~\bibnamefont {Taniguchi}}, \bibinfo {author} {\bibfnamefont {I.~V.}\
  \bibnamefont {Borzenets}}, \bibinfo {author} {\bibfnamefont {F.}~\bibnamefont
  {Amet}},\ and\ \bibinfo {author} {\bibfnamefont {G.}~\bibnamefont
  {Finkelstein}},\ }\bibfield  {title} {\bibinfo {title} {Supercurrent flow in
  multiterminal graphene {J}osephson junctions},\ }\href
  {https://doi.org/10.1021/acs.nanolett.8b04330} {\bibfield  {journal}
  {\bibinfo  {journal} {Nano Lett.}\ }\textbf {\bibinfo {volume} {19}},\
  \bibinfo {pages} {1039} (\bibinfo {year} {2019})}\BibitemShut {NoStop}%
\bibitem [{\citenamefont {Graziano}\ \emph {et~al.}(2020)\citenamefont
  {Graziano}, \citenamefont {Lee}, \citenamefont {Pendharkar}, \citenamefont
  {Palmstr\o{}m},\ and\ \citenamefont {Pribiag}}]{Graziano2020}%
  \BibitemOpen
  \bibfield  {author} {\bibinfo {author} {\bibfnamefont {G.~V.}\ \bibnamefont
  {Graziano}}, \bibinfo {author} {\bibfnamefont {J.~S.}\ \bibnamefont {Lee}},
  \bibinfo {author} {\bibfnamefont {M.}~\bibnamefont {Pendharkar}}, \bibinfo
  {author} {\bibfnamefont {C.~J.}\ \bibnamefont {Palmstr\o{}m}},\ and\ \bibinfo
  {author} {\bibfnamefont {V.~S.}\ \bibnamefont {Pribiag}},\ }\bibfield
  {title} {\bibinfo {title} {Transport studies in a gate-tunable three-terminal
  {J}osephson junction},\ }\href {https://doi.org/10.1103/PhysRevB.101.054510}
  {\bibfield  {journal} {\bibinfo  {journal} {Phys. Rev. B}\ }\textbf {\bibinfo
  {volume} {101}},\ \bibinfo {pages} {054510} (\bibinfo {year}
  {2020})}\BibitemShut {NoStop}%
\bibitem [{\citenamefont {Pankratova}\ \emph {et~al.}(2020)\citenamefont
  {Pankratova}, \citenamefont {Lee}, \citenamefont {Kuzmin}, \citenamefont
  {Wickramasinghe}, \citenamefont {Mayer}, \citenamefont {Yuan}, \citenamefont
  {Vavilov}, \citenamefont {Shabani},\ and\ \citenamefont
  {Manucharyan}}]{Pankratova2020}%
  \BibitemOpen
  \bibfield  {author} {\bibinfo {author} {\bibfnamefont {N.}~\bibnamefont
  {Pankratova}}, \bibinfo {author} {\bibfnamefont {H.}~\bibnamefont {Lee}},
  \bibinfo {author} {\bibfnamefont {R.}~\bibnamefont {Kuzmin}}, \bibinfo
  {author} {\bibfnamefont {K.}~\bibnamefont {Wickramasinghe}}, \bibinfo
  {author} {\bibfnamefont {W.}~\bibnamefont {Mayer}}, \bibinfo {author}
  {\bibfnamefont {J.}~\bibnamefont {Yuan}}, \bibinfo {author} {\bibfnamefont
  {M.~G.}\ \bibnamefont {Vavilov}}, \bibinfo {author} {\bibfnamefont
  {J.}~\bibnamefont {Shabani}},\ and\ \bibinfo {author} {\bibfnamefont {V.~E.}\
  \bibnamefont {Manucharyan}},\ }\bibfield  {title} {\bibinfo {title}
  {Multiterminal {J}osephson effect},\ }\href
  {https://doi.org/10.1103/PhysRevX.10.031051} {\bibfield  {journal} {\bibinfo
  {journal} {Phys. Rev. X}\ }\textbf {\bibinfo {volume} {10}},\ \bibinfo
  {pages} {031051} (\bibinfo {year} {2020})}\BibitemShut {NoStop}%
\bibitem [{\citenamefont {Arnault}\ \emph {et~al.}(2021)\citenamefont
  {Arnault}, \citenamefont {Larson}, \citenamefont {Seredinski}, \citenamefont
  {Zhao}, \citenamefont {Idris}, \citenamefont {McConnell}, \citenamefont
  {Watanabe}, \citenamefont {Taniguchi}, \citenamefont {Borzenets},
  \citenamefont {Amet},\ and\ \citenamefont {Finkelstein}}]{Arnault2021}%
  \BibitemOpen
  \bibfield  {author} {\bibinfo {author} {\bibfnamefont {E.~G.}\ \bibnamefont
  {Arnault}}, \bibinfo {author} {\bibfnamefont {T.~F.~Q.}\ \bibnamefont
  {Larson}}, \bibinfo {author} {\bibfnamefont {A.}~\bibnamefont {Seredinski}},
  \bibinfo {author} {\bibfnamefont {L.}~\bibnamefont {Zhao}}, \bibinfo {author}
  {\bibfnamefont {S.}~\bibnamefont {Idris}}, \bibinfo {author} {\bibfnamefont
  {A.}~\bibnamefont {McConnell}}, \bibinfo {author} {\bibfnamefont
  {K.}~\bibnamefont {Watanabe}}, \bibinfo {author} {\bibfnamefont
  {T.}~\bibnamefont {Taniguchi}}, \bibinfo {author} {\bibfnamefont
  {I.}~\bibnamefont {Borzenets}}, \bibinfo {author} {\bibfnamefont
  {F.}~\bibnamefont {Amet}},\ and\ \bibinfo {author} {\bibfnamefont
  {G.}~\bibnamefont {Finkelstein}},\ }\bibfield  {title} {\bibinfo {title}
  {Multiterminal inverse {AC} {J}osephson effect},\ }\href
  {https://doi.org/10.1021/acs.nanolett.1c03474} {\bibfield  {journal}
  {\bibinfo  {journal} {Nano Lett.}\ }\textbf {\bibinfo {volume} {21}},\
  \bibinfo {pages} {9668} (\bibinfo {year} {2021})}\BibitemShut {NoStop}%
\bibitem [{\citenamefont {Graziano}\ \emph {et~al.}(2022)\citenamefont
  {Graziano}, \citenamefont {Gupta}, \citenamefont {Pendharkar}, \citenamefont
  {Dong}, \citenamefont {Dempsey}, \citenamefont {Palmstrøm},\ and\
  \citenamefont {Pribiag}}]{Graziano2022}%
  \BibitemOpen
  \bibfield  {author} {\bibinfo {author} {\bibfnamefont {G.~V.}\ \bibnamefont
  {Graziano}}, \bibinfo {author} {\bibfnamefont {M.}~\bibnamefont {Gupta}},
  \bibinfo {author} {\bibfnamefont {M.}~\bibnamefont {Pendharkar}}, \bibinfo
  {author} {\bibfnamefont {J.~T.}\ \bibnamefont {Dong}}, \bibinfo {author}
  {\bibfnamefont {C.~P.}\ \bibnamefont {Dempsey}}, \bibinfo {author}
  {\bibfnamefont {C.}~\bibnamefont {Palmstrøm}},\ and\ \bibinfo {author}
  {\bibfnamefont {V.~S.}\ \bibnamefont {Pribiag}},\ }\bibfield  {title}
  {\bibinfo {title} {Selective control of conductance modes in multi-terminal
  {J}osephson junctions},\ }\href {https://doi.org/10.1038/s41467-022-33682-2}
  {\bibfield  {journal} {\bibinfo  {journal} {Nat. Commun.}\ }\textbf {\bibinfo
  {volume} {13}},\ \bibinfo {pages} {5933} (\bibinfo {year}
  {2022})}\BibitemShut {NoStop}%
\bibitem [{\citenamefont {Gupta}\ \emph {et~al.}(2023)\citenamefont {Gupta},
  \citenamefont {Graziano}, \citenamefont {Pendharkar}, \citenamefont {Dong},
  \citenamefont {Dempsey}, \citenamefont {Palmstrøm},\ and\ \citenamefont
  {Pribiag}}]{Gupta2023}%
  \BibitemOpen
  \bibfield  {author} {\bibinfo {author} {\bibfnamefont {M.}~\bibnamefont
  {Gupta}}, \bibinfo {author} {\bibfnamefont {G.~V.}\ \bibnamefont {Graziano}},
  \bibinfo {author} {\bibfnamefont {M.}~\bibnamefont {Pendharkar}}, \bibinfo
  {author} {\bibfnamefont {J.~T.}\ \bibnamefont {Dong}}, \bibinfo {author}
  {\bibfnamefont {C.~P.}\ \bibnamefont {Dempsey}}, \bibinfo {author}
  {\bibfnamefont {C.}~\bibnamefont {Palmstrøm}},\ and\ \bibinfo {author}
  {\bibfnamefont {V.~S.}\ \bibnamefont {Pribiag}},\ }\bibfield  {title}
  {\bibinfo {title} {Gate-tunable superconducting diode effect in a
  three-terminal {J}osephson device},\ }\href
  {https://doi.org/10.1038/s41467-023-38856-0} {\bibfield  {journal} {\bibinfo
  {journal} {Nat. Commun.}\ }\textbf {\bibinfo {volume} {14}},\ \bibinfo
  {pages} {3078} (\bibinfo {year} {2023})}\BibitemShut {NoStop}%
\bibitem [{\citenamefont {Coraiola}\ \emph
  {et~al.}(2024{\natexlab{a}})\citenamefont {Coraiola}, \citenamefont
  {Svetogorov}, \citenamefont {Haxell}, \citenamefont {Sabonis}, \citenamefont
  {Hinderling}, \citenamefont {ten Kate}, \citenamefont {Cheah}, \citenamefont
  {Krizek}, \citenamefont {Schott}, \citenamefont {Wegscheider}, \citenamefont
  {Cuevas}, \citenamefont {Belzig},\ and\ \citenamefont
  {Nichele}}]{Coraiola2024}%
  \BibitemOpen
  \bibfield  {author} {\bibinfo {author} {\bibfnamefont {M.}~\bibnamefont
  {Coraiola}}, \bibinfo {author} {\bibfnamefont {A.~E.}\ \bibnamefont
  {Svetogorov}}, \bibinfo {author} {\bibfnamefont {D.~Z.}\ \bibnamefont
  {Haxell}}, \bibinfo {author} {\bibfnamefont {D.}~\bibnamefont {Sabonis}},
  \bibinfo {author} {\bibfnamefont {M.}~\bibnamefont {Hinderling}}, \bibinfo
  {author} {\bibfnamefont {S.~C.}\ \bibnamefont {ten Kate}}, \bibinfo {author}
  {\bibfnamefont {E.}~\bibnamefont {Cheah}}, \bibinfo {author} {\bibfnamefont
  {F.}~\bibnamefont {Krizek}}, \bibinfo {author} {\bibfnamefont
  {R.}~\bibnamefont {Schott}}, \bibinfo {author} {\bibfnamefont
  {W.}~\bibnamefont {Wegscheider}}, \bibinfo {author} {\bibfnamefont {J.~C.}\
  \bibnamefont {Cuevas}}, \bibinfo {author} {\bibfnamefont {W.}~\bibnamefont
  {Belzig}},\ and\ \bibinfo {author} {\bibfnamefont {F.}~\bibnamefont
  {Nichele}},\ }\bibfield  {title} {\bibinfo {title} {Flux-tunable {J}osephson
  diode effect in a hybrid four-terminal {J}osephson junction},\ }\href
  {https://doi.org/10.1021/acsnano.4c01642} {\bibfield  {journal} {\bibinfo
  {journal} {ACS Nano}\ }\textbf {\bibinfo {volume} {18}},\ \bibinfo {pages}
  {9221–9231} (\bibinfo {year} {2024}{\natexlab{a}})}\BibitemShut {NoStop}%
\bibitem [{\citenamefont {Pillet}\ \emph {et~al.}(2019)\citenamefont {Pillet},
  \citenamefont {Benzoni}, \citenamefont {Griesmar}, \citenamefont {Smirr},\
  and\ \citenamefont {Girit}}]{Pillet2019}%
  \BibitemOpen
  \bibfield  {author} {\bibinfo {author} {\bibfnamefont {J.-D.}\ \bibnamefont
  {Pillet}}, \bibinfo {author} {\bibfnamefont {V.}~\bibnamefont {Benzoni}},
  \bibinfo {author} {\bibfnamefont {J.}~\bibnamefont {Griesmar}}, \bibinfo
  {author} {\bibfnamefont {J.-L.}\ \bibnamefont {Smirr}},\ and\ \bibinfo
  {author} {\bibfnamefont {{\c{C}}.~O.}\ \bibnamefont {Girit}},\ }\bibfield
  {title} {\bibinfo {title} {Nonlocal {J}osephson effect in {A}ndreev
  molecules},\ }\href {https://doi.org/10.1021/acs.nanolett.9b02686} {\bibfield
   {journal} {\bibinfo  {journal} {Nano Lett.}\ }\textbf {\bibinfo {volume}
  {19}},\ \bibinfo {pages} {7138} (\bibinfo {year} {2019})}\BibitemShut
  {NoStop}%
\bibitem [{\citenamefont {Kornich}\ \emph {et~al.}(2019)\citenamefont
  {Kornich}, \citenamefont {Barakov},\ and\ \citenamefont
  {Nazarov}}]{Kornich2019}%
  \BibitemOpen
  \bibfield  {author} {\bibinfo {author} {\bibfnamefont {V.}~\bibnamefont
  {Kornich}}, \bibinfo {author} {\bibfnamefont {H.~S.}\ \bibnamefont
  {Barakov}},\ and\ \bibinfo {author} {\bibfnamefont {Y.~V.}\ \bibnamefont
  {Nazarov}},\ }\bibfield  {title} {\bibinfo {title} {Fine energy splitting of
  overlapping {A}ndreev bound states in multiterminal superconducting
  nanostructures},\ }\href {https://doi.org/10.1103/PhysRevResearch.1.033004}
  {\bibfield  {journal} {\bibinfo  {journal} {Phys. Rev. Res.}\ }\textbf
  {\bibinfo {volume} {1}},\ \bibinfo {pages} {033004} (\bibinfo {year}
  {2019})}\BibitemShut {NoStop}%
\bibitem [{\citenamefont {Keliri}\ and\ \citenamefont
  {Dou\ifmmode~\mbox{\c{c}}\else \c{c}\fi{}ot}(2023)}]{Keliri2023}%
  \BibitemOpen
  \bibfield  {author} {\bibinfo {author} {\bibfnamefont {A.}~\bibnamefont
  {Keliri}}\ and\ \bibinfo {author} {\bibfnamefont {B.}~\bibnamefont
  {Dou\ifmmode~\mbox{\c{c}}\else \c{c}\fi{}ot}},\ }\bibfield  {title} {\bibinfo
  {title} {Driven {A}ndreev molecule},\ }\href
  {https://doi.org/10.1103/PhysRevB.107.094505} {\bibfield  {journal} {\bibinfo
   {journal} {Phys. Rev. B}\ }\textbf {\bibinfo {volume} {107}},\ \bibinfo
  {pages} {094505} (\bibinfo {year} {2023})}\BibitemShut {NoStop}%
\bibitem [{\citenamefont {Kocsis}\ \emph {et~al.}(2023)\citenamefont {Kocsis},
  \citenamefont {Scher\"{u}bl}, \citenamefont {F\"{u}l\"{o}p}, \citenamefont
  {Makk},\ and\ \citenamefont {Csonka}}]{Kocsis2023}%
  \BibitemOpen
  \bibfield  {author} {\bibinfo {author} {\bibfnamefont {M.}~\bibnamefont
  {Kocsis}}, \bibinfo {author} {\bibfnamefont {Z.}~\bibnamefont
  {Scher\"{u}bl}}, \bibinfo {author} {\bibfnamefont {G.}~\bibnamefont
  {F\"{u}l\"{o}p}}, \bibinfo {author} {\bibfnamefont {P.}~\bibnamefont
  {Makk}},\ and\ \bibinfo {author} {\bibfnamefont {S.}~\bibnamefont {Csonka}},\
  }\href {https://arxiv.org/abs/2303.14842} {\bibinfo {title} {Strong nonlocal
  tuning of the current-phase relation of a quantum dot based {A}ndreev
  molecule}} (\bibinfo {year} {2023}),\ \Eprint
  {https://arxiv.org/abs/2303.14842} {arXiv:2303.14842} \BibitemShut {NoStop}%
\bibitem [{\citenamefont {Johannsen}\ and\ \citenamefont
  {Schrade}(2024)}]{Johannsen2024}%
  \BibitemOpen
  \bibfield  {author} {\bibinfo {author} {\bibfnamefont {P.~D.}\ \bibnamefont
  {Johannsen}}\ and\ \bibinfo {author} {\bibfnamefont {C.}~\bibnamefont
  {Schrade}},\ }\href {https://arxiv.org/abs/2404.12430} {\bibinfo {title}
  {{Fermionic quantum simulation on Andreev bound state superlattices}}}
  (\bibinfo {year} {2024}),\ \Eprint {https://arxiv.org/abs/2404.12430}
  {arXiv:2404.12430 [cond-mat.mes-hall]} \BibitemShut {NoStop}%
\bibitem [{\citenamefont {Matsuo}\ \emph {et~al.}(2022)\citenamefont {Matsuo},
  \citenamefont {Lee}, \citenamefont {Chang}, \citenamefont {Sato},
  \citenamefont {Ueda}, \citenamefont {Palmstr{\o}m},\ and\ \citenamefont
  {Tarucha}}]{Matsuo2022}%
  \BibitemOpen
  \bibfield  {author} {\bibinfo {author} {\bibfnamefont {S.}~\bibnamefont
  {Matsuo}}, \bibinfo {author} {\bibfnamefont {J.~S.}\ \bibnamefont {Lee}},
  \bibinfo {author} {\bibfnamefont {C.-Y.}\ \bibnamefont {Chang}}, \bibinfo
  {author} {\bibfnamefont {Y.}~\bibnamefont {Sato}}, \bibinfo {author}
  {\bibfnamefont {K.}~\bibnamefont {Ueda}}, \bibinfo {author} {\bibfnamefont
  {C.~J.}\ \bibnamefont {Palmstr{\o}m}},\ and\ \bibinfo {author} {\bibfnamefont
  {S.}~\bibnamefont {Tarucha}},\ }\bibfield  {title} {\bibinfo {title}
  {Observation of nonlocal {J}osephson effect on double {InAs} nanowires},\
  }\href {https://doi.org/10.1038/s42005-022-00994-0} {\bibfield  {journal}
  {\bibinfo  {journal} {Commun. Phys.}\ }\textbf {\bibinfo {volume} {5}},\
  \bibinfo {pages} {221} (\bibinfo {year} {2022})}\BibitemShut {NoStop}%
\bibitem [{\citenamefont {Haxell}\ \emph {et~al.}(2023)\citenamefont {Haxell},
  \citenamefont {Coraiola}, \citenamefont {Hinderling}, \citenamefont {ten
  Kate}, \citenamefont {Sabonis}, \citenamefont {Svetogorov}, \citenamefont
  {Belzig}, \citenamefont {Cheah}, \citenamefont {Krizek}, \citenamefont
  {Schott}, \citenamefont {Wegscheider},\ and\ \citenamefont
  {Nichele}}]{Haxell2023}%
  \BibitemOpen
  \bibfield  {author} {\bibinfo {author} {\bibfnamefont {D.~Z.}\ \bibnamefont
  {Haxell}}, \bibinfo {author} {\bibfnamefont {M.}~\bibnamefont {Coraiola}},
  \bibinfo {author} {\bibfnamefont {M.}~\bibnamefont {Hinderling}}, \bibinfo
  {author} {\bibfnamefont {S.~C.}\ \bibnamefont {ten Kate}}, \bibinfo {author}
  {\bibfnamefont {D.}~\bibnamefont {Sabonis}}, \bibinfo {author} {\bibfnamefont
  {A.~E.}\ \bibnamefont {Svetogorov}}, \bibinfo {author} {\bibfnamefont
  {W.}~\bibnamefont {Belzig}}, \bibinfo {author} {\bibfnamefont
  {E.}~\bibnamefont {Cheah}}, \bibinfo {author} {\bibfnamefont
  {F.}~\bibnamefont {Krizek}}, \bibinfo {author} {\bibfnamefont
  {R.}~\bibnamefont {Schott}}, \bibinfo {author} {\bibfnamefont
  {W.}~\bibnamefont {Wegscheider}},\ and\ \bibinfo {author} {\bibfnamefont
  {F.}~\bibnamefont {Nichele}},\ }\bibfield  {title} {\bibinfo {title}
  {Demonstration of the nonlocal {J}osephson effect in {A}ndreev molecules},\
  }\href {https://doi.org/10.1021/acs.nanolett.3c02066} {\bibfield  {journal}
  {\bibinfo  {journal} {Nano Lett.}\ }\textbf {\bibinfo {volume} {23}},\
  \bibinfo {pages} {7532–7538} (\bibinfo {year} {2023})}\BibitemShut
  {NoStop}%
\bibitem [{\citenamefont {Matsuo}\ \emph
  {et~al.}(2023{\natexlab{a}})\citenamefont {Matsuo}, \citenamefont {Imoto},
  \citenamefont {Yokoyama}, \citenamefont {Sato}, \citenamefont {Lindemann},
  \citenamefont {Gronin}, \citenamefont {Gardner}, \citenamefont {Manfra},\
  and\ \citenamefont {Tarucha}}]{Matsuo2023b}%
  \BibitemOpen
  \bibfield  {author} {\bibinfo {author} {\bibfnamefont {S.}~\bibnamefont
  {Matsuo}}, \bibinfo {author} {\bibfnamefont {T.}~\bibnamefont {Imoto}},
  \bibinfo {author} {\bibfnamefont {T.}~\bibnamefont {Yokoyama}}, \bibinfo
  {author} {\bibfnamefont {Y.}~\bibnamefont {Sato}}, \bibinfo {author}
  {\bibfnamefont {T.}~\bibnamefont {Lindemann}}, \bibinfo {author}
  {\bibfnamefont {S.}~\bibnamefont {Gronin}}, \bibinfo {author} {\bibfnamefont
  {G.~C.}\ \bibnamefont {Gardner}}, \bibinfo {author} {\bibfnamefont {M.~J.}\
  \bibnamefont {Manfra}},\ and\ \bibinfo {author} {\bibfnamefont
  {S.}~\bibnamefont {Tarucha}},\ }\bibfield  {title} {\bibinfo {title} {Phase
  engineering of anomalous {J}osephson effect derived from {A}ndreev
  molecules},\ }\href {https://doi.org/10.1126/sciadv.adj3698} {\bibfield
  {journal} {\bibinfo  {journal} {Sci. Adv.}\ }\textbf {\bibinfo {volume}
  {9}},\ \bibinfo {pages} {eadj369} (\bibinfo {year}
  {2023}{\natexlab{a}})}\BibitemShut {NoStop}%
\bibitem [{\citenamefont {Prosko}\ \emph {et~al.}(2024)\citenamefont {Prosko},
  \citenamefont {Huisman}, \citenamefont {Kulesh}, \citenamefont {Xiao},
  \citenamefont {Thomas}, \citenamefont {Manfra},\ and\ \citenamefont
  {Goswami}}]{Prosko2024}%
  \BibitemOpen
  \bibfield  {author} {\bibinfo {author} {\bibfnamefont {C.~G.}\ \bibnamefont
  {Prosko}}, \bibinfo {author} {\bibfnamefont {W.~D.}\ \bibnamefont {Huisman}},
  \bibinfo {author} {\bibfnamefont {I.}~\bibnamefont {Kulesh}}, \bibinfo
  {author} {\bibfnamefont {D.}~\bibnamefont {Xiao}}, \bibinfo {author}
  {\bibfnamefont {C.}~\bibnamefont {Thomas}}, \bibinfo {author} {\bibfnamefont
  {M.~J.}\ \bibnamefont {Manfra}},\ and\ \bibinfo {author} {\bibfnamefont
  {S.}~\bibnamefont {Goswami}},\ }\bibfield  {title} {\bibinfo {title}
  {Flux-tunable {J}osephson effect in a four-terminal junction},\ }\href
  {https://doi.org/10.1103/PhysRevB.110.064518} {\bibfield  {journal} {\bibinfo
   {journal} {Phys. Rev. B}\ }\textbf {\bibinfo {volume} {110}},\ \bibinfo
  {pages} {064518} (\bibinfo {year} {2024})}\BibitemShut {NoStop}%
\bibitem [{\citenamefont {Deutscher}\ and\ \citenamefont
  {Feinberg}(2000)}]{Deutscher2000}%
  \BibitemOpen
  \bibfield  {author} {\bibinfo {author} {\bibfnamefont {G.}~\bibnamefont
  {Deutscher}}\ and\ \bibinfo {author} {\bibfnamefont {D.}~\bibnamefont
  {Feinberg}},\ }\bibfield  {title} {\bibinfo {title} {Coupling
  superconducting-ferromagnetic point contacts by {A}ndreev reflections},\
  }\href {https://doi.org/10.1063/1.125796} {\bibfield  {journal} {\bibinfo
  {journal} {Appl. Phys. Lett.}\ }\textbf {\bibinfo {volume} {76}},\ \bibinfo
  {pages} {487–489} (\bibinfo {year} {2000})}\BibitemShut {NoStop}%
\bibitem [{\citenamefont {Ohnmacht}\ \emph {et~al.}(2024)\citenamefont
  {Ohnmacht}, \citenamefont {Coraiola}, \citenamefont {{Garc{\'i}a-Esteban}},
  \citenamefont {Sabonis}, \citenamefont {Nichele}, \citenamefont {Belzig},\
  and\ \citenamefont {Cuevas}}]{Ohnmacht2024}%
  \BibitemOpen
  \bibfield  {author} {\bibinfo {author} {\bibfnamefont {D.~C.}\ \bibnamefont
  {Ohnmacht}}, \bibinfo {author} {\bibfnamefont {M.}~\bibnamefont {Coraiola}},
  \bibinfo {author} {\bibfnamefont {J.~J.}\ \bibnamefont
  {{Garc{\'i}a-Esteban}}}, \bibinfo {author} {\bibfnamefont {D.}~\bibnamefont
  {Sabonis}}, \bibinfo {author} {\bibfnamefont {F.}~\bibnamefont {Nichele}},
  \bibinfo {author} {\bibfnamefont {W.}~\bibnamefont {Belzig}},\ and\ \bibinfo
  {author} {\bibfnamefont {J.~C.}\ \bibnamefont {Cuevas}},\ }\bibfield  {title}
  {\bibinfo {title} {Quartet tomography in multiterminal {{Josephson}}
  junctions},\ }\href {https://doi.org/10.1103/PhysRevB.109.L241407} {\bibfield
   {journal} {\bibinfo  {journal} {Phys. Rev. B}\ }\textbf {\bibinfo {volume}
  {109}},\ \bibinfo {pages} {L241407} (\bibinfo {year} {2024})}\BibitemShut
  {NoStop}%
\bibitem [{\citenamefont {Kitaev}(2001)}]{Kitaev2001}%
  \BibitemOpen
  \bibfield  {author} {\bibinfo {author} {\bibfnamefont {A.~Y.}\ \bibnamefont
  {Kitaev}},\ }\bibfield  {title} {\bibinfo {title} {Unpaired {M}ajorana
  fermions in quantum wires},\ }\href
  {https://doi.org/10.1070/1063-7869/44/10s/s29} {\bibfield  {journal}
  {\bibinfo  {journal} {Phys.-Uspekhi}\ }\textbf {\bibinfo {volume} {44}},\
  \bibinfo {pages} {131} (\bibinfo {year} {2001})}\BibitemShut {NoStop}%
\bibitem [{\citenamefont {Sau}\ and\ \citenamefont {Sarma}(2012)}]{Sau2012}%
  \BibitemOpen
  \bibfield  {author} {\bibinfo {author} {\bibfnamefont {J.~D.}\ \bibnamefont
  {Sau}}\ and\ \bibinfo {author} {\bibfnamefont {S.~D.}\ \bibnamefont
  {Sarma}},\ }\bibfield  {title} {\bibinfo {title} {Realizing a robust
  practical {M}ajorana chain in a quantum-dot-superconductor linear array},\
  }\href {https://doi.org/10.1038/ncomms1966} {\bibfield  {journal} {\bibinfo
  {journal} {Nat. Commun.}\ }\textbf {\bibinfo {volume} {3}},\ \bibinfo {pages}
  {964} (\bibinfo {year} {2012})}\BibitemShut {NoStop}%
\bibitem [{\citenamefont {Leijnse}\ and\ \citenamefont
  {Flensberg}(2012)}]{Leijnse2012}%
  \BibitemOpen
  \bibfield  {author} {\bibinfo {author} {\bibfnamefont {M.}~\bibnamefont
  {Leijnse}}\ and\ \bibinfo {author} {\bibfnamefont {K.}~\bibnamefont
  {Flensberg}},\ }\bibfield  {title} {\bibinfo {title} {Parity qubits and poor
  man's {M}ajorana bound states in double quantum dots},\ }\href
  {https://doi.org/10.1103/PhysRevB.86.134528} {\bibfield  {journal} {\bibinfo
  {journal} {Phys. Rev. B}\ }\textbf {\bibinfo {volume} {86}},\ \bibinfo
  {pages} {134528} (\bibinfo {year} {2012})}\BibitemShut {NoStop}%
\bibitem [{\citenamefont {Fulga}\ \emph {et~al.}(2013)\citenamefont {Fulga},
  \citenamefont {Haim}, \citenamefont {Akhmerov},\ and\ \citenamefont
  {Oreg}}]{Fulga2013}%
  \BibitemOpen
  \bibfield  {author} {\bibinfo {author} {\bibfnamefont {I.~C.}\ \bibnamefont
  {Fulga}}, \bibinfo {author} {\bibfnamefont {A.}~\bibnamefont {Haim}},
  \bibinfo {author} {\bibfnamefont {A.~R.}\ \bibnamefont {Akhmerov}},\ and\
  \bibinfo {author} {\bibfnamefont {Y.}~\bibnamefont {Oreg}},\ }\bibfield
  {title} {\bibinfo {title} {Adaptive tuning of {M}ajorana fermions in a
  quantum dot chain},\ }\href {https://doi.org/10.1088/1367-2630/15/4/045020}
  {\bibfield  {journal} {\bibinfo  {journal} {New J. Phys.}\ }\textbf {\bibinfo
  {volume} {15}},\ \bibinfo {pages} {045020} (\bibinfo {year}
  {2013})}\BibitemShut {NoStop}%
\bibitem [{\citenamefont {Liu}\ \emph {et~al.}(2022)\citenamefont {Liu},
  \citenamefont {Wang}, \citenamefont {Dvir},\ and\ \citenamefont
  {Wimmer}}]{Liu2022}%
  \BibitemOpen
  \bibfield  {author} {\bibinfo {author} {\bibfnamefont {C.-X.}\ \bibnamefont
  {Liu}}, \bibinfo {author} {\bibfnamefont {G.}~\bibnamefont {Wang}}, \bibinfo
  {author} {\bibfnamefont {T.}~\bibnamefont {Dvir}},\ and\ \bibinfo {author}
  {\bibfnamefont {M.}~\bibnamefont {Wimmer}},\ }\bibfield  {title} {\bibinfo
  {title} {Tunable superconducting coupling of quantum dots via {A}ndreev bound
  states in semiconductor-superconductor nanowires},\ }\href
  {https://doi.org/10.1103/PhysRevLett.129.267701} {\bibfield  {journal}
  {\bibinfo  {journal} {Phys. Rev. Lett.}\ }\textbf {\bibinfo {volume} {129}},\
  \bibinfo {pages} {267701} (\bibinfo {year} {2022})}\BibitemShut {NoStop}%
\bibitem [{\citenamefont {Wang}\ \emph {et~al.}(2022)\citenamefont {Wang},
  \citenamefont {Dvir}, \citenamefont {Mazur}, \citenamefont {Liu},
  \citenamefont {van Loo}, \citenamefont {ten Haaf}, \citenamefont {Bordin},
  \citenamefont {Gazibegovic}, \citenamefont {Badawy}, \citenamefont {Bakkers},
  \citenamefont {Wimmer},\ and\ \citenamefont {Kouwenhoven}}]{Wang2022}%
  \BibitemOpen
  \bibfield  {author} {\bibinfo {author} {\bibfnamefont {G.}~\bibnamefont
  {Wang}}, \bibinfo {author} {\bibfnamefont {T.}~\bibnamefont {Dvir}}, \bibinfo
  {author} {\bibfnamefont {G.~P.}\ \bibnamefont {Mazur}}, \bibinfo {author}
  {\bibfnamefont {C.-X.}\ \bibnamefont {Liu}}, \bibinfo {author} {\bibfnamefont
  {N.}~\bibnamefont {van Loo}}, \bibinfo {author} {\bibfnamefont {S.~L.~D.}\
  \bibnamefont {ten Haaf}}, \bibinfo {author} {\bibfnamefont {A.}~\bibnamefont
  {Bordin}}, \bibinfo {author} {\bibfnamefont {S.}~\bibnamefont {Gazibegovic}},
  \bibinfo {author} {\bibfnamefont {G.}~\bibnamefont {Badawy}}, \bibinfo
  {author} {\bibfnamefont {E.~P. A.~M.}\ \bibnamefont {Bakkers}}, \bibinfo
  {author} {\bibfnamefont {M.}~\bibnamefont {Wimmer}},\ and\ \bibinfo {author}
  {\bibfnamefont {L.~P.}\ \bibnamefont {Kouwenhoven}},\ }\bibfield  {title}
  {\bibinfo {title} {Singlet and triplet {C}ooper pair splitting in hybrid
  superconducting nanowires},\ }\href
  {https://doi.org/10.1038/s41586-022-05352-2} {\bibfield  {journal} {\bibinfo
  {journal} {Nature}\ }\textbf {\bibinfo {volume} {612}},\ \bibinfo {pages}
  {448–453} (\bibinfo {year} {2022})}\BibitemShut {NoStop}%
\bibitem [{\citenamefont {Bordin}\ \emph {et~al.}(2023)\citenamefont {Bordin},
  \citenamefont {Wang}, \citenamefont {Liu}, \citenamefont {ten Haaf},
  \citenamefont {van Loo}, \citenamefont {Mazur}, \citenamefont {Xu},
  \citenamefont {van Driel}, \citenamefont {Zatelli}, \citenamefont
  {Gazibegovic}, \citenamefont {Badawy}, \citenamefont {Bakkers}, \citenamefont
  {Wimmer}, \citenamefont {Kouwenhoven},\ and\ \citenamefont
  {Dvir}}]{Bordin2023}%
  \BibitemOpen
  \bibfield  {author} {\bibinfo {author} {\bibfnamefont {A.}~\bibnamefont
  {Bordin}}, \bibinfo {author} {\bibfnamefont {G.}~\bibnamefont {Wang}},
  \bibinfo {author} {\bibfnamefont {C.-X.}\ \bibnamefont {Liu}}, \bibinfo
  {author} {\bibfnamefont {S.~L.~D.}\ \bibnamefont {ten Haaf}}, \bibinfo
  {author} {\bibfnamefont {N.}~\bibnamefont {van Loo}}, \bibinfo {author}
  {\bibfnamefont {G.~P.}\ \bibnamefont {Mazur}}, \bibinfo {author}
  {\bibfnamefont {D.}~\bibnamefont {Xu}}, \bibinfo {author} {\bibfnamefont
  {D.}~\bibnamefont {van Driel}}, \bibinfo {author} {\bibfnamefont
  {F.}~\bibnamefont {Zatelli}}, \bibinfo {author} {\bibfnamefont
  {S.}~\bibnamefont {Gazibegovic}}, \bibinfo {author} {\bibfnamefont
  {G.}~\bibnamefont {Badawy}}, \bibinfo {author} {\bibfnamefont {E.~P. A.~M.}\
  \bibnamefont {Bakkers}}, \bibinfo {author} {\bibfnamefont {M.}~\bibnamefont
  {Wimmer}}, \bibinfo {author} {\bibfnamefont {L.~P.}\ \bibnamefont
  {Kouwenhoven}},\ and\ \bibinfo {author} {\bibfnamefont {T.}~\bibnamefont
  {Dvir}},\ }\bibfield  {title} {\bibinfo {title} {Tunable crossed {A}ndreev
  reflection and elastic cotunneling in hybrid nanowires},\ }\href
  {https://doi.org/10.1103/PhysRevX.13.031031} {\bibfield  {journal} {\bibinfo
  {journal} {Phys. Rev. X}\ }\textbf {\bibinfo {volume} {13}},\ \bibinfo
  {pages} {031031} (\bibinfo {year} {2023})}\BibitemShut {NoStop}%
\bibitem [{\citenamefont {Bordin}\ \emph
  {et~al.}(2024{\natexlab{a}})\citenamefont {Bordin}, \citenamefont {Li},
  \citenamefont {van Driel}, \citenamefont {Wolff}, \citenamefont {Wang},
  \citenamefont {ten Haaf}, \citenamefont {Wang}, \citenamefont {van Loo},
  \citenamefont {Kouwenhoven},\ and\ \citenamefont {Dvir}}]{Bordin2024}%
  \BibitemOpen
  \bibfield  {author} {\bibinfo {author} {\bibfnamefont {A.}~\bibnamefont
  {Bordin}}, \bibinfo {author} {\bibfnamefont {X.}~\bibnamefont {Li}}, \bibinfo
  {author} {\bibfnamefont {D.}~\bibnamefont {van Driel}}, \bibinfo {author}
  {\bibfnamefont {J.~C.}\ \bibnamefont {Wolff}}, \bibinfo {author}
  {\bibfnamefont {Q.}~\bibnamefont {Wang}}, \bibinfo {author} {\bibfnamefont
  {S.~L.~D.}\ \bibnamefont {ten Haaf}}, \bibinfo {author} {\bibfnamefont
  {G.}~\bibnamefont {Wang}}, \bibinfo {author} {\bibfnamefont {N.}~\bibnamefont
  {van Loo}}, \bibinfo {author} {\bibfnamefont {L.~P.}\ \bibnamefont
  {Kouwenhoven}},\ and\ \bibinfo {author} {\bibfnamefont {T.}~\bibnamefont
  {Dvir}},\ }\bibfield  {title} {\bibinfo {title} {Crossed {A}ndreev reflection
  and elastic cotunneling in three quantum dots coupled by superconductors},\
  }\href {https://doi.org/10.1103/PhysRevLett.132.056602} {\bibfield  {journal}
  {\bibinfo  {journal} {Phys. Rev. Lett.}\ }\textbf {\bibinfo {volume} {132}},\
  \bibinfo {pages} {056602} (\bibinfo {year} {2024}{\natexlab{a}})}\BibitemShut
  {NoStop}%
\bibitem [{\citenamefont {van Driel}\ \emph {et~al.}(2024)\citenamefont {van
  Driel}, \citenamefont {Roovers}, \citenamefont {Zatelli}, \citenamefont
  {Bordin}, \citenamefont {Wang}, \citenamefont {van Loo}, \citenamefont
  {Wolff}, \citenamefont {Mazur}, \citenamefont {Gazibegovic}, \citenamefont
  {Badawy}, \citenamefont {Bakkers}, \citenamefont {Kouwenhoven},\ and\
  \citenamefont {Dvir}}]{vanDriel2024}%
  \BibitemOpen
  \bibfield  {author} {\bibinfo {author} {\bibfnamefont {D.}~\bibnamefont {van
  Driel}}, \bibinfo {author} {\bibfnamefont {B.}~\bibnamefont {Roovers}},
  \bibinfo {author} {\bibfnamefont {F.}~\bibnamefont {Zatelli}}, \bibinfo
  {author} {\bibfnamefont {A.}~\bibnamefont {Bordin}}, \bibinfo {author}
  {\bibfnamefont {G.}~\bibnamefont {Wang}}, \bibinfo {author} {\bibfnamefont
  {N.}~\bibnamefont {van Loo}}, \bibinfo {author} {\bibfnamefont {J.~C.}\
  \bibnamefont {Wolff}}, \bibinfo {author} {\bibfnamefont {G.~P.}\ \bibnamefont
  {Mazur}}, \bibinfo {author} {\bibfnamefont {S.}~\bibnamefont {Gazibegovic}},
  \bibinfo {author} {\bibfnamefont {G.}~\bibnamefont {Badawy}}, \bibinfo
  {author} {\bibfnamefont {E.~P.}\ \bibnamefont {Bakkers}}, \bibinfo {author}
  {\bibfnamefont {L.~P.}\ \bibnamefont {Kouwenhoven}},\ and\ \bibinfo {author}
  {\bibfnamefont {T.}~\bibnamefont {Dvir}},\ }\bibfield  {title} {\bibinfo
  {title} {Charge sensing the parity of an {A}ndreev molecule},\ }\href
  {https://doi.org/10.1103/PRXQuantum.5.020301} {\bibfield  {journal} {\bibinfo
   {journal} {PRX Quantum}\ }\textbf {\bibinfo {volume} {5}},\ \bibinfo {pages}
  {020301} (\bibinfo {year} {2024})}\BibitemShut {NoStop}%
\bibitem [{\citenamefont {Bordin}\ \emph
  {et~al.}(2024{\natexlab{b}})\citenamefont {Bordin}, \citenamefont {Evertsz'},
  \citenamefont {Steffensen}, \citenamefont {Dvir}, \citenamefont {Mazur},
  \citenamefont {van Driel}, \citenamefont {van Loo}, \citenamefont {Wolff},
  \citenamefont {Bakkers}, \citenamefont {Yeyati},\ and\ \citenamefont
  {Kouwenhoven}}]{Bordin2024b}%
  \BibitemOpen
  \bibfield  {author} {\bibinfo {author} {\bibfnamefont {A.}~\bibnamefont
  {Bordin}}, \bibinfo {author} {\bibfnamefont {F.~J.~B.}\ \bibnamefont
  {Evertsz'}}, \bibinfo {author} {\bibfnamefont {G.~O.}\ \bibnamefont
  {Steffensen}}, \bibinfo {author} {\bibfnamefont {T.}~\bibnamefont {Dvir}},
  \bibinfo {author} {\bibfnamefont {G.~P.}\ \bibnamefont {Mazur}}, \bibinfo
  {author} {\bibfnamefont {D.}~\bibnamefont {van Driel}}, \bibinfo {author}
  {\bibfnamefont {N.}~\bibnamefont {van Loo}}, \bibinfo {author} {\bibfnamefont
  {J.~C.}\ \bibnamefont {Wolff}}, \bibinfo {author} {\bibfnamefont {E.~P.
  A.~M.}\ \bibnamefont {Bakkers}}, \bibinfo {author} {\bibfnamefont {A.~L.}\
  \bibnamefont {Yeyati}},\ and\ \bibinfo {author} {\bibfnamefont {L.~P.}\
  \bibnamefont {Kouwenhoven}},\ }\href {https://arxiv.org/abs/2402.19284}
  {\bibinfo {title} {Supercurrent through an {A}ndreev trimer}} (\bibinfo
  {year} {2024}{\natexlab{b}}),\ \Eprint {https://arxiv.org/abs/2402.19284}
  {arXiv:2402.19284 [cond-mat.mes-hall]} \BibitemShut {NoStop}%
\bibitem [{\citenamefont {Dvir}\ \emph {et~al.}(2023)\citenamefont {Dvir},
  \citenamefont {Wang}, \citenamefont {van Loo}, \citenamefont {Liu},
  \citenamefont {Mazur}, \citenamefont {Bordin}, \citenamefont {ten Haaf},
  \citenamefont {Wang}, \citenamefont {van Driel}, \citenamefont {Zatelli},
  \citenamefont {Li}, \citenamefont {Malinowski}, \citenamefont {Gazibegovic},
  \citenamefont {Badawy}, \citenamefont {Bakkers}, \citenamefont {Wimmer},\
  and\ \citenamefont {Kouwenhoven}}]{Dvir2023}%
  \BibitemOpen
  \bibfield  {author} {\bibinfo {author} {\bibfnamefont {T.}~\bibnamefont
  {Dvir}}, \bibinfo {author} {\bibfnamefont {G.}~\bibnamefont {Wang}}, \bibinfo
  {author} {\bibfnamefont {N.}~\bibnamefont {van Loo}}, \bibinfo {author}
  {\bibfnamefont {C.-X.}\ \bibnamefont {Liu}}, \bibinfo {author} {\bibfnamefont
  {G.~P.}\ \bibnamefont {Mazur}}, \bibinfo {author} {\bibfnamefont
  {A.}~\bibnamefont {Bordin}}, \bibinfo {author} {\bibfnamefont {S.~L.~D.}\
  \bibnamefont {ten Haaf}}, \bibinfo {author} {\bibfnamefont {J.-Y.}\
  \bibnamefont {Wang}}, \bibinfo {author} {\bibfnamefont {D.}~\bibnamefont {van
  Driel}}, \bibinfo {author} {\bibfnamefont {F.}~\bibnamefont {Zatelli}},
  \bibinfo {author} {\bibfnamefont {X.}~\bibnamefont {Li}}, \bibinfo {author}
  {\bibfnamefont {F.~K.}\ \bibnamefont {Malinowski}}, \bibinfo {author}
  {\bibfnamefont {S.}~\bibnamefont {Gazibegovic}}, \bibinfo {author}
  {\bibfnamefont {G.}~\bibnamefont {Badawy}}, \bibinfo {author} {\bibfnamefont
  {E.~P. A.~M.}\ \bibnamefont {Bakkers}}, \bibinfo {author} {\bibfnamefont
  {M.}~\bibnamefont {Wimmer}},\ and\ \bibinfo {author} {\bibfnamefont {L.~P.}\
  \bibnamefont {Kouwenhoven}},\ }\bibfield  {title} {\bibinfo {title}
  {Realization of a minimal {K}itaev chain in coupled quantum dots},\ }\href
  {https://doi.org/10.1038/s41586-022-05585-1} {\bibfield  {journal} {\bibinfo
  {journal} {Nature}\ }\textbf {\bibinfo {volume} {614}},\ \bibinfo {pages}
  {445} (\bibinfo {year} {2023})}\BibitemShut {NoStop}%
\bibitem [{\citenamefont {ten Haaf}\ \emph {et~al.}(2024)\citenamefont {ten
  Haaf}, \citenamefont {Wang}, \citenamefont {Bozkurt}, \citenamefont {Liu},
  \citenamefont {Kulesh}, \citenamefont {Kim}, \citenamefont {Xiao},
  \citenamefont {Thomas}, \citenamefont {Manfra}, \citenamefont {Dvir},
  \citenamefont {Wimmer},\ and\ \citenamefont {Goswami}}]{tenHaaf2024}%
  \BibitemOpen
  \bibfield  {author} {\bibinfo {author} {\bibfnamefont {S.~L.~D.}\
  \bibnamefont {ten Haaf}}, \bibinfo {author} {\bibfnamefont {Q.}~\bibnamefont
  {Wang}}, \bibinfo {author} {\bibfnamefont {A.~M.}\ \bibnamefont {Bozkurt}},
  \bibinfo {author} {\bibfnamefont {C.-X.}\ \bibnamefont {Liu}}, \bibinfo
  {author} {\bibfnamefont {I.}~\bibnamefont {Kulesh}}, \bibinfo {author}
  {\bibfnamefont {P.}~\bibnamefont {Kim}}, \bibinfo {author} {\bibfnamefont
  {D.}~\bibnamefont {Xiao}}, \bibinfo {author} {\bibfnamefont {C.}~\bibnamefont
  {Thomas}}, \bibinfo {author} {\bibfnamefont {M.~J.}\ \bibnamefont {Manfra}},
  \bibinfo {author} {\bibfnamefont {T.}~\bibnamefont {Dvir}}, \bibinfo {author}
  {\bibfnamefont {M.}~\bibnamefont {Wimmer}},\ and\ \bibinfo {author}
  {\bibfnamefont {S.}~\bibnamefont {Goswami}},\ }\bibfield  {title} {\bibinfo
  {title} {{A two-site Kitaev chain in a two-dimensional electron gas}},\
  }\href {https://doi.org/10.1038/s41586-024-07434-9} {\bibfield  {journal}
  {\bibinfo  {journal} {Nature}\ }\textbf {\bibinfo {volume} {630}},\ \bibinfo
  {pages} {329–334} (\bibinfo {year} {2024})}\BibitemShut {NoStop}%
\bibitem [{\citenamefont {Coraiola}\ \emph {et~al.}(2023)\citenamefont
  {Coraiola}, \citenamefont {Haxell}, \citenamefont {Sabonis}, \citenamefont
  {Weisbrich}, \citenamefont {Svetogorov}, \citenamefont {Hinderling},
  \citenamefont {ten Kate}, \citenamefont {Cheah}, \citenamefont {Krizek},
  \citenamefont {Schott}, \citenamefont {Wegscheider}, \citenamefont {Cuevas},
  \citenamefont {Belzig},\ and\ \citenamefont {Nichele}}]{Coraiola2023}%
  \BibitemOpen
  \bibfield  {author} {\bibinfo {author} {\bibfnamefont {M.}~\bibnamefont
  {Coraiola}}, \bibinfo {author} {\bibfnamefont {D.~Z.}\ \bibnamefont
  {Haxell}}, \bibinfo {author} {\bibfnamefont {D.}~\bibnamefont {Sabonis}},
  \bibinfo {author} {\bibfnamefont {H.}~\bibnamefont {Weisbrich}}, \bibinfo
  {author} {\bibfnamefont {A.~E.}\ \bibnamefont {Svetogorov}}, \bibinfo
  {author} {\bibfnamefont {M.}~\bibnamefont {Hinderling}}, \bibinfo {author}
  {\bibfnamefont {S.~C.}\ \bibnamefont {ten Kate}}, \bibinfo {author}
  {\bibfnamefont {E.}~\bibnamefont {Cheah}}, \bibinfo {author} {\bibfnamefont
  {F.}~\bibnamefont {Krizek}}, \bibinfo {author} {\bibfnamefont
  {R.}~\bibnamefont {Schott}}, \bibinfo {author} {\bibfnamefont
  {W.}~\bibnamefont {Wegscheider}}, \bibinfo {author} {\bibfnamefont {J.~C.}\
  \bibnamefont {Cuevas}}, \bibinfo {author} {\bibfnamefont {W.}~\bibnamefont
  {Belzig}},\ and\ \bibinfo {author} {\bibfnamefont {F.}~\bibnamefont
  {Nichele}},\ }\bibfield  {title} {\bibinfo {title} {Phase-engineering the
  {A}ndreev band structure of a three-terminal {J}osephson junction},\ }\href
  {https://doi.org/10.1038/s41467-023-42356-6} {\bibfield  {journal} {\bibinfo
  {journal} {Nat. Commun.}\ }\textbf {\bibinfo {volume} {14}},\ \bibinfo
  {pages} {6784} (\bibinfo {year} {2023})}\BibitemShut {NoStop}%
\bibitem [{\citenamefont {Matsuo}\ \emph
  {et~al.}(2023{\natexlab{b}})\citenamefont {Matsuo}, \citenamefont {Imoto},
  \citenamefont {Yokoyama}, \citenamefont {Sato}, \citenamefont {Lindemann},
  \citenamefont {Gronin}, \citenamefont {Gardner}, \citenamefont {Nakosai},
  \citenamefont {Tanaka}, \citenamefont {Manfra},\ and\ \citenamefont
  {Tarucha}}]{Matsuo2023}%
  \BibitemOpen
  \bibfield  {author} {\bibinfo {author} {\bibfnamefont {S.}~\bibnamefont
  {Matsuo}}, \bibinfo {author} {\bibfnamefont {T.}~\bibnamefont {Imoto}},
  \bibinfo {author} {\bibfnamefont {T.}~\bibnamefont {Yokoyama}}, \bibinfo
  {author} {\bibfnamefont {Y.}~\bibnamefont {Sato}}, \bibinfo {author}
  {\bibfnamefont {T.}~\bibnamefont {Lindemann}}, \bibinfo {author}
  {\bibfnamefont {S.}~\bibnamefont {Gronin}}, \bibinfo {author} {\bibfnamefont
  {G.~C.}\ \bibnamefont {Gardner}}, \bibinfo {author} {\bibfnamefont
  {S.}~\bibnamefont {Nakosai}}, \bibinfo {author} {\bibfnamefont
  {Y.}~\bibnamefont {Tanaka}}, \bibinfo {author} {\bibfnamefont {M.~J.}\
  \bibnamefont {Manfra}},\ and\ \bibinfo {author} {\bibfnamefont
  {S.}~\bibnamefont {Tarucha}},\ }\bibfield  {title} {\bibinfo {title}
  {Phase-dependent {A}ndreev molecules and superconducting gap closing in
  coherently-coupled {J}osephson junctions},\ }\bibfield  {journal} {\bibinfo
  {journal} {Nat. Commun.}\ }\textbf {\bibinfo {volume} {14}},\ \href
  {https://doi.org/10.1038/s41467-023-44111-3} {10.1038/s41467-023-44111-3}
  (\bibinfo {year} {2023}{\natexlab{b}})\BibitemShut {NoStop}%
\bibitem [{\citenamefont {van Heck}\ \emph {et~al.}(2014)\citenamefont {van
  Heck}, \citenamefont {Mi},\ and\ \citenamefont {Akhmerov}}]{vanHeck2014}%
  \BibitemOpen
  \bibfield  {author} {\bibinfo {author} {\bibfnamefont {B.}~\bibnamefont {van
  Heck}}, \bibinfo {author} {\bibfnamefont {S.}~\bibnamefont {Mi}},\ and\
  \bibinfo {author} {\bibfnamefont {A.~R.}\ \bibnamefont {Akhmerov}},\
  }\bibfield  {title} {\bibinfo {title} {Single fermion manipulation via
  superconducting phase differences in multiterminal {J}osephson junctions},\
  }\href {https://doi.org/10.1103/PhysRevB.90.155450} {\bibfield  {journal}
  {\bibinfo  {journal} {Phys. Rev. B}\ }\textbf {\bibinfo {volume} {90}},\
  \bibinfo {pages} {155450} (\bibinfo {year} {2014})}\BibitemShut {NoStop}%
\bibitem [{\citenamefont {Coraiola}\ \emph
  {et~al.}(2024{\natexlab{b}})\citenamefont {Coraiola}, \citenamefont {Haxell},
  \citenamefont {Sabonis}, \citenamefont {Hinderling}, \citenamefont {Kate},
  \citenamefont {Cheah}, \citenamefont {Krizek}, \citenamefont {Schott},
  \citenamefont {Wegscheider},\ and\ \citenamefont {Nichele}}]{Coraiola2024b}%
  \BibitemOpen
  \bibfield  {author} {\bibinfo {author} {\bibfnamefont {M.}~\bibnamefont
  {Coraiola}}, \bibinfo {author} {\bibfnamefont {D.~Z.}\ \bibnamefont
  {Haxell}}, \bibinfo {author} {\bibfnamefont {D.}~\bibnamefont {Sabonis}},
  \bibinfo {author} {\bibfnamefont {M.}~\bibnamefont {Hinderling}}, \bibinfo
  {author} {\bibfnamefont {S.~C.~t.}\ \bibnamefont {Kate}}, \bibinfo {author}
  {\bibfnamefont {E.}~\bibnamefont {Cheah}}, \bibinfo {author} {\bibfnamefont
  {F.}~\bibnamefont {Krizek}}, \bibinfo {author} {\bibfnamefont
  {R.}~\bibnamefont {Schott}}, \bibinfo {author} {\bibfnamefont
  {W.}~\bibnamefont {Wegscheider}},\ and\ \bibinfo {author} {\bibfnamefont
  {F.}~\bibnamefont {Nichele}},\ }\bibfield  {title} {\bibinfo {title}
  {Spin-degeneracy breaking and parity transitions in three-terminal
  {J}osephson junctions},\ }\href {https://doi.org/10.1103/PhysRevX.14.031024}
  {\bibfield  {journal} {\bibinfo  {journal} {Phys. Rev. X}\ }\textbf {\bibinfo
  {volume} {14}},\ \bibinfo {pages} {031024} (\bibinfo {year}
  {2024}{\natexlab{b}})}\BibitemShut {NoStop}%
\bibitem [{\citenamefont {Lee}(2022)}]{Lee2022}%
  \BibitemOpen
  \bibfield  {author} {\bibinfo {author} {\bibfnamefont {H.}~\bibnamefont
  {Lee}},\ }\emph {\bibinfo {title} {Supercurrent and {A}ndreev bound states in
  multi-terminal {J}osephson junctions}},\ \href@noop {} {Ph.D. thesis},\
  \bibinfo  {school} {University of Maryland} (\bibinfo {year}
  {2022})\BibitemShut {NoStop}%
\bibitem [{\citenamefont {Xie}\ \emph {et~al.}(2017)\citenamefont {Xie},
  \citenamefont {Vavilov},\ and\ \citenamefont {Levchenko}}]{Xie2017}%
  \BibitemOpen
  \bibfield  {author} {\bibinfo {author} {\bibfnamefont {H.-Y.}\ \bibnamefont
  {Xie}}, \bibinfo {author} {\bibfnamefont {M.~G.}\ \bibnamefont {Vavilov}},\
  and\ \bibinfo {author} {\bibfnamefont {A.}~\bibnamefont {Levchenko}},\
  }\bibfield  {title} {\bibinfo {title} {Topological {A}ndreev bands in
  three-terminal {J}osephson junctions},\ }\href
  {https://doi.org/10.1103/PhysRevB.96.161406} {\bibfield  {journal} {\bibinfo
  {journal} {Phys. Rev. B}\ }\textbf {\bibinfo {volume} {96}},\ \bibinfo
  {pages} {161406} (\bibinfo {year} {2017})}\BibitemShut {NoStop}%
\bibitem [{\citenamefont {Meyer}\ and\ \citenamefont
  {Houzet}(2017)}]{Meyer2017}%
  \BibitemOpen
  \bibfield  {author} {\bibinfo {author} {\bibfnamefont {J.~S.}\ \bibnamefont
  {Meyer}}\ and\ \bibinfo {author} {\bibfnamefont {M.}~\bibnamefont {Houzet}},\
  }\bibfield  {title} {\bibinfo {title} {Nontrivial {C}hern numbers in
  three-terminal {J}osephson junctions},\ }\href
  {https://doi.org/10.1103/PhysRevLett.119.136807} {\bibfield  {journal}
  {\bibinfo  {journal} {Phys. Rev. Lett.}\ }\textbf {\bibinfo {volume} {119}},\
  \bibinfo {pages} {136807} (\bibinfo {year} {2017})}\BibitemShut {NoStop}%
\bibitem [{\citenamefont {Shabani}\ \emph {et~al.}(2016)\citenamefont
  {Shabani}, \citenamefont {Kjaergaard}, \citenamefont {Suominen},
  \citenamefont {Kim}, \citenamefont {Nichele}, \citenamefont {Pakrouski},
  \citenamefont {Stankevic}, \citenamefont {Lutchyn}, \citenamefont
  {Krogstrup}, \citenamefont {Feidenhans'l}, \citenamefont {Kraemer},
  \citenamefont {Nayak}, \citenamefont {Troyer}, \citenamefont {Marcus},\ and\
  \citenamefont {Palmstr\o{}m}}]{Shabani2016}%
  \BibitemOpen
  \bibfield  {author} {\bibinfo {author} {\bibfnamefont {J.}~\bibnamefont
  {Shabani}}, \bibinfo {author} {\bibfnamefont {M.}~\bibnamefont {Kjaergaard}},
  \bibinfo {author} {\bibfnamefont {H.~J.}\ \bibnamefont {Suominen}}, \bibinfo
  {author} {\bibfnamefont {Y.}~\bibnamefont {Kim}}, \bibinfo {author}
  {\bibfnamefont {F.}~\bibnamefont {Nichele}}, \bibinfo {author} {\bibfnamefont
  {K.}~\bibnamefont {Pakrouski}}, \bibinfo {author} {\bibfnamefont
  {T.}~\bibnamefont {Stankevic}}, \bibinfo {author} {\bibfnamefont {R.~M.}\
  \bibnamefont {Lutchyn}}, \bibinfo {author} {\bibfnamefont {P.}~\bibnamefont
  {Krogstrup}}, \bibinfo {author} {\bibfnamefont {R.}~\bibnamefont
  {Feidenhans'l}}, \bibinfo {author} {\bibfnamefont {S.}~\bibnamefont
  {Kraemer}}, \bibinfo {author} {\bibfnamefont {C.}~\bibnamefont {Nayak}},
  \bibinfo {author} {\bibfnamefont {M.}~\bibnamefont {Troyer}}, \bibinfo
  {author} {\bibfnamefont {C.~M.}\ \bibnamefont {Marcus}},\ and\ \bibinfo
  {author} {\bibfnamefont {C.~J.}\ \bibnamefont {Palmstr\o{}m}},\ }\bibfield
  {title} {\bibinfo {title} {Two-dimensional epitaxial
  superconductor{\textendash}semiconductor heterostructures: {A} platform for
  topological superconducting networks},\ }\href
  {https://doi.org/10.1103/PhysRevB.93.155402} {\bibfield  {journal} {\bibinfo
  {journal} {Phys. Rev. B}\ }\textbf {\bibinfo {volume} {93}},\ \bibinfo
  {pages} {155402} (\bibinfo {year} {2016})}\BibitemShut {NoStop}%
\bibitem [{\citenamefont {Cheah}\ \emph {et~al.}(2023)\citenamefont {Cheah},
  \citenamefont {Haxell}, \citenamefont {Schott}, \citenamefont {Zeng},
  \citenamefont {Paysen}, \citenamefont {ten Kate}, \citenamefont {Coraiola},
  \citenamefont {Landstetter}, \citenamefont {Zadeh}, \citenamefont {Trampert},
  \citenamefont {Sousa}, \citenamefont {Riel}, \citenamefont {Nichele},
  \citenamefont {Wegscheider},\ and\ \citenamefont {Krizek}}]{Cheah2023}%
  \BibitemOpen
  \bibfield  {author} {\bibinfo {author} {\bibfnamefont {E.}~\bibnamefont
  {Cheah}}, \bibinfo {author} {\bibfnamefont {D.~Z.}\ \bibnamefont {Haxell}},
  \bibinfo {author} {\bibfnamefont {R.}~\bibnamefont {Schott}}, \bibinfo
  {author} {\bibfnamefont {P.}~\bibnamefont {Zeng}}, \bibinfo {author}
  {\bibfnamefont {E.}~\bibnamefont {Paysen}}, \bibinfo {author} {\bibfnamefont
  {S.~C.}\ \bibnamefont {ten Kate}}, \bibinfo {author} {\bibfnamefont
  {M.}~\bibnamefont {Coraiola}}, \bibinfo {author} {\bibfnamefont
  {M.}~\bibnamefont {Landstetter}}, \bibinfo {author} {\bibfnamefont {A.~B.}\
  \bibnamefont {Zadeh}}, \bibinfo {author} {\bibfnamefont {A.}~\bibnamefont
  {Trampert}}, \bibinfo {author} {\bibfnamefont {M.}~\bibnamefont {Sousa}},
  \bibinfo {author} {\bibfnamefont {H.}~\bibnamefont {Riel}}, \bibinfo {author}
  {\bibfnamefont {F.}~\bibnamefont {Nichele}}, \bibinfo {author} {\bibfnamefont
  {W.}~\bibnamefont {Wegscheider}},\ and\ \bibinfo {author} {\bibfnamefont
  {F.}~\bibnamefont {Krizek}},\ }\bibfield  {title} {\bibinfo {title} {Control
  over epitaxy and the role of the {InAs/Al} interface in hybrid
  two-dimensional electron gas systems},\ }\href
  {https://doi.org/10.1103/PhysRevMaterials.7.073403} {\bibfield  {journal}
  {\bibinfo  {journal} {Phys. Rev. Mater.}\ }\textbf {\bibinfo {volume} {7}},\
  \bibinfo {pages} {073403} (\bibinfo {year} {2023})}\BibitemShut {NoStop}%
\bibitem [{SM()}]{SM}%
  \BibitemOpen
  \href@noop {} {}\bibinfo {note} {See Supplemental Material for additional
  supporting data and details on the theoretical model
  (Figs.~S1--S8).}\BibitemShut {Stop}%
\bibitem [{\citenamefont {Wang}\ and\ \citenamefont {Zhang}(2012)}]{Wang2012}%
  \BibitemOpen
  \bibfield  {author} {\bibinfo {author} {\bibfnamefont {Z.}~\bibnamefont
  {Wang}}\ and\ \bibinfo {author} {\bibfnamefont {S.-C.}\ \bibnamefont
  {Zhang}},\ }\bibfield  {title} {\bibinfo {title} {Simplified topological
  invariants for interacting insulators},\ }\href
  {https://doi.org/10.1103/PhysRevX.2.031008} {\bibfield  {journal} {\bibinfo
  {journal} {Phys. Rev. X}\ }\textbf {\bibinfo {volume} {2}},\ \bibinfo {pages}
  {031008} (\bibinfo {year} {2012})}\BibitemShut {NoStop}%
\bibitem [{\citenamefont {Wang}\ and\ \citenamefont {Yan}(2013)}]{Wang2013}%
  \BibitemOpen
  \bibfield  {author} {\bibinfo {author} {\bibfnamefont {Z.}~\bibnamefont
  {Wang}}\ and\ \bibinfo {author} {\bibfnamefont {B.}~\bibnamefont {Yan}},\
  }\bibfield  {title} {\bibinfo {title} {Topological {{Hamiltonian}} as an
  exact tool for topological invariants},\ }\href
  {https://doi.org/10.1088/0953-8984/25/15/155601} {\bibfield  {journal}
  {\bibinfo  {journal} {J. Phys.: Condens. Matter}\ }\textbf {\bibinfo {volume}
  {25}},\ \bibinfo {pages} {155601} (\bibinfo {year} {2013})}\BibitemShut
  {NoStop}%
\bibitem [{\citenamefont {Gavensky}\ \emph {et~al.}(2023)\citenamefont
  {Gavensky}, \citenamefont {Usaj},\ and\ \citenamefont
  {Balseiro}}]{Gavensky2023}%
  \BibitemOpen
  \bibfield  {author} {\bibinfo {author} {\bibfnamefont {L.~P.}\ \bibnamefont
  {Gavensky}}, \bibinfo {author} {\bibfnamefont {G.}~\bibnamefont {Usaj}},\
  and\ \bibinfo {author} {\bibfnamefont {C.~A.}\ \bibnamefont {Balseiro}},\
  }\bibfield  {title} {\bibinfo {title} {Multi-terminal {{Josephson}}
  junctions: {{A}} road to topological flux networks},\ }\href
  {https://doi.org/10.1209/0295-5075/acb2f6} {\bibfield  {journal} {\bibinfo
  {journal} {EPL}\ }\textbf {\bibinfo {volume} {141}},\ \bibinfo {pages}
  {36001} (\bibinfo {year} {2023})}\BibitemShut {NoStop}%
\bibitem [{\citenamefont {Fukui}\ \emph {et~al.}(2005)\citenamefont {Fukui},
  \citenamefont {Hatsugai},\ and\ \citenamefont {Suzuki}}]{Fukui2005}%
  \BibitemOpen
  \bibfield  {author} {\bibinfo {author} {\bibfnamefont {T.}~\bibnamefont
  {Fukui}}, \bibinfo {author} {\bibfnamefont {Y.}~\bibnamefont {Hatsugai}},\
  and\ \bibinfo {author} {\bibfnamefont {H.}~\bibnamefont {Suzuki}},\
  }\bibfield  {title} {\bibinfo {title} {Chern {{Numbers}} in {{Discretized
  Brillouin Zone}}: {{Efficient Method}} of {{Computing}} ({{Spin}}) {{Hall
  Conductances}}},\ }\href {https://doi.org/10.1143/JPSJ.74.1674} {\bibfield
  {journal} {\bibinfo  {journal} {J. Phys. Soc. Jpn.}\ }\textbf {\bibinfo
  {volume} {74}},\ \bibinfo {pages} {1674} (\bibinfo {year}
  {2005})}\BibitemShut {NoStop}%
\bibitem [{\citenamefont {Hasan}\ and\ \citenamefont {Kane}(2010)}]{Hasan2010}%
  \BibitemOpen
  \bibfield  {author} {\bibinfo {author} {\bibfnamefont {M.~Z.}\ \bibnamefont
  {Hasan}}\ and\ \bibinfo {author} {\bibfnamefont {C.~L.}\ \bibnamefont
  {Kane}},\ }\bibfield  {title} {\bibinfo {title} {Colloquium: Topological
  insulators},\ }\href {https://doi.org/10.1103/RevModPhys.82.3045} {\bibfield
  {journal} {\bibinfo  {journal} {Rev. Mod. Phys.}\ }\textbf {\bibinfo {volume}
  {82}},\ \bibinfo {pages} {3045} (\bibinfo {year} {2010})}\BibitemShut
  {NoStop}%
\bibitem [{\citenamefont {Larsen}\ \emph {et~al.}(2015)\citenamefont {Larsen},
  \citenamefont {Petersson}, \citenamefont {Kuemmeth}, \citenamefont
  {Jespersen}, \citenamefont {Krogstrup}, \citenamefont {Nyg\aa{}rd},\ and\
  \citenamefont {Marcus}}]{Larsen2015}%
  \BibitemOpen
  \bibfield  {author} {\bibinfo {author} {\bibfnamefont {T.~W.}\ \bibnamefont
  {Larsen}}, \bibinfo {author} {\bibfnamefont {K.~D.}\ \bibnamefont
  {Petersson}}, \bibinfo {author} {\bibfnamefont {F.}~\bibnamefont {Kuemmeth}},
  \bibinfo {author} {\bibfnamefont {T.~S.}\ \bibnamefont {Jespersen}}, \bibinfo
  {author} {\bibfnamefont {P.}~\bibnamefont {Krogstrup}}, \bibinfo {author}
  {\bibfnamefont {J.}~\bibnamefont {Nyg\aa{}rd}},\ and\ \bibinfo {author}
  {\bibfnamefont {C.~M.}\ \bibnamefont {Marcus}},\ }\bibfield  {title}
  {\bibinfo {title} {Semiconductor-nanowire-based superconducting qubit},\
  }\href {https://doi.org/10.1103/PhysRevLett.115.127001} {\bibfield  {journal}
  {\bibinfo  {journal} {Phys. Rev. Lett.}\ }\textbf {\bibinfo {volume} {115}},\
  \bibinfo {pages} {127001} (\bibinfo {year} {2015})}\BibitemShut {NoStop}%
\bibitem [{\citenamefont {de~Lange}\ \emph {et~al.}(2015)\citenamefont
  {de~Lange}, \citenamefont {van Heck}, \citenamefont {Bruno}, \citenamefont
  {van Woerkom}, \citenamefont {Geresdi}, \citenamefont {Plissard},
  \citenamefont {Bakkers}, \citenamefont {Akhmerov},\ and\ \citenamefont
  {DiCarlo}}]{deLange2015}%
  \BibitemOpen
  \bibfield  {author} {\bibinfo {author} {\bibfnamefont {G.}~\bibnamefont
  {de~Lange}}, \bibinfo {author} {\bibfnamefont {B.}~\bibnamefont {van Heck}},
  \bibinfo {author} {\bibfnamefont {A.}~\bibnamefont {Bruno}}, \bibinfo
  {author} {\bibfnamefont {D.~J.}\ \bibnamefont {van Woerkom}}, \bibinfo
  {author} {\bibfnamefont {A.}~\bibnamefont {Geresdi}}, \bibinfo {author}
  {\bibfnamefont {S.~R.}\ \bibnamefont {Plissard}}, \bibinfo {author}
  {\bibfnamefont {E.~P. A.~M.}\ \bibnamefont {Bakkers}}, \bibinfo {author}
  {\bibfnamefont {A.~R.}\ \bibnamefont {Akhmerov}},\ and\ \bibinfo {author}
  {\bibfnamefont {L.}~\bibnamefont {DiCarlo}},\ }\bibfield  {title} {\bibinfo
  {title} {Realization of microwave quantum circuits using hybrid
  superconducting-semiconducting nanowire {J}osephson elements},\ }\href
  {https://doi.org/10.1103/PhysRevLett.115.127002} {\bibfield  {journal}
  {\bibinfo  {journal} {Phys. Rev. Lett.}\ }\textbf {\bibinfo {volume} {115}},\
  \bibinfo {pages} {127002} (\bibinfo {year} {2015})}\BibitemShut {NoStop}%
\bibitem [{\citenamefont {Zellekens}\ \emph {et~al.}(2022)\citenamefont
  {Zellekens}, \citenamefont {Deacon}, \citenamefont {Perla}, \citenamefont
  {Gr\"{u}tzmacher}, \citenamefont {Lepsa}, \citenamefont {Sch\"{a}pers},\ and\
  \citenamefont {Ishibashi}}]{Zellekens2022}%
  \BibitemOpen
  \bibfield  {author} {\bibinfo {author} {\bibfnamefont {P.}~\bibnamefont
  {Zellekens}}, \bibinfo {author} {\bibfnamefont {R.~S.}\ \bibnamefont
  {Deacon}}, \bibinfo {author} {\bibfnamefont {P.}~\bibnamefont {Perla}},
  \bibinfo {author} {\bibfnamefont {D.}~\bibnamefont {Gr\"{u}tzmacher}},
  \bibinfo {author} {\bibfnamefont {M.~I.}\ \bibnamefont {Lepsa}}, \bibinfo
  {author} {\bibfnamefont {T.}~\bibnamefont {Sch\"{a}pers}},\ and\ \bibinfo
  {author} {\bibfnamefont {K.}~\bibnamefont {Ishibashi}},\ }\bibfield  {title}
  {\bibinfo {title} {Microwave spectroscopy of {A}ndreev states in {InAs}
  nanowire-based hybrid junctions using a flip-chip layout},\ }\href
  {https://doi.org/10.1038/s42005-022-01035-6} {\bibfield  {journal} {\bibinfo
  {journal} {Commun. Phys.}\ }\textbf {\bibinfo {volume} {5}},\ \bibinfo
  {pages} {267} (\bibinfo {year} {2022})}\BibitemShut {NoStop}%
\bibitem [{\citenamefont {Hinderling}\ \emph {et~al.}(2023)\citenamefont
  {Hinderling}, \citenamefont {Sabonis}, \citenamefont {Paredes}, \citenamefont
  {Haxell}, \citenamefont {Coraiola}, \citenamefont {ten Kate}, \citenamefont
  {Cheah}, \citenamefont {Krizek}, \citenamefont {Schott}, \citenamefont
  {Wegscheider},\ and\ \citenamefont {Nichele}}]{Hinderling2023}%
  \BibitemOpen
  \bibfield  {author} {\bibinfo {author} {\bibfnamefont {M.}~\bibnamefont
  {Hinderling}}, \bibinfo {author} {\bibfnamefont {D.}~\bibnamefont {Sabonis}},
  \bibinfo {author} {\bibfnamefont {S.}~\bibnamefont {Paredes}}, \bibinfo
  {author} {\bibfnamefont {D.}~\bibnamefont {Haxell}}, \bibinfo {author}
  {\bibfnamefont {M.}~\bibnamefont {Coraiola}}, \bibinfo {author}
  {\bibfnamefont {S.}~\bibnamefont {ten Kate}}, \bibinfo {author}
  {\bibfnamefont {E.}~\bibnamefont {Cheah}}, \bibinfo {author} {\bibfnamefont
  {F.}~\bibnamefont {Krizek}}, \bibinfo {author} {\bibfnamefont
  {R.}~\bibnamefont {Schott}}, \bibinfo {author} {\bibfnamefont
  {W.}~\bibnamefont {Wegscheider}},\ and\ \bibinfo {author} {\bibfnamefont
  {F.}~\bibnamefont {Nichele}},\ }\bibfield  {title} {\bibinfo {title}
  {Flip-chip-based microwave spectroscopy of {A}ndreev bound states in a planar
  {J}osephson junction},\ }\href
  {https://doi.org/10.1103/PhysRevApplied.19.054026} {\bibfield  {journal}
  {\bibinfo  {journal} {Phys. Rev. Appl.}\ }\textbf {\bibinfo {volume} {19}},\
  \bibinfo {pages} {054026} (\bibinfo {year} {2023})}\BibitemShut {NoStop}%
\bibitem [{\citenamefont {Matute-Ca\~nadas}\ \emph {et~al.}(2024)\citenamefont
  {Matute-Ca\~nadas}, \citenamefont {Tosi},\ and\ \citenamefont
  {Yeyati}}]{MatuteCanadas2024}%
  \BibitemOpen
  \bibfield  {author} {\bibinfo {author} {\bibfnamefont {F.}~\bibnamefont
  {Matute-Ca\~nadas}}, \bibinfo {author} {\bibfnamefont {L.}~\bibnamefont
  {Tosi}},\ and\ \bibinfo {author} {\bibfnamefont {A.~L.}\ \bibnamefont
  {Yeyati}},\ }\bibfield  {title} {\bibinfo {title} {Quantum circuits with
  multiterminal {J}osephson-{A}ndreev junctions},\ }\href
  {https://doi.org/10.1103/PRXQuantum.5.020340} {\bibfield  {journal} {\bibinfo
   {journal} {PRX Quantum}\ }\textbf {\bibinfo {volume} {5}},\ \bibinfo {pages}
  {020340} (\bibinfo {year} {2024})}\BibitemShut {NoStop}%
\bibitem [{\citenamefont {Matute-Ca\~nadas}\ \emph {et~al.}(2022)\citenamefont
  {Matute-Ca\~nadas}, \citenamefont {Metzger}, \citenamefont {Park},
  \citenamefont {Tosi}, \citenamefont {Krogstrup}, \citenamefont {Nyg\aa{}rd},
  \citenamefont {Goffman}, \citenamefont {Urbina}, \citenamefont {Pothier},\
  and\ \citenamefont {Yeyati}}]{Yeyati2022}%
  \BibitemOpen
  \bibfield  {author} {\bibinfo {author} {\bibfnamefont {F.~J.}\ \bibnamefont
  {Matute-Ca\~nadas}}, \bibinfo {author} {\bibfnamefont {C.}~\bibnamefont
  {Metzger}}, \bibinfo {author} {\bibfnamefont {S.}~\bibnamefont {Park}},
  \bibinfo {author} {\bibfnamefont {L.}~\bibnamefont {Tosi}}, \bibinfo {author}
  {\bibfnamefont {P.}~\bibnamefont {Krogstrup}}, \bibinfo {author}
  {\bibfnamefont {J.}~\bibnamefont {Nyg\aa{}rd}}, \bibinfo {author}
  {\bibfnamefont {M.~F.}\ \bibnamefont {Goffman}}, \bibinfo {author}
  {\bibfnamefont {C.}~\bibnamefont {Urbina}}, \bibinfo {author} {\bibfnamefont
  {H.}~\bibnamefont {Pothier}},\ and\ \bibinfo {author} {\bibfnamefont {A.~L.}\
  \bibnamefont {Yeyati}},\ }\bibfield  {title} {\bibinfo {title} {Signatures of
  interactions in the andreev spectrum of nanowire josephson junctions},\
  }\href {https://doi.org/10.1103/PhysRevLett.128.197702} {\bibfield  {journal}
  {\bibinfo  {journal} {Phys. Rev. Lett.}\ }\textbf {\bibinfo {volume} {128}},\
  \bibinfo {pages} {197702} (\bibinfo {year} {2022})}\BibitemShut {NoStop}%
\end{thebibliography}%

\twocolumngrid
\onecolumngrid
\clearpage

\begin{center}
    \large \textbf{{\Large Supplemental Material}} \\
    \vspace{0.5em}
\end{center}

\newcounter{myc} 
\renewcommand{\thefigure}{S.\arabic{myc}}
\setcounter{section}{0} 
\renewcommand{\thesection}{\Roman{section}} 


\section{Gate dependence of differential conductance}\label{Section1}
\begin{figure*}[h!]
	\centering
	\includegraphics[width=0.5\linewidth]{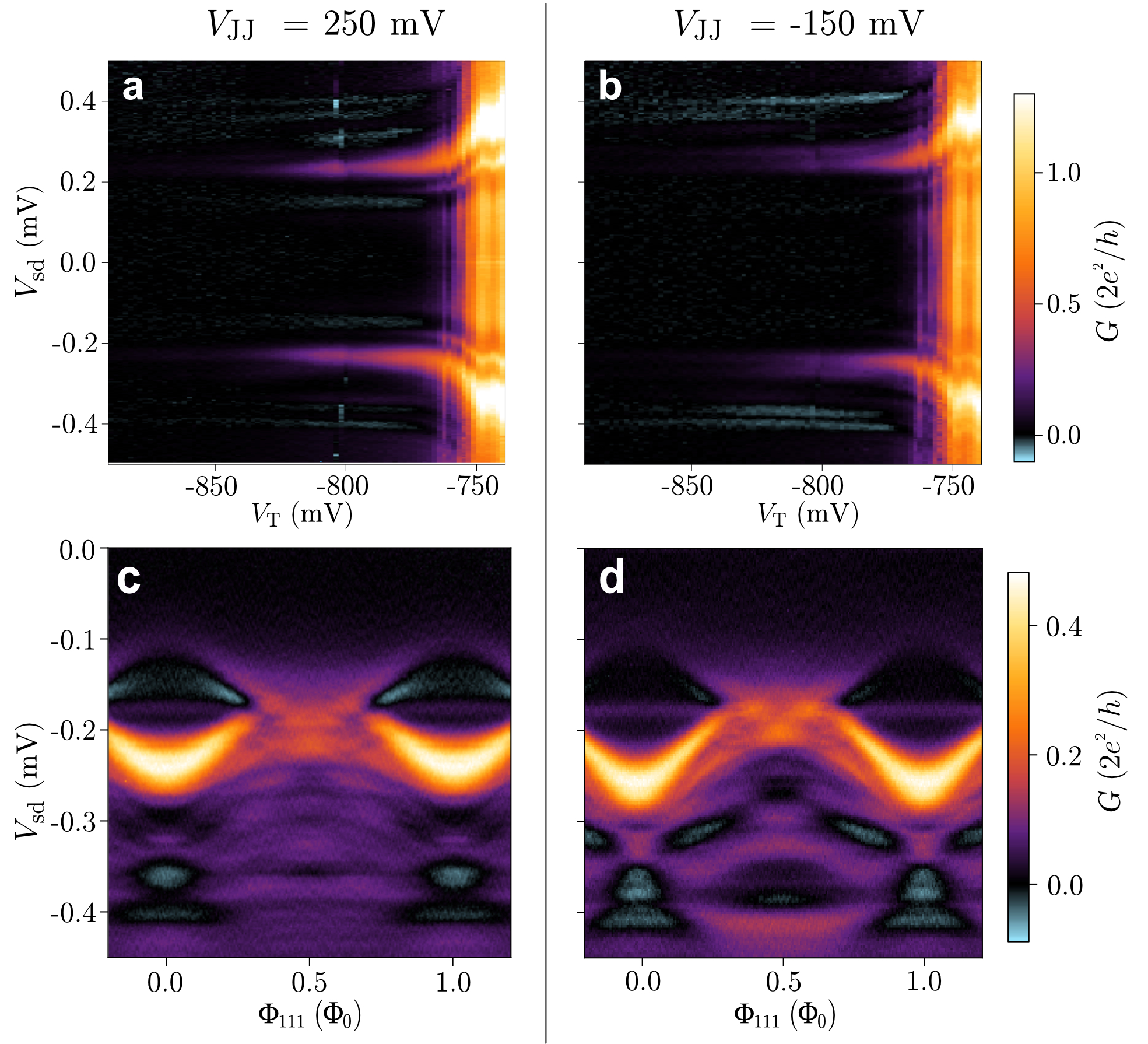}
	\caption{(a,b) Differential conductance $G$ as a function of voltage bias $V_{\mathrm{sd}}$ and gate voltage $V_{\mathrm{T}} = V_{\mathrm{TL}} = V_{\mathrm{TR}}$, for $V_{\mathrm{JJ}}=250~\mathrm{mV}$ (a) and $V_{\mathrm{JJ}}=-150~\mathrm{mV}$ (b). (c,d) Tunneling spectra measured along the flux direction $\Phi_{111}$ (see definition in the Main Text), for $V_{\mathrm{JJ}}=250~\mathrm{mV}$ (c) and $V_{\mathrm{JJ}}=-150~\mathrm{mV}$ (d).}
	\label{figS1}
\end{figure*}

The transmission between the superconducting probe and the four-terminal Josephson junction (4TJJ) is tuned by the gate voltages $V_{\rm{TL}}$ and $V_{\rm{TR}}$ (see Fig.~1 of the Main Text). Figure \ref{figS1} shows the dependence of the differential conductance spectrum on $V_{\rm{T}} = V_{\rm{TL}}$ = $V_{\rm{TR}}$ for two values of the gate voltage $V_{\rm{JJ}}$: $250~\mathrm{mV}$ (a) and $-150~\mathrm{mV}$ (b). Both spectra reveal a transport gap of $\sim 0.35~\mathrm{mV}$ and differential conductance peaks at $\sim \pm 0.25~\mathrm{mV}$ that correspond to the Andreev bound states (ABSs). By setting more negative $V_{\rm{T}}$ values, the overall differential conductance $G$ decreases, until the tunneling barrier is closed. All data presented in the Main Text and Supplemental Material are measured in the tunneling regime, where the normal-state differential conductance (at $|V_{\rm{sd}}|\approx 0.45~\rm{mV}$) is $\sim 5\%$ of the conductance quantum $G_0 = 2e^2/h$. The other gate voltages are kept fixed at $V_{\rm{H}} = 250~\rm{mV}$ and $V_{\rm{JJ}} = -150~\rm{mV}$, apart from the datasets in Fig.~1(c-e) of the Main Text, which are measured at $V_{\rm{JJ}}  = 250~\rm{mV}$. Figure~\ref{figS1}(c,d) shows the tunneling conductance spectra measured in these two gate configurations along the flux direction $\Phi_{111}$ (we use the crystallographic notation defined in the Main Text). Both measurements reveal the dispersion of three hybridized ABSs forming a tri-Andreev molecule, as discussed in Sec.~VI of the Main Text. More low-transmission states are present at low energies in Fig.~\ref{figS1}(c) compared to (d), since electrons are accumulated in the InAs quantum well (QW) between the terminals for $V_{\rm{JJ}} = 250~\rm{mV}$. In addition, the repulsion between these states and the high-transmission ABSs reduces the energy range of the dispersion of the high-transmission ABSs in Fig.~\ref{figS1}(c) with respect to the more depleted configuration of Fig.~\ref{figS1}(d).

\setcounter{myc}{1}

\section{Current-flux remapping using 3$\times$3 mutual inductance matrix}
\begin{figure*}[h!]
	\centering
	\includegraphics[width=\linewidth]{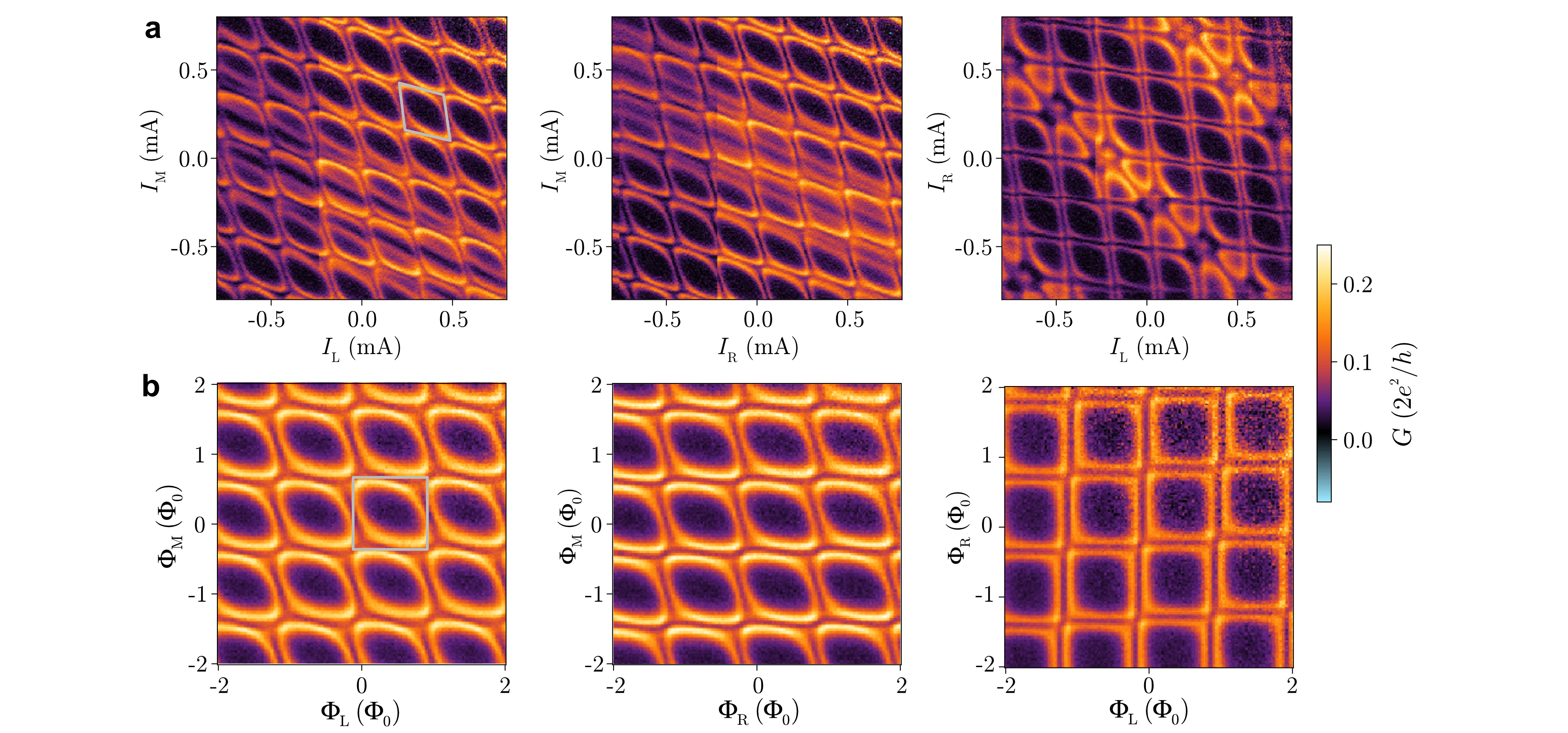}
	\caption{(a) Differential conductance maps measured at $V_{\rm{sd}} = -0.2~\rm{mV}$ as a function of each pair of flux-line currents $I_{\rm{L}}$, $I_{\rm{M}}$, and $I_{\rm{R}}$. The flux unit cell (grey polygon) is used to estimate the mutual inductance matrix $M$ and compensate the cross-coupling between loops and flux lines. (b) As in (a), with $G$ measured as a function of magnetic fluxes $\Phi_{\rm{L}}$, $\Phi_{\rm{M}}$, and $\Phi_{\rm{R}}$ after the first extraction of $M$ from (a), showing that deviation of the unit cell from a square shape is significantly reduced. The process is iterated two more times to obtain fully orthogonal maps as in Fig.~1(e) of the Main Text.}
	\label{figS2}
\end{figure*}

The three loops and the three flux lines control the three superconducting phase differences across each terminal pair. However, the magnetic field generated by one flux line also threads the other loops, thus affecting all phase differences. Due to this cross-coupling, the current-current maps displayed in Fig.~\ref{figS2}(a) represent skewed cuts of the 3D flux space, i.e., they are not along the faces of the flux unit cell. To achieve independent flux control, the cross-coupling has be corrected by determining the $3\times3$ mutual inductance matrix $M$, defined in Eq.~1 of the Main Text. Each matrix element $M_{ij}$ represents the mutual inductance between the superconducting loop $i$ and the flux-line current $I_j$ ($i,j=\rm{L,M,R}$). The elements of $M$ are derived from Fig.~\ref{figS2}(a), where each of the three maps allows one to extract 4 matrix elements (e.g., the $I_{\rm{L}}$-$I_{\rm{M}}$ map contains information on $M_{\rm{LL}}$, $M_{\rm{LM}}$, $M_{\rm{ML}}$ and $M_{\rm{MM}}$). Reference~\cite{Coraiola2023} provides more details on the extraction procedure. To minimize the error on the matrix elements, we iterate this process at least three times, until the relative variation in any matrix element between two consecutive iterations is smaller than $0.5 \%$. Figure~\ref{figS2}(b) shows the conductance maps measured as a function of the magnetic fluxes extracted in the first iteration, where the cross-coupling is already strongly reduced. After three iterations, the extracted $M$ is:
\begin{equation}
M
=
\begin{pmatrix}
4.053 & 0.430 & 0.291 \\
0.875 & 3.786 & 0.883 \\
0.316 & 0.424 & 4.031
\end{pmatrix}
\rm{pH,}
\end{equation}
that is utilized in all the datasets of the Main Text where one or more magnetic fluxes are swept.

\setcounter{myc}{2}

\section{Dependence of tunneling spectrum on individual fluxes}
\begin{figure*}[h!]
	\centering
	\includegraphics[width=\linewidth]{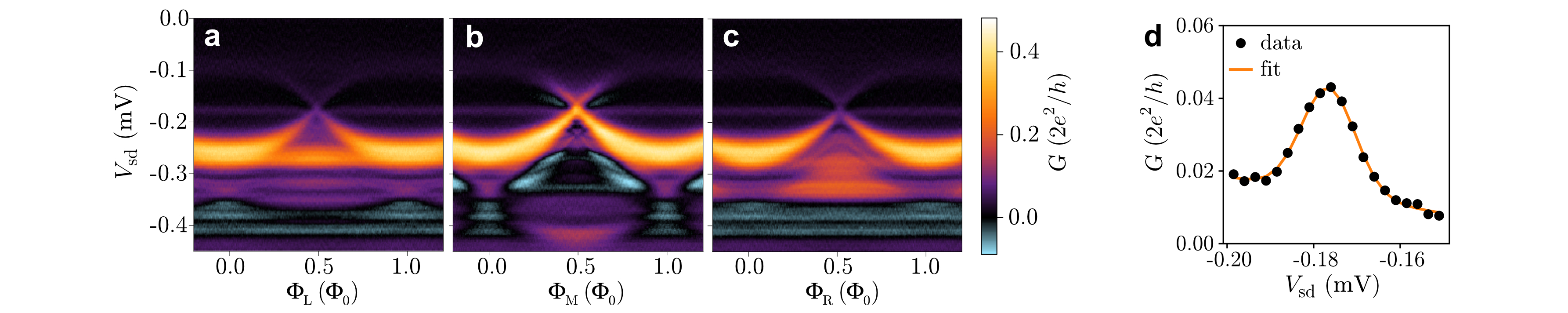}
	\caption{(a-c) Tunneling spectra of the two-terminal Andreev bound states, measured by sweeping one magnetic flux while keeping the other two to zero. All the three states show a similar dispersion and high transmission. (d) Fit of the conductance peak associated to multiple Andreev reflection, measured at $\Phi_{\rm{L}}=\Phi_{\rm{M}} = \Phi_{\rm{R}}=0$, using a Gaussian function and a linear background. The full width at half maximum of the Gaussian is $15~\mu \rm{V}$ and gives an estimate of the energy resolution of tunneling spectroscopy in our device.}
	\label{figS3}
\end{figure*}

The four-terminal device under study behaves as an effective two-terminal system when one flux is swept and the other two are kept at zero. Figure~\ref{figS3}(a-c) shows the energy dispersion as a function of $\Phi_{\rm{L}}$, $\Phi_{\rm{M}}$ and $\Phi_{\rm{R}}$, respectively, in these effective two-terminal configurations. The three high-transmission ABSs span similar energy ranges. In all three cases, low-transmission modes are visible at bias voltages $V_{\rm{sd}}$ below approximately $-0.3~\rm{mV}$. The Andreev dispersion along $\Phi_{\rm{M}}$ has a larger spectral weight than the other two, likely because the T2-T3 junction is positioned directly in front of the tunneling probe.

We estimate the energy resolution from the full width at half maximum (FWHM) of the multiple Andreev reflection (MAR) peak at $V_{\rm{sd}} = -0.175~\rm{mV}$, as displayed in Fig.~\ref{figS3}(d) for $\Phi_{\rm{L}}=\Phi_{\rm{M}} = \Phi_{\rm{R}}=0$. The data is well fit by a Gaussian function having a $\rm{FWHM} = 15~\mu \rm{V}$ with a linear background. This indicates that the resolution of our probe is comparable to state-of-the-art tunneling probes using gate-tunable constrictions and significantly higher to those employing insulating barriers.

\clearpage

\setcounter{myc}{3}

\section{Tunneling spectra along $\Phi_{10\overline{1}}$ and $\Phi_{1\overline{1}0}$}
\begin{figure*}[h!]
	\centering
	\includegraphics[width=0.6\linewidth]{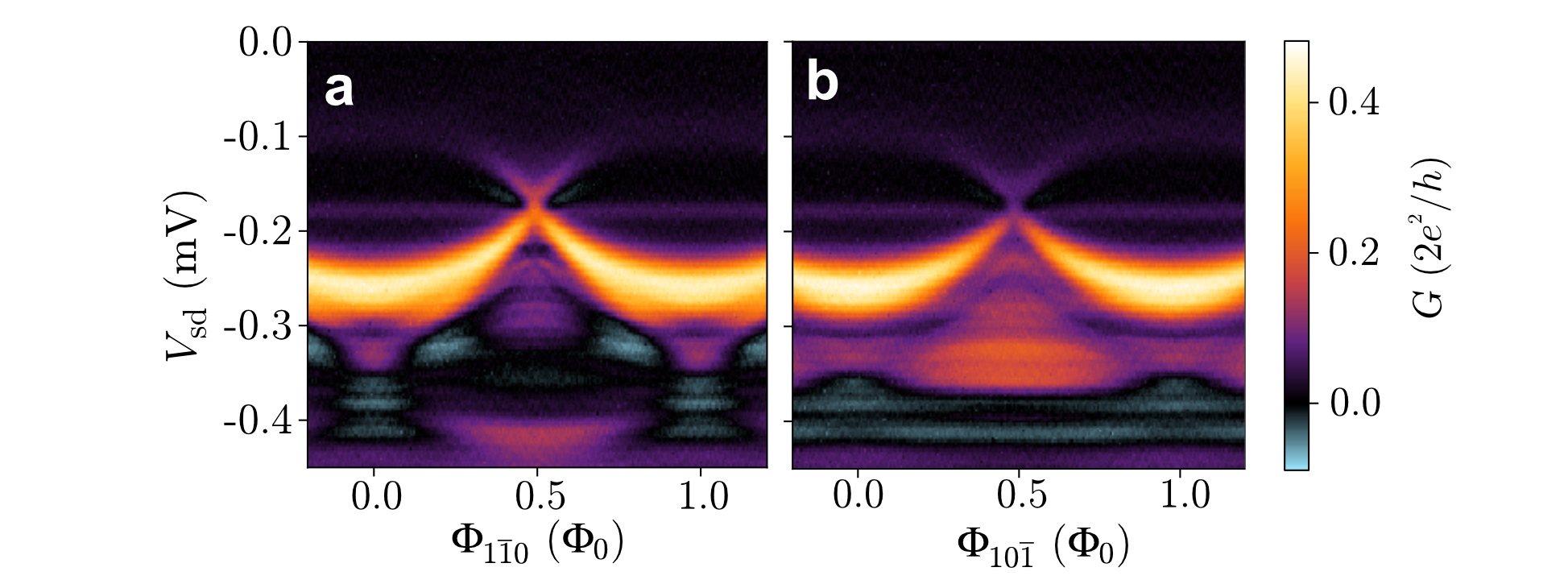}
	\caption{(a,b) Tunneling spectra measured along the flux directions $\Phi_{1\bar{1}0}$ and $\Phi_{10\bar{1}}$, respectively. Along these directions, the energy splitting induced by the hybridization between two Andreev bound states is expected to be of the order of the minigap, which, in turn, is smaller than the experimental resolution.}
	\label{figS4}
\end{figure*}

The hybridization between two ABSs is expected to produce different avoided crossings when the two phases are swept in the same or opposite directions \cite{Pillet2019}. An energy splitting is resolved along $\Phi_{101}$ and $\Phi_{110}$ [see Fig.~2(f,g) of the Main Text]. In Fig.~\ref{figS4}, we report the spectra measured along the two orthogonal directions, namely $\phi_{10\overline{1}}$ and $\phi_{1\overline{1}0}$, where the energy splitting is expected to be of the order of the minigap (that is, the energy gap between the hole-like and electron-like branches of a two-terminal ABS due to its finite transparency). Since the minigap in our device is smaller than the spectral broadening, such splitting is not resolved.
The anisotropy of the ABS dispersion along the two orthogonal directions observed in our 4TJJ is also in agreement with previous studies of bi-Andreev molecules \cite{Coraiola2023}.

\setcounter{myc}{4}

\section{Comparison between $\mid$L$\rangle$ and $\mid$R$\rangle$}

We observe that the ABS formed between the terminals T1-T2 $\mid$L$\rangle$ has a slightly larger transparency than the state $\mid$R$\rangle$ formed between T3-T4. In Fig.~\ref{figS3}, we show the $V_{\rm{sd}}$ dependence of the $\Phi_{\rm{L}}$-$\Phi_{\rm{R}}$ map measured at $\Phi_{\rm{M}} = 0$. At  $V_{\rm{sd}} = -175~\mu$V, the map displays a vertical and a horizontal conductance line, representing the maxima of $\mid$L$\rangle$ and $\mid$R$\rangle$, respectively. By setting $V_{\rm{sd}}$ more negative, these lines gradually split, as the dispersions of both ABSs are cut twice at voltages below their maxima. The vertical line representing $\mid$L$\rangle$ splits earlier (at $V_{\rm{sd}} = -183~\mu$V) than the horizontal line, $\mid$R$\rangle$, which instead develops a clear splitting only at $-186~\mu$V. Therefore, we conclude that $\mid$L$\rangle$ has a slightly larger transparency compared to $\mid$R$\rangle$, thus explaining the splitting observed in Fig.~3(a) and the asymmetry between the two diagonals in Fig.~5(e) of the Main Text.

\clearpage

\setcounter{myc}{5}
\begin{figure*}[h!]
	\centering
	\includegraphics[width=\linewidth]{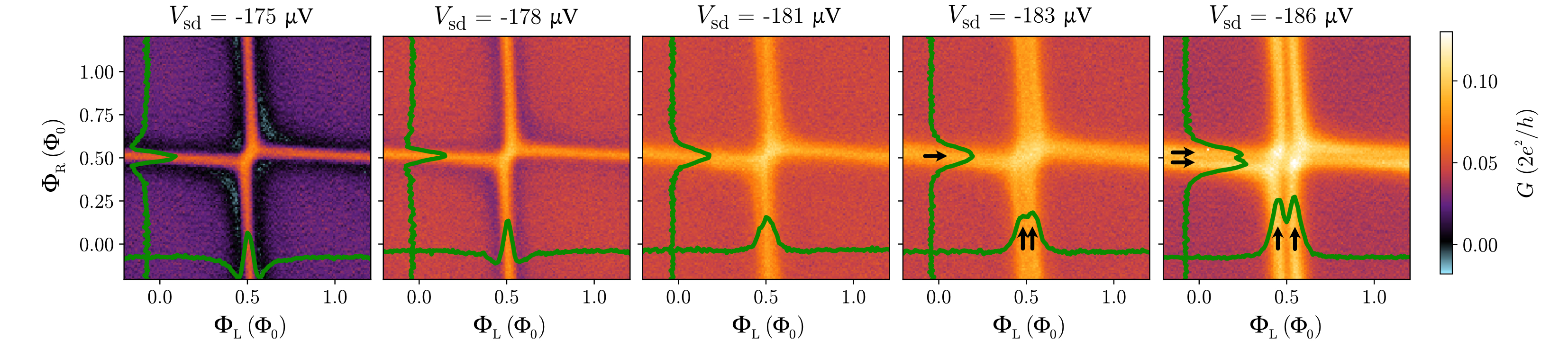}
	\caption{Dependence of the conductance map $\Phi_{\rm{L}}$-$\Phi_{\rm{R}}$ on the voltage bias $V_{\rm{sd}}$, measured at $\Phi_{\rm{M}} = 0$. As the bias is decreased, the vertical line start splitting at $-183~\mu$V, while the splitting of the horizontal line is clearly resolved only at slightly lower voltage ($-186~\mu$V). This indicates a slightly different transparency between the $\mid$L$\rangle$ and $\mid$R$\rangle$ Andreev bound states.}
	\label{figS5}
\end{figure*}

\section{Theoretical model}

This section is meant to further discuss the model used to explain the experimental findings in the Main Text. We simulate the ABS spectrum using a minimal model displayed in Fig.~4(a) of the Main Text, comprising four superconducting leads with superconducting phases $\phi_j$ ($j = 1,2,3,4$), respectively, and three quantum dots. Because of gauge invariance, we specify the following three superconducting phase differences $\phi_{\rm L} = \phi_2-\phi_1$, $\phi_{\rm M} = \phi_3-\phi_2$ and $\phi_{\rm R} = \phi_4-\phi_3$, in accordance with the three magnetic fluxes $\Phi_{\rm L,M,R}$ which are varied in the experiment. We choose three dots motivated by the observation of three high-energy ABSs in the experiment. In particular, the geometry is chosen such that each quantum dot gives rise to two-terminal ABSs between leads 1-2, 2-3 and 3-4, which disperse with their corresponding phase differences $\phi_{\rm L}$, $\phi_{\rm M}$ and $\phi_{\rm R}$, respectively, when all other phase differences are zero.

We consider a coupling strength $\Gamma$ between the leads and the dots [see Fig.~4(a) of the Main Text], and we ignore any possible anisotropy between the couplings. This choice is justified by the following arguments: first, we strive for a model with the least number of parameters that still quantitatively captures the experimental features. Secondly, as the model is suitable for comparison with the experiment, any asymmetry introduced between the couplings would be arbitrary. Lastly, the topological aspects of the system discussed in the Main Text are not due to any symmetry of parameters (including the couplings), meaning that they are stable under small variations of the parameters. Therefore, the symmetric model leads to a more insightful discussion about the topology and also captures the physics of a system with small anisotropy.

The experimental spectra [see, for example, Fig.~2(a,b) of the Main Text and Fig.~\ref{figS3}] highlight three high-transmission ABSs ($\tau \lesssim 1$), and additional lower transmission ABSs. Numerous low-transmission states occupy the spectrum near the superconducting gap edge and, due to level repulsion, these states spread over a finite energy range \cite{Yeyati2022}. This, in turn, effectively pushes the high-transmission states further from the gap edge. Modeling the many low-transmission states and their level repulsion requires an unreasonable number of fitting parameters and goes beyond the scope of this work. Instead, we proceed with an effective model where each dot has two energy levels, corresponding to the high- and low-transmission states, respectively. The repulsion of the high-transmission states from the gap edge is accounted for by renormalizing the coupling strength to the superconducting leads. In addition, we do not include level repulsion between the two dot levels, meaning that the transport through each combination of levels is considered as independent. The on-site energy $\epsilon$ of each dot level determines the effective transmission of the two-terminal JJs formed between pairs of leads. As all couplings $\Gamma$ are chosen to be equal, the effective transmission of the two-terminal JJs is $1$ for $\epsilon=0$ (where the two-terminal ABSs reach zero energy) and $< 1$ for $|\epsilon|>0$ (where a finite minigap opens between the ABSs). The on-site energies of the low- and high-transmission channels in each dot are chosen such that $|\epsilon_{\rm low}|>|\epsilon_{\rm high}|>0$.
The dots are symmetrically coupled to each other with the hybridization strength $t$. We note that the experiments show a weaker avoided crossing between the $|\rm{L}\rangle$ and $|\rm{R}\rangle$ ABSs compared to $|\rm{L}\rangle$-$|\rm{M}\rangle$ and $|\rm{M}\rangle$-$|\rm{R}\rangle$, which might naively be attributed to a weaker coupling between the left and right dots. However, the geometry of the model is sufficient to explain the weaker avoided crossing, without introducing different hybridization strengths.

The coupling strengths $\Gamma = 0.37\Delta$ are chosen in accordance with the reduced energy range of the ABS dispersion, experimentally observed in Fig.~2(a,b) of the Main Text. The on-site energies for the high- and low-transmission ABSs are $\epsilon_{\rm high} = 0.005\Delta$ and $\epsilon_{\rm low} = 1.5\Delta$, respectively, while $t= 0.12\Delta$. We observe that the exact value of $\epsilon_{\rm high}$ does not affect the form of the corresponding ABSs significantly, as long as the effective two-terminal transmission is $\lesssim 1$. As level repulsion between high- and low-transmission states is negligible close to zero energy, the low-transmission ABSs do not influence the topological properties of the system.

\begin{figure}
	\centering
	\includegraphics[width=0.5\linewidth]{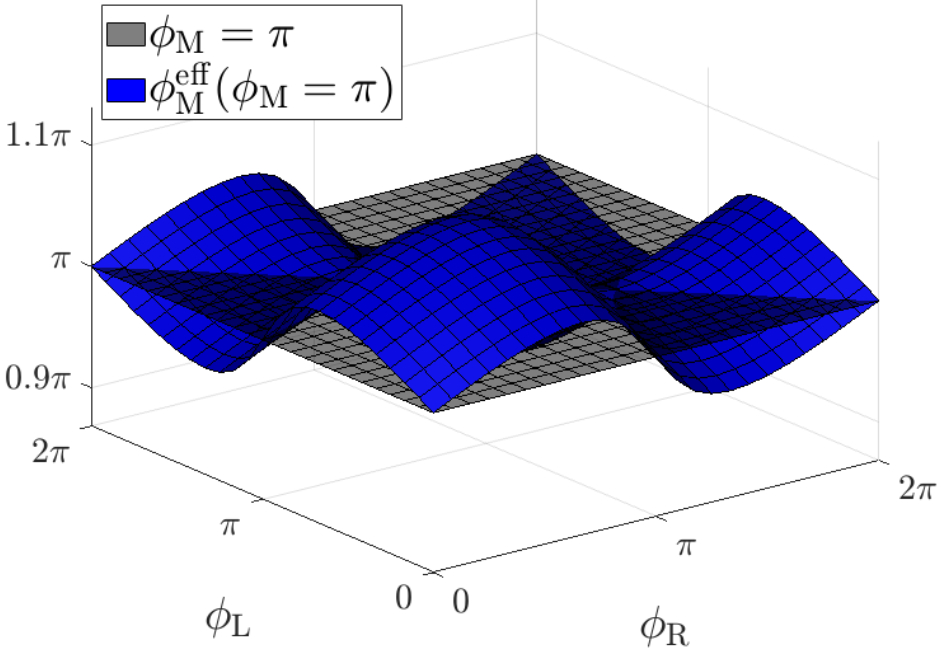}
	\caption{Cuts of the phase space at constant $\phi_{\rm M}=\pi$ (gray) and the surface of the corresponding effective phase $\phi_{\rm M}^{\rm eff}(\phi_{\rm M}= \pi)$ (blue). Due to the mutual inductive coupling in the phases (fluxes) (see Sec.~V of the Main Text), the effective phase (flux) surface is not constant.}
	\label{FigS6}
\end{figure}

We compute the density of states (DOS) of the central region as a function of energy and of the superconducting phases $\phi_{\rm L,M,R}$ (or equivalently the bare magnetic fluxes $\Phi_{\rm L,M,R}$). Whereas the DOS is portrayed as a function of bare phases (bare fluxes), it depends on the effective phases (effective fluxes), as discussed in Sec.~V of the Main Text. Thus, the final DOS used to compare with the experimental findings reads as
\begin{equation}
    \rho\left(\phi_{\rm L}^{\rm eff}(\phi_{\rm L},\phi_{\rm M}),\phi_{\rm R}^{\rm eff}(\phi_{\rm R},\phi_{\rm M}),\phi_{\rm M}^{\rm eff}(\phi_{\rm M},\phi_{\rm L},\phi_{\rm R})\right). 
\end{equation}
In the main text, the DOS is portrayed as a function of magnetic fluxes by setting $\phi_{i} = 2\pi\Phi_{i}/\Phi_0$ ($i = {\rm L,M,R}$). In the following, we illustrate the conversion from superconducting phases (magnetic fluxes) to effective phases (effective magnetic fluxes). We plot a surface of constant phase $\phi_{\rm M}=\pi$ and a surface of the corresponding effective phase $\phi_{\rm M}^{\rm eff}(\phi_{\rm M}= \pi)$ in the phase space (Fig.~\ref{FigS6}). The former is simply a plane parallel to the $\phi_{\rm L}$-$\phi_{\rm R}$ plane, while the latter has a more complex shape in the phase space.

\setcounter{myc}{6}

\section{Green's function formalism and topological Hamiltonian} 

In the following, we provide the microscopic model from which we infer the central region Green's function and from that the topological Hamiltonian used in Sec.~VII of the Main Text. As discussed in the previous section, our system comprises four superconducting leads coupled to three quantum dots, each of them hosting two energy states representing the high- and low-transmission ABSs. Since these two energy levels are not interacting with each other, we can independently compute the DOS derived from each ABS manifold. The Hamiltonian for one set of levels reads 
\begin{align}
	H=&\sum_{j=1,2,3,4} H_{(\text{s},j)}+H_{\rm dd} \\
	&+\sum_{j= 1,2,3,4/\alpha = {\rm L,M,R}}H_{( {\rm s},j)-({\rm dd},\alpha)},
\end{align}
	with $H_{(\text{s},j)}$ the Hamiltonian of the $j$th superconducting lead, $H_{\rm dd}$ the Hamiltonian describing the three quantum dots labelled as $\alpha \in \{\mathrm{L},{\rm M},\mathrm{R}\}$ and $H_{( {\rm s},j)-({\rm dd},\alpha)}$ representing the interaction between the leads and the quantum dots.
	
	The first term takes the following BCS form:
	\begin{align}
		H_{(\mathrm{s},j)} 
		= 
		\sum_{\bm{k}\sigma}\epsilon_{\bm{k}} c_{j\bm{k}\sigma}^\dagger c_{j\bm{k}\sigma}^{\phantom\dagger}
		+
		\sum_{\bm{k}} \Bigl( \Delta e^{i\phi_{j}}c_{j\bm{k}\uparrow}^\dagger c_{j(-\bm{k})\downarrow}^\dagger+ \mathrm{h.c.} \Bigr),
	\end{align}
	where $\phi_j$ is the superconducting phase of the $j$th terminal, $c_{j\bm{k}\sigma}^\dagger$ the creation operator of an electron in the $j$th lead with momentum $\bm{k}$, spin $\sigma$ and superconducting gap $\Delta$ (which is chosen the same for each terminal). Considering only one spinful energy state per quantum dot, $H_{\mathrm{dd}}$ reads
	\begin{align}
		H_{\mathrm{dd}}
		&=
		\sum_{ \alpha = \mathrm{L,M,R}} \sum_{ \sigma} \epsilon_\alpha d_{\alpha\sigma}^\dagger d_{\alpha\sigma}^{\phantom\dagger} +
		\sum_{\alpha\neq\beta={\rm L,M,R}}\sum_{\sigma} t_{\alpha \beta} \bigl( d_{\mathrm{\beta}\sigma}^\dagger d_{\mathrm{\alpha}\sigma}^{\phantom\dagger} + \mathrm{h.c.} \bigr)   ,
		\label{eq:bareDoubleDot}
	\end{align} 
	where $t_{\alpha\beta}$ is the hopping between dots $\alpha$ and $\beta$, and $d_{\alpha\sigma}^\dagger$ is the creation operator of an electron on the dot $\alpha$ with energy $\epsilon_\alpha$ and spin $\sigma =  \uparrow,\downarrow$.
	The coupling between the energy levels and the superconducting terminals is described by following tunneling Hamiltonian:
	\begin{align}
		H_{(\text{s},j)-\rm{(dd,\alpha)}} &= \sum_{\bm{k}\sigma, \alpha} v_{j,\alpha}d_{\alpha\sigma}^\dagger c_{j\bm{k}\sigma}^{\phantom\dagger}+\mathrm{h.c.} \, ,
	\end{align}
	with the hopping amplitude $v_{j,\alpha}$ between the $j$th superconducting lead and the $\alpha$th energy level.
	The Dyson equation for the central region Green's function reads
	\begin{align}
		G_{\text{dd}}^{\rm r/a}=g_{\text{dd}}^{\rm r/a}+g_{\text{dd}}^{\rm r/a}\Sigma^{\rm r/a} G_{\text{dd}}^{\rm r/a}
	\end{align}
	where $G_{\text{dd}}^{\rm r/a}$	are the retarded/advanced dressed Green's functions, $g_{\text{dd}}^{\rm r/a}$ the unpertubed Green's functions, and $\Sigma^{\rm r/a}$ the self energies due to the coupling to the superconducting leads, which are defined as:
	\begin{equation}
		\Sigma^{\rm r/a} =\sum_{j = 1,2,3,4/ \alpha = \rm L,M,R}V_{j,\alpha }^\dagger g_{(\rm{s},j)}^{\rm r/a}V_{j,\alpha }.
	\end{equation}
	The bare Green's function of the superconducting leads in spin-Nambu space is given by
	\begin{align}
		g_{(\rm{s},j)}(E) = - \frac{\pi N_0}{\sqrt{\Delta^2-E^2}} \,  \sigma_0 \otimes  \left(E\tau_0 + \Delta e^{i\phi_j\tau_3}\tau_1 \right),
	\end{align}
	with the Pauli matrices $\sigma_j$, $\tau_j$ in spin and Nambu space respectively, $N_0$ is the DOS in the normal state, and $g_{\mathrm{sj}}^{r/a}=g_{\mathrm{sj}}(E\pm i\eta)$ with the Dynes' parameter $\eta$. Since the hoppings do not depend on the quasimomentum $\bm{k}$, we have integrated out $\bm{k}$ to obtain an effective Green's function of the superconducting leads.
	The unperturbed Green's function of the dot is given by
	\begin{align}
		g_{\text{dd}}^{\rm r/a}(E)=(E\pm i\eta-H_{\text{dd}})^{-1}.
	\end{align}
	The different coupling terms $V_{j,\alpha}$ read
	\begin{align}
		V_{1, \rm L} &\equiv v_{1, \rm L}\sigma_0\otimes\tau_3 \\
		V_{2, \rm L} &\equiv v_{2, \rm L}\sigma_0\otimes\tau_3 \\
		V_{2, \rm M} &\equiv v_{2, \rm M}\sigma_0\otimes\tau_3 \\
		V_{3, \rm M} &\equiv v_{3, \rm M}\sigma_0\otimes\tau_3 \\
		V_{3, \rm R} &\equiv v_{3, \rm R}\sigma_0\otimes\tau_3 \\
		V_{4, \rm R} &\equiv v_{4, \rm R}\sigma_0\otimes\tau_3,
	\end{align}
	and all other components are zero. 
	The DOS is obtained by
	\begin{equation}
		\rho = -{\rm Im} \ {\rm tr}\  G_{\rm dd}^{\rm r}.
	\end{equation}
	Finally, the total DOS is the sum of DOS of the high- and low-transmission ABSs:
	\begin{equation}
		\rho_{\rm total} = \rho_{\rm high}+\rho_{\rm low}.
	\end{equation}
	For the zero energy limit $E = 0$, the topological information of the system is contained in the so-called topological Hamiltonian $H_{\rm top}$, which reads
	\begin{align}
		H_{\text{top}}=G_{\rm dd}^{-1}(E = 0) = H_{\rm dd}-\Sigma(E = 0).
	\end{align}
	The Green's functions of the superconductors become $g_{\text{s},j}=-\pi N_0 \, ( \sigma_0 \otimes   e^{i\phi_j\tau_3}\tau_1 )$ in the case of $E = 0$. Additionally, upon introducing the couplings $\Gamma_{j,\alpha} = \pi N_0v_{j,\alpha}^2$, we obtain the following self energy:
		\begin{align}
			\Sigma(E = 0) = (\sigma_0\otimes \tau_1)\begin{pmatrix}
				\Gamma_{1,\rm L}e^{i \phi_1\tau_3}+\Gamma_{2,\rm L} e^{i \phi_2\tau_3} & 0 & 0 \\ 
				0 &\Gamma_{2, \rm M}e^{i \phi_2\tau_3}+\Gamma_{3, \rm M} e^{i \phi_3\tau_3} & 0 \\
				0 & 0 & \Gamma_{3, \rm R}e^{i \phi_3\tau_3}+\Gamma_{4, \rm R} e^{i \phi_4\tau_3}
			\end{pmatrix}.
		\end{align}
	which then results in the topological Hamiltonian:
		\begin{equation}
			H_{\text{top}} = \sigma_0\begin{pmatrix}
				\epsilon_{\rm L}\tau_3- \tau_1(\Gamma_{1, \rm L}e^{i \phi_1\tau_3}+\Gamma_{2, \rm L} e^{i \phi_2\tau_3}) & t_{\rm LM}\tau_3 & t_{\rm LR}\tau_3 \\ 
				t_{\rm LM}\tau_3 &\epsilon_{\rm M}\tau_3-\tau_1(\Gamma_{2, \rm M}e^{i \phi_2\tau_3}+\Gamma_{3, \rm M} e^{i \phi_3\tau_3}) & t_{\rm MR}\tau_3 \\
				t_{\rm LR}\tau_3 & t_{\rm MR}\tau_3 & \epsilon_{\rm R} \tau_3 - \tau_1(\Gamma_{3, \rm R}e^{i \phi_3\tau_3}+\Gamma_{4, \rm R} e^{i \phi_4\tau_3})
			\end{pmatrix}.
		\end{equation}

	We neglect the contributions from the low-transmission ABSs, as they do not influence the topology, thus $H_{\rm top}= H_{\rm top, high}$. For $E = 0$, the superconducting gap $\Delta$ drops out of the expression for the superconducting Green's functions and thus does not appear in the topological Hamiltonian. Therefore, we choose one coupling $\Gamma_{\rm L}$ as the unit of energy and normalize all parameters with respect to it. In agreement with the DOS calculations, where all couplings $\Gamma_{j,\alpha} = \Gamma$, hybridizations $t_{\alpha,\beta} = t$ and dot energies $\epsilon_{\alpha}= \epsilon$ are equal, the topological Hamiltonian only depends on two parameters: the dimensionless dot-hybridization $\tilde{t}\equiv t/\Gamma$ and the dimensionless on-site energy $\tilde{\epsilon} = \epsilon/\Gamma$.
	
	\setcounter{myc}{7}
	\begin{figure}
		\centering
		\includegraphics[width=1\linewidth]{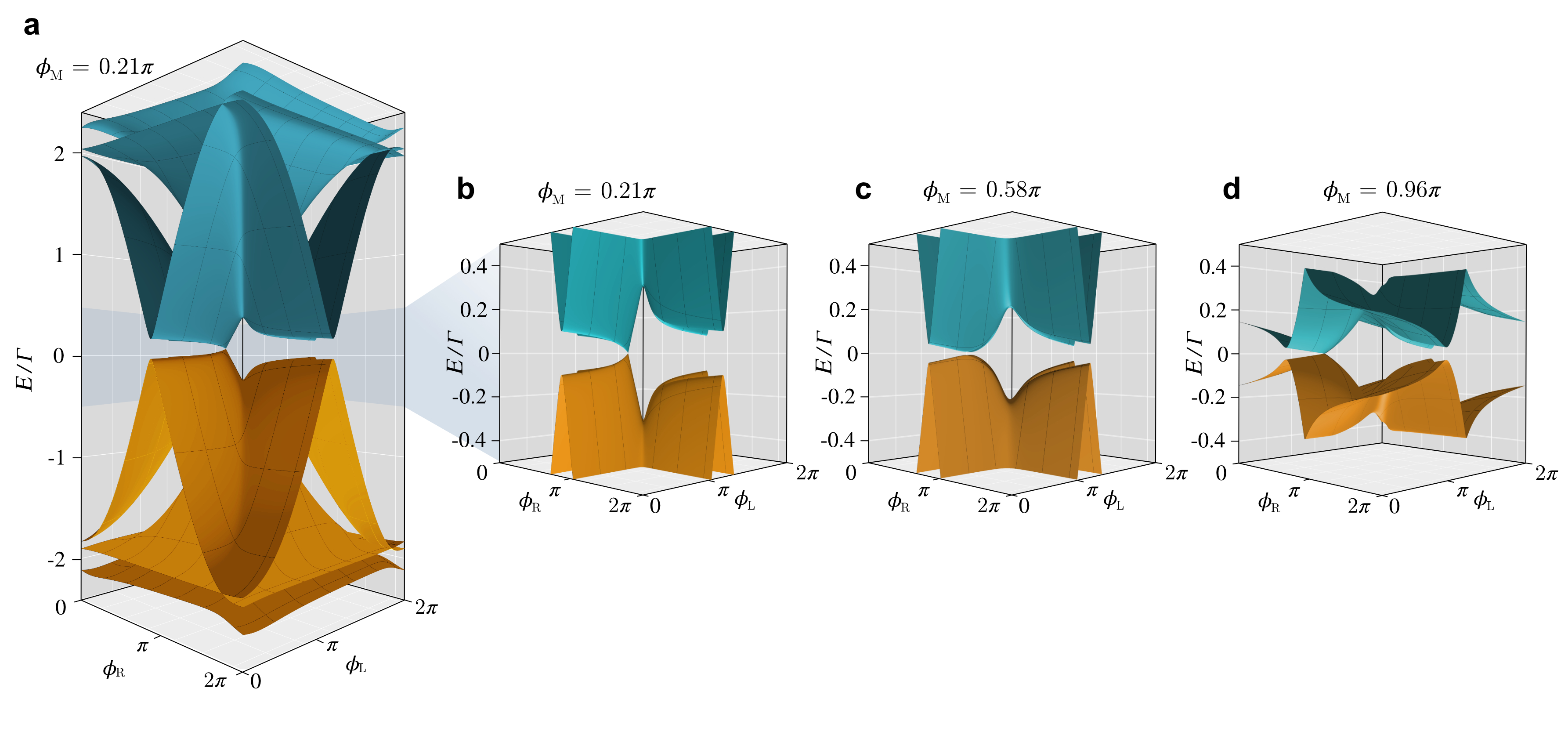}
		\caption{(a) Eigenenergies of the topological Hamiltonian $H_{\rm top}$ calculated as a function of the phase differences $\phi_{\rm L}$ and $\phi_{\rm R}$ for fixed $\phi_{\rm M}=0.21 \pi$. The energy gap between the electron- and hole-like branches vanishes at the Weyl node. (b) Enlargement of (a) around zero energy, highlighting the Weyl node dispersion. (c) As in (b), but for $\phi_{\rm M}=0.58 \pi$, revealing a topological gap between the bands. (d) As in (b), but for $\phi_{\rm M}=0.96 \pi$, showing a second Weyl node with opposite topological charge where the gap closes. All calculations were performed with the input parameters $\tilde{\epsilon} \approx 0.01$ and $\tilde{t} \approx 0.32$, in agreement with the simulations of the spectra in Figs.~4 and 5. The spectra closely resemble the corresponding bands extracted from the maxima of the density of states in Fig.~6(a-d) of the Main Text, respectively.}
		\label{FigS7}
	\end{figure}
	
	The ABS energies plotted in Fig.~6 of the Main Text are extracted by the local maxima of $\rho_{\rm{high}}$ when the Dynes' parameter goes to zero, $\eta \to 0$. We find that these ABSs exhibit zero-energy crossings and corresponding Weyl nodes when the Chern number changes, as illustrated in the Main Text. For completeness, we show in Fig.~\ref{FigS7}(a-d) the energy spectrum of the topological Hamiltonian $H_{\rm top}$ that closely matches the one extracted from $\rho_{\rm{high}}$ shown in Fig.~6(a-d) of the Main Text. The two numerical methods provide very similar band structures that are in one-to-one correspondence at exactly zero energy. Therefore, the topological Hamiltonian is sufficient to determine topological properties of the system.

\section{Second device}
\begin{figure}
    \centering
    \includegraphics[width=\linewidth]{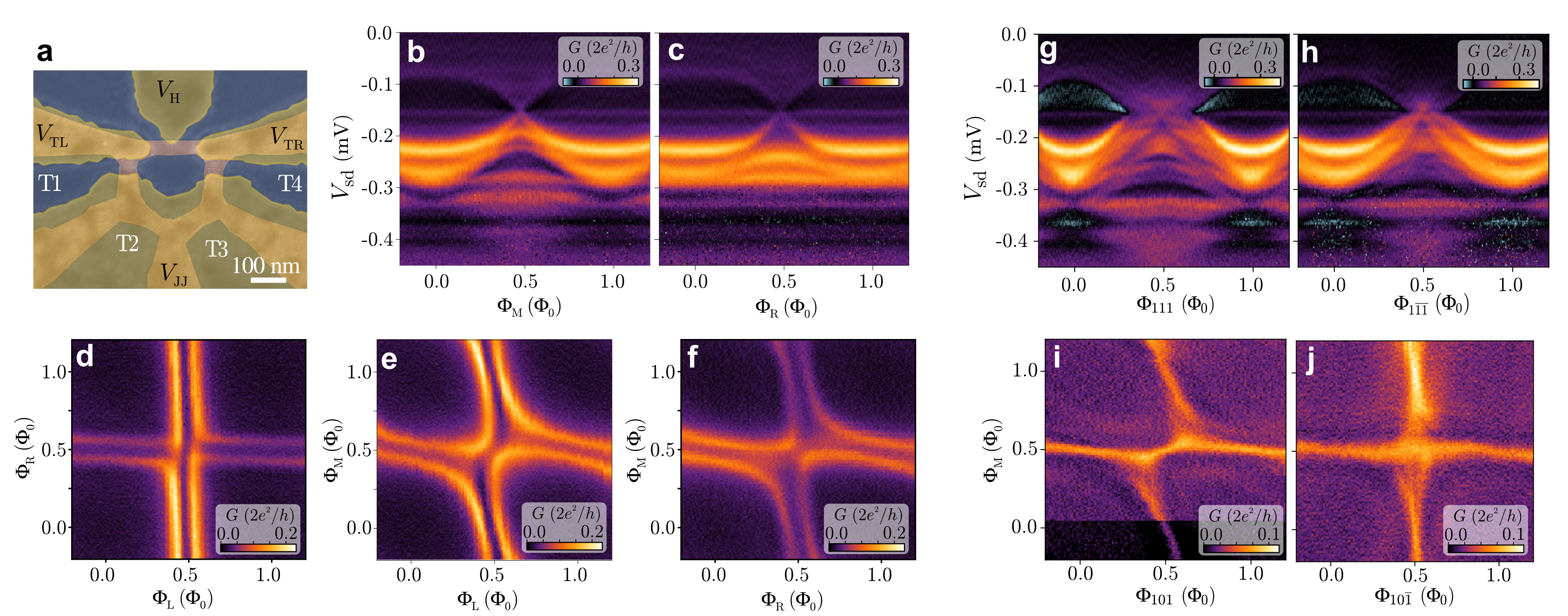}
    \caption{(a) False-colored scanning electron micrograph of Device 2, showing a slightly different scattering area compared to Device 1 (shown in the Main Text). (b,c) Tunneling spectra of two-terminal Andreev bound states (ABSs) as a function of magnetic flux $\Phi_{\rm{M}}$ for $\Phi_{\rm{L}}=\Phi_{\rm{R}}=0$ (b), and as a function of $\Phi_{\rm{R}}$ for $\Phi_{\rm{L}}=\Phi_{\rm{M}}=0$ (c). (d-f) Flux-flux maps measured at $V_{\rm{sd}} = -0.182~\rm{mV}$, revealing avoided crossings induced by the hybridization between ABSs. (g,h) Tunneling spectra along the cube diagonal directions $\Phi_{111}$ (g) and $\Phi_{1\bar{1}\bar{1}}$ (h), where all three ABSs hybridize as discussed in Sec.~VI of the Main Text. (i,j) Flux-flux maps measured at $V_{\rm{sd}} = -0.157$~meV along two cube diagonal planes, $\Phi_{\rm{M}}-\Phi_{\rm{101}}$ and $\Phi_{\rm{M}}-\Phi_{\rm{10\bar{1}}}$, respectively. As in the datasets showing in Fig.~5(a,e) of the Main Text for Device 1, the Andreev band structure measured in Device 2 shows the characteristic features indicating the formation of a tri-Andreev molecule.}
    \label{FigS8}
\end{figure}

In this section, we present tunneling spectroscopy measurements performed on a second device. The geometry of Device 2 is similar to that discussed in the Main Text for Device 1, but with a slightly different scattering area, as shown in Fig.~\ref{FigS8}(a). The aluminum island at the center of the scattering region has dimension $260 \times 200~\rm{nm}^2$, larger than in Device 1. Also, the gate tuning the InAs region between the junctions, energized by the voltage $V_{\rm{JJ}}$, has a different shape, partially covering the superconducting island. Gate voltages are set to $V_{\rm{H}} = 110~\rm{mV}$, $V_{\rm{JJ}} = -350~\rm{mV}$ and $V_{\rm{TL}} = V_{\rm{TR}} = -1.419~\mathrm{V}$ to form a tunneling barrier between the superconducting probe and the 4TJJ. Figure \ref{FigS8}(b,c) shows the ABS dispersion of $\mid$M$\rangle$ and $\mid$R$\rangle$, respectively. A larger number of states is visible in this device compared to Device 1 [see Fig. 2(a,b) of the Main Text]. Nevertheless, discrete states are still visible in the spectrum, with one of them having near-unity transparency and reaching energy close to zero at $0.5\Phi_0$. The superconducting probe gap is slightly smaller, $0.157~\rm{meV}$ compared to $0.175$~meV in Device 1. Cutting the energy spectrum at $V_{\rm{sd}} = -0.182$~meV [Fig.~\ref{FigS8}(d-f)], we observe flux-flux conductance maps with avoided crossings at the center of the map as a result of the hybridization between two ABSs, as also reported in Fig.~2(d,e) of the Main Text.

We also verified the formation of a tri-Andreev molecule by measuring the energy spectrum along the cube diagonal directions $\Phi_{111}$ and $\Phi_{1\bar{1}\bar{1}}$, as shown in Fig.~\ref{FigS7}(g,h). Both dispersions are similar to those discussed in the Main Text [see Fig.~5(c,g)]. Moreover, the maps measured along the cube diagonal planes in Fig.~\ref{FigS8}(i,j) show the same characteristic signatures reported in Fig.~5(a,e) of the Main Text for Device 1.

\setcounter{myc}{8}

\end{document}